\documentclass{lmcs}
\pdfoutput=1

\usepackage{lastpage}
\lmcsdoi{20}{4}{13}
\lmcsheading{}{\pageref{LastPage}}{}{}%
{May~16,~2022}{Nov.~12,~2024}{}


\usepackage[utf8]{inputenc}
\usepackage{amssymb,amsmath,amstext}
\usepackage{hyperref, graphicx}
\usepackage{array}
\usepackage{tabularx}

\theoremstyle{definition}
\theoremstyle{definition}
\theoremstyle{plain}
\theoremstyle{definition}
\theoremstyle{definition}
\theoremstyle{definition}

\newcommand{\mr}{\mathrm}

\newcommand{\mb}{\mathbf}

\newcommand{\es}{\emptyset}

\newcommand{\uhr}{\upharpoonright}

\newcommand{\nts}{\negthickspace}
\newcommand{\uhrc}{\nts \upharpoonright \nts}

\newcommand{\mcA}{\mathcal{A}}
\newcommand{\mcB}{\mathcal{B}}

\newcommand{\mcG}{\mathcal{G}}

\newcommand{\mcL}{\mathcal{L}}

\newcommand{\mff}{\mathfrak{f}}
\newcommand{\mfg}{\mathfrak{g}}

\newcommand{\mbE}{\mathbf{E}}
\newcommand{\mbF}{\mathbf{F}}

\newcommand{\mbW}{\mathbf{W}}
\newcommand{\mbX}{\mathbf{X}}
\newcommand{\mbY}{\mathbf{Y}}
\newcommand{\mbZ}{\mathbf{Z}}

\newcommand{\mbbG}{\mathbb{G}}
\newcommand{\mbbP}{\mathbb{P}}
\newcommand{\mbbN}{\mathbb{N}}
\newcommand{\mbbR}{\mathbb{R}}

\newcommand{\msfP}{\mathsf{P}}
\newcommand{\msfC}{\mathsf{C}}

\newcommand{\rng}{\mathrm{rng}}

\begin{document}

\title[Asymptotic expressivity of inference frameworks]{On the relative asymptotic expressivity of inference frameworks\rsuper*}
\titlecomment{{\lsuper*} ACM Class F.4.1. The research leading to this publication was partially conducted during a visit of the second author to the University of Uppsala, enabled by LMUExcellent and funded by the Federal Ministry of Education and Research (BMBF) and the Free State of Bavaria under the Excellence Strategy of the Federal Government and the Länder.}

\author[V.~Koponen]{Vera Koponen\lmcsorcid{0000-0002-9838-3403}}[a]
\author[F.~Weitk\"amper]{Felix Weitk\"amper\lmcsorcid{0000-0002-3895-8279}}[b]

\address{Department of Mathematics, Uppsala University, Sweden.}
\email{vera.koponen@math.uu.se}

\address{Institut für Informatik,
  Ludwig-Maximilians-Universität München, Munich, Germany.}
\email{felix.weitkaemper@lmu.de}

\begin{abstract}
We consider logics with truth values in the unit interval $[0,1]$.
Such logics are used to define queries and to define probability distributions.
In this context
the notion of almost sure equivalence of formulas is generalized to the notion of asymptotic equivalence.
We prove two new results about the asymptotic equivalence of formulas where each result has a convergence law as a corollary.
These results as well as several older results can be formulated as results about the relative asymptotic expressivity
of inference frameworks. 
An inference framework $\mbF$ is a class of pairs $(\mbbP, L)$, where 
$\mbbP = (\mbbP_n : n = 1, 2, 3, \ldots)$, $\mbbP_n$ are probability distributions on the set $\mbW_n$ of
all $\sigma$-structures with domain $\{1, \ldots, n\}$ (where $\sigma$ is a first-order signature)
and $L$ is a logic with truth values in the unit interval $[0, 1]$.
An inference framework $\mbF'$ is asymptotically at least as expressive as an
inference framework $\mbF$ if for every $(\mbbP, L) \in \mbF$ there is $(\mbbP', L') \in \mbF'$ such that 
$\mbbP$ is asymptotically total variation equivalent to $\mbbP'$ and for every $\varphi(\bar{x}) \in L$ there is $\varphi'(\bar{x}) \in L'$ such that 
$\varphi'(\bar{x})$ is asymptotically equivalent to $\varphi(\bar{x})$ with respect to $\mbbP$.
This relation is a preorder.
If, in addition, $\mbF$ is at least as expressive as $\mbF'$ then we say that $\mbF$ and $\mbF'$ are asymptotically equally 
expressive.
Our third contribution is to systematize the new results of this paper and several previous results in order to get a
preorder on a number of inference systems that are of relevance in the context of machine learning and artificial intelligence.
\end{abstract}

\maketitle

\section{Introduction}\label{Introduction}

\noindent
In modern artificial intelligence, logics with (truth) values in the unit inverval $[0, 1]$ 
are used not only as query languages, but also to define the probability distributions with respect to which queries are evaluated.
Such probability distributions can be defined by formalisms called {\em probabilistic graphical models (PGMs)} 
which are determined by
a finite graph (with a vertex for each relation symbol) and formulas that express (conditional) probabilities of individual relations.
We can view a formula $\varphi(x_1 \dots, x_n)$  in a probability logic $L(\sigma)$ over a signature $\sigma$ as a function taking a $\sigma$-structure $\mathcal{A}$ and a tuple $a_1, \dots, a_n \in A$ as input and returning a real number in $[0,1]$ as output (see Definition \ref{definition of logic}); this number
will be denoted $\mcA(\varphi(a_1, \ldots, a_n))$ and called the {\em value} of 
$\varphi(a_1, \ldots, a_n)$ in $\mcA$.
Then for any finite relational $\sigma$ we can define an {\em $L(\sigma)$-network} 
as a directed acyclic graph with vertex set $\sigma$, and for every vertex $R \in \sigma$ of arity $n$, an $L(\sigma)$-formula 
$\theta_R(x_1, \dots x_n)$ using only relation symbols among the parents of $R$ in the directed acyclic graph (see Definition \ref{definition of L-network}).
For any relation $R \in \sigma$, any finite domain $A$ and tuple $\bar{a}$ of elements from $A$ such that
its length matches the arity of $R$, the conditional probabilitity of $R(a_1, \dots, a_n)$, 
given the interpretation of all the parents of $R$, is specified by the value of 
$\theta_R(a_1, \ldots, a_n)$. 
Let $\sigma$ be a finite and relational signature and let $\mbW_n$ denote the set of all $\sigma$-structures with
domain $[n] = \{1, \ldots, n\}$.
In the way indicated above each $L(\sigma)$-network now defines a probability distribution on $\mbW_n$ for each $n$
(see Definition~\ref{the probability distribution induced by an L-network}). 

We will mostly work with a logic that we call $PLA^+$, or $PLA^+(\sigma)$ if we want to indicate the signature used,
which uses aggregation functions instead of quantifiers (see Definition~\ref{definition of aggregation function}). 
In this logic we can express all queries (on finite structures) that can be expressed by first-order logic
because the aggregation functions max and min can be used to express existential and universal quantification.
Of course, there are many more aggregation functions, for example the average, 
also called arithmetic mean (of a finite sequence of reals from $[0, 1]$).
With such aggregation functions $PLA^+$ can express, for example, each stage of the definitions of PageRank and SimRank
(more about this in Example~\ref{SimRank}).
We will use $PLA^+$ both as a query language and as a language to specify probability distributions on $\mbW_n$
via $PLA^+(\sigma)$-networks.

$PLA^+$ is a very expressive query language, and $PLA^+$-networks can define a great variety of probability distributions.
For this reason we do not expect to be able to prove general theorems for all $PLA^+$-queries
and all $PLA^+$-networks. 
Therefore we look for sublogics, say $L$ and $L'$, of $PLA^+$ such that 
(a) these logics are expressive enough to be useful, and (b)
we can prove general results
for probability distributions defined by $L$-networks and queries expressed by $L'$.
These sublogics will be defined mainly by restricting $PLA^+$ to formulas that only use
aggregation functions that have certain ``nice'' properties.
The most extreme case is to ban all aggregation functions, so let us call a $PLA^+$-formula
which does not use any aggregation function {\em aggregation-free}.

Let $\mbbG$ denote a $PLA^+(\sigma)$-network  and let $\mbbP_n$ denote the probability distribution on $\mbW_n$
that is determined by $\mbbG$. 
Our first main result, Theorem~\ref{elimination of strongly admissible aggregation functions}, 
is that if every formula associated to $\mbbG$ contains only 
continuous aggregation functions (in the sense of Definition~\ref{alternative definition of admissible function}),
then every $PLA^+(\sigma)$-formula that contains only continuous aggregation functions is 
asymptotically equivalent to an aggregation-free formula with respect to the sequence of distributions
$\mbbP = (\mbbP_n : n = ,1, 2, 3, \ldots)$.
The notion of ``asymptotic equivalence'' 
(Definition~\ref{definition of asymptotically equivalent formulas}) 
is a generalization of the notion of ``almost sure equivalence'' that
makes sense for logics with more than two truth values. Intuitively speaking, two formulas are
asymptotically equivalent if, with high probability, their values are almost the same.
Moreover, as stated in Corollary~\ref{corollary to main results}, 
if $\varphi(\bar{x})$ is a $PLA^+(\sigma)$ formula and $\mbbG$ is as above then
$\varphi(\bar{x})$ satisfies a convergence law with respect to $\mbbP$.
In the more technical Corollary~\ref{equivalence to an aggregation-free network}
we show how the mentioned results can be used to approximate probabilities of queries without 
any reference to the domain size.

Our second main result still considers a $PLA^+(\sigma)$-network $\mbbG$ such that all 
associated formulas use only continuous aggregation functions. However, as query language, we
use the two-valued {\em conditional probability logic (CPL)} introduced in \cite{Kop20}.
$CPL$ is an extension of first-order logic which allows constructions that can express statements like
``the relative frequency of $\bar{x}$ that satisfy $\varphi_1(\bar{x})$
among $\bar{x}$ that satisfy $\varphi_2(\bar{x})$ is at least (a constant) $c$, or alternatively,
is at least as large as $c$ plus the relative frequency of $\bar{x}$ that satisfy $\psi_1(\bar{x})$ among
$\bar{x}$ that satisfy $\psi_2(\bar{x})$'' where $c$ is a non-negative real. 
Here the conditions expressed by $\varphi_1(\bar{x})$ et cetera may
themselves be $CPL$-formulas, so this construction can be nested and viewed as a kind of quantification.
The second main result, Theorem~\ref{quantifier elimination of safe formulas}, 
is that if the  $CPL(\sigma)$-formula $\varphi(\bar{x})$ is {\em safe (with respect to $\mbbG$)} 
then $\varphi(\bar{x})$ is almost surely equivalent
to a quantifier-free first-order formula.
The condition that $\varphi(\bar{x})$ is {\em safe} roughly means that in every relative frequency statement as
described above that occurs in $\varphi(\bar{x})$, the constant $c$ does not belong to a certain finite set of numbers
that is determined by the syntactic structure of $\varphi(\bar{x})$ and by $\mbbG$.
We obtain two corollaries. The first is a convergence law for safe $CPL$-formulas.
The second roughly says that there is a $PLA^+(\sigma)$-network $\tilde{\mbbG}$ such that all associated $PLA^+$-
formulas are aggregation-free, and for every safe $CPL$-formula $\varphi(\bar{x})$ and for sufficiently large $n$,
the probability  that a sequence of parameters satisfies $\varphi(\bar{x})$ under the distribution $\tilde{\mbbP}_n$ determined by $\tilde{\mbbG}$ 
approximates the corresponding
probability under $\mbbP_n$ arbitrarily closely.
This implies that the first probability can be estimated, with as high accuracy as we like for large enough $n$, 
in time which is independent from the domain size.

\begin{figure}[h!]

\caption{
\label{map of inference frameworks}
{\small The inference frameworks are defined in Definition~\ref{concrete inference frameworks}.
The symbol $\simeq$ denotes asymptotic equal expressivity. 
A path upwards means that the the upper inference framework is asymptotically more expressive (i.e. $\prec$ holds).
The absence of a path ``upwards'' between two inference framewords means that the inference frameworks are incomparable 
with respect to asymptotic expressivity.}
}
\bigskip
\begin{center}
\includegraphics[scale=0.9]{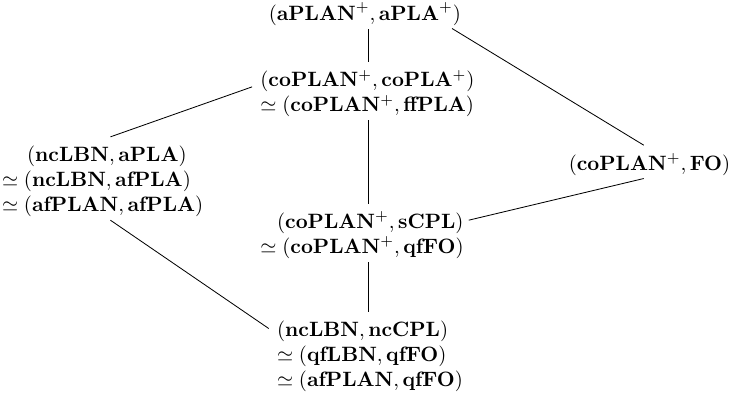}
\end{center}

\end{figure}

Our third contribution, 
Theorem~\ref{partial order of inference frameworks}
illustrated by 
Figure~\ref{map of inference frameworks},
is to systematize
the new results of this article and those in \cite{Kop20, KW1, SS}
by means of the notions of inference framework and 
relative asymptotic expressivity of inference frameworks.
\footnote{We did not include the inference framework to which the main results in \cite{Jae98a} 
apply because
we have not determined where it belongs relative to the inference frameworks of
Theorem~\ref{partial order of inference frameworks} or 
Figure~\ref{map of inference frameworks}, but we suspect that it belongs to the
equivalence class in the bottom of Figure~\ref{map of inference frameworks}.}
An \emph{inference framework} (for a signature $\sigma$) is a class $\mbF$ of pairs  $(\mathbb{P}, L)$ 
where $\mbbP = (\mbbP_n : n \in \mbbN^+)$, each $\mbbP_n$ is a probability distribution on $\mbW_n$,
and
$L$ is a logic (the associated query language) which uses the signature $\sigma$
(see Definition \ref{definition of inference framework}).
Note that we allow $L$ to depend on $\mathbb{P}$. 
The reason is that it allows us make finer distinctions between queries which are ``easy'' to evaluate
with respect to a given sequence $\mbbP$ of probability distributions and queries which are
``hard'' to evaluate with respect to the same $\mbbP$.

The asymptotic expressivity of an inference framework should now be studied on both components
of its pairs $(\mbbP, L$).
We call an inference framework
$\mbF'$  {\em asymptotically at least as expressive} as another inference framework $\mbF$ if
for every $(\mbbP, L) \in \mbF$ there is $(\mbbP', L') \in \mbF'$ such that 
$\mbbP$ is asymptotically total variation equivalent to $\mbbP'$ 
(see Definition~\ref{definitions of asymptotic equivalences})
and for every $\varphi(\bar{x}) \in L$ there is $\varphi'(\bar{x}) \in L'$ such that 
$\varphi'(\bar{x})$ is asymptotically equivalent to $\varphi(\bar{x})$ with respect to $\mbbP$ (or equivalently $\mbbP'$, see Definition \ref{definition of relative expressibility of inference frameworks}).
If in addition, $\mbF$ is asymptotically at least as expressive as $\mbF'$ then we say that they 
are {\em asymptotically equally expressive}.

In the discussion above we have already implicitly seen examples of inference frameworks and
results stating that two inference frameworks are asymptotically equally expressive.
For example, if $coPLA^+$ is the set of all $PLA^+$-formulas that use only
continuous aggregation functions and $\mbF$ is the set of all pairs $(\mbbP, coPLA^+)$ such that $\mbbP$ is defined by a
$coPLA^+$-network, then Theorem~\ref{elimination of strongly admissible aggregation functions}
implies that $\mbF$ is asymptotically equally expressive as the inference framework $\mbF'$
consisting of all pairs $(\mbbP', afPLA)$ where $\mbbP'$ is defined by a $coPLA^+$-network and $afPLA$
is set of all aggregation-free formulas in $PLA^+$.

Naturally one can question whether our notion of ``asymptotically at least as expressive'' is
the most relevant one. In particular, the asumption about asymptotic total variation equivalence
may appear to be too strong, since the logic(s) used in a particular inference framework
may not be able to ``define'' all subsets of
the probability space $\mbW_n$.
But we believe that a useful notion of ``asymptotically at least as expressive'' should be {\em transitive}
and all other seemingly reasonable candidates that we have considered
(except the one mentioned above and in Definition~\ref{definition of relative expressibility of inference frameworks} below)
turned out {\em not} to be transitive.

\vspace{-5mm}

\subsection*{Related work}

Researchers in statistical relational artificial intelligence,
a branch of artificial intelligence and machine learning
(see e.g. \cite{DKNP, GT, KMG, VKNP} for introductions to the field), have developed several different formalisms to specify 
probabilistic graphical models on an abstract relational level. They include relational Bayesian networks \cite{Jae97},
relational logistic regression \cite{KBKNP}, 
Bayesian logic programs \cite{KdR}, and lifted Bayesian networks \cite{Kop20}. 
Each of these formalisms can be viewed as a $L(\sigma)$-network for a suitable logic $L$ and signature 
$\sigma$.
In fact, each of the mentioned formalisms can be viewed as a $PLA^+(\sigma)$-network.
Besides notational differences, the main difference between $PLA^+$ and the probability logic in \cite{Jae98a} is
that $PLA^+$ allows for more general connectives.

As before, let $\sigma$ be a finite and relational signature, $\mbW_n$ the set of all $\sigma$-structures with
domain $[n]$, and $\mbbP_n$ a probability distribution on $\mbW_n$.
Jaeger's main result in \cite{Jae98a} can now be formulated as saying that if $ecPLA^+$ is set of 
$PLA^+$-formulas which use only {\em exponentially convergent} aggregation functions
and if $\mbbP_n$ is defined by an $ecPLA^+(\sigma)$-network, then the probability
of every first-order query converges as $n \to \infty$.
The aggregation functions noisy-or, maximum and minimum are exponentially convergent, 
but not the arithmetic or geometric means.
On the other hand the arithmetic and geometric means as well as (for $\alpha > 0$) $length_\alpha$ are continuous, but not
exponentially convergent,
where for every sequence of reals $\bar{r} = (r_1, \ldots, r_m)$, $length_\alpha(\bar{r}) = m^{-\alpha}$.
Therefore Theorem~\ref{elimination of strongly admissible aggregation functions}
and its corollaries apply to different sequences of probability distributions and different queries than the 
main result in \cite{Jae98a}.

The main results in \cite{KW1}
can be stated as follows:
Suppose that $\mbbP_n$ is defined by a lifted Bayesian network for $\sigma$ (as defined in \cite{Kop20})
and let $aPLA(\sigma)$ be the set of $PLA^+(\sigma)$-formulas in which only
{\em admissible} aggregation functions are allowed and which also satisfies two other conditions
(see Definitions~\ref{alternative definition of admissible function} and~\ref{restrictions of PLA+}).
Then every $aPLA(\sigma)$-formula is {\em asymptotically equivalent} 
to an aggregation-free formula, and every $aPLA(\sigma)$-formula satisfies a convergence law (as $n\to\infty$).
Admissibility is a weak form of continuity (so continuity implies admissibility) satisfied by the aggregation functions 
maximum and minimum.
Every sequence of distributions defined by a lifted Bayesian network is asymptotically total variation equivalent
to a sequence of distributions defined by a $coPLA^+$-network, but not vice versa.
So Theorem~\ref{elimination of strongly admissible aggregation functions} is, with respect to probability distributions, 
more general than
the main result in \cite{KW1}, but not with respect to queries. 
Moreover, Theorem~\ref{elimination of strongly admissible aggregation functions} 
cannot be generalized to hold for $aPLA$-queries, or for first-order queries.
The reason is as follows.
First, the aggregation function $length_\alpha$ is continuous, so if $R$ is a binary relation symbol and $\alpha \in (0, 1)$, 
then there is a $coPLA^+$-network such that the formula $R(x, y)$ has probability $n^{-\alpha}$ 
(for any pair, independently of other pairs) where $n$ is the domain size.
By results on random graphs by Shelah and Spencer \cite{SS}, if $\alpha$ is rational then 
there is a first-order sentence such that its probability
does not converge as $n \to \infty$. Since every first-order formula is equivalent to some $aPLA$-formula we cannot
have convergence for all $aPLA$-formulas.

From \cite[Theorem~3.16]{Kop20} it follows that 
lifted Bayesian networks only can define probability distributions $\mbbP_n$ where for each $R \in \sigma$
the probability of $R(\bar{x})$ is either constantly 0 or tends to a constant $c > 0$ as $n \to \infty$.
It seems like the same is true for relational Bayesian networks as in \cite{Jae98a} for the following reason:
Suppose for a contradiction that a sequence of probability distributions $\mbbP_n$ is defined by a relational Bayesian network and
that the probability that a pair of parameters satisfies an atomic formula $R(x, y)$ is $n^{-\alpha}$ 
(independently of other pairs) where $\alpha \in (0, 1)$ is rational and $n$ is the
size of the domain. By\cite{SS} again there is a first-order sentence such that its probability
does not converge as $n \to \infty$, but this contradicts the main result in \cite{Jae98a}.

In \cite{Kop20} Koponen studied so-called lifted Bayesian networks, for producing probability distributions,
and a fragment of conditional probability logic (CPL) which extends first-order logic and 
avoids certain ``critical'' parameters as a query language. 
Theorem~\ref{quantifier elimination of safe formulas} 
considers another fragment of CPL 
and more general, up to asymptotic total variation equivalence,
probability distributions. The fragment of CPL considered in 
Theorem~\ref{quantifier elimination of safe formulas}  avoids first-order quantifiers and 
``unsafe'' parameters in ``conditional proportion statements''.

Other works than \cite{Jae98a, Kop20, KW1} about the asymptotics of logics 
with respect to  probability distributions defined by probabilistic graphical models
are less related to the results presented here. They include the following publications:
\cite{CM, MBGS, PBKKN, Wei21, Wei24, WRF23}.

\subsection*{Organization}

This article is organized as follows.
Section~\ref{Preliminaries}
fixes some basic notation and terminology, and states a result that will be used later.
Section~\ref{Connectives and aggregation functions} defines the notions of {\em connective} and
{\em aggregation function} that we will use and also defines the notion of {\em (strongly) admissible}
aggregation function.
Section~\ref{logics} defines the general notion of a {\em logic} that we will use, as well as the particular logics that will be considered later.
Section~\ref{PGM} defines the notion of $L(\sigma)$-network and
states our first main result, 
Theorem~\ref{elimination of strongly admissible aggregation functions}, and a corollary.
It also recalls the notion of lifted Bayesian network from \cite{Kop20} and the main 
results from \cite{Kop20} and \cite{KW1} (which are used in Theorem~\ref{partial order of inference frameworks}).
Section~\ref{asymptotic elimination of aggregation functions}
proves a key technical result  about asymptotic elimination of strongly admissible aggregation functions,
Proposition~\ref{saturation implies elimination of strongly admissible functions},
which is 
used in Section~\ref{proof of main theorem}
to prove Theorem~\ref{elimination of strongly admissible aggregation functions}.
Section~\ref{proof of main theorem} completes the proof of
Theorem~\ref{elimination of strongly admissible aggregation functions}
by, roughly speaking, proving that the preconditions of 
Proposition~\ref{saturation implies elimination of strongly admissible functions}
are satisfied.
In Section~\ref{elimination of safe CPL-formulas}
we introduce the notion of a {\em safe} $CPL$-formula and state and prove
our second main result,
Theorem~\ref{quantifier elimination of safe formulas}.
Then Section~\ref{inference frameworks}
defines the notions of {\em inference framework} and
{\em asymptotically at least as expressive (inference framework)} and states and proves
our last result, Theorem~\ref{partial order of inference frameworks}.
Finally, Section~\ref{Conclusion} briefly recalls what we have done.

\section{Preliminaries}\label{Preliminaries}

\noindent
We let $\mbbN$ be the set of all non-negative integers and we let $\mbbN^+$ be the set of positive integers.
Finite sequences/tuples of objects are denoted by $\bar{a}, \bar{b}, \bar{r}, \bar{x}$, et cetera.
(Typically the objects of the sequence are numbers, elements from the domain of a structure or logical variables.)

For a finite sequence $\bar{a}$, $|\bar{a}|$ denotes the length of the sequence and $\rng(\bar{a})$ denotes the set of elements
occurring in the sequence $\bar{a}$.
For a set $A$, $|A|$ denotes its cardinality.
If $A$ is a set and $n \in \mbbN^+$, then $A^n$ is the set of all finite sequences of length $n$ of elements from $A$ and
$A^{<\omega}$ is the set of all finite nonempty sequences of elements from $A$, so
$A^{<\omega} = \bigcup_{n\in\mbbN^+} A^n$
The set $\{1, \ldots, n\}$ will be denoted by  $[n]$.

Structures in the sense of first-order logic will be denoted by calligraphic letters such as $\mcA, \mcB, \ldots$. Unless otherwise specified, their domains/universes
are denoted by the corresponding non-calligraphic letter $A, B, \ldots$ (see for example \cite{EF} for basics about structures in the sense of first-order logic and (finite) model theory).
We will consider different logics, but the semantics are always based on first-order structures.
If $\sigma$ is a (first-order) {\em signature}, also called {\em vocabulary}, then a {\em $\sigma$-structure} is a first-order structure in which all symbols in
$\sigma$ have an interpretation (and, for counting reasons, we assume that no other symbols are interpreted in a $\sigma$-structure).
A signature $\sigma$ is {\em finite and relational} if it is finite and contains only relation symbols.
If $R$ is a relation symbol of a signature $\sigma$, its intepretation in a $\sigma$-structure $\mcA$ is denoted by $R^\mcA$.
If $\sigma' \subset \sigma$ are signatures and $\mcA$ is a $\sigma$-structure then the reduct of $\mcA$ to 
$\sigma'$ is denoted by $\mcA \uhrc \sigma'$.
For more about notation concerning logical concepts, see Section~\ref{logics}.

By {\em directed acyclic graph (DAG)} we mean a directed graph without loops or directed cycles.
Let $G = (V, G)$ be a DAG. 
If $v \in V$ then $\mr{par}(v)$ denotes the set of {\em parents of} $v$, that is, the set of all vertices $w \in V$ such 
that $(w, v) \in E$.
For $v \in V$ we define the {\em maximal path rank} of $v$, 
denoted $\mr{mp}(v)$, to be the maximal integer $n > 0$
such that there is a directed path $v_0, \ldots, v_n \in V$ (meaning that $(v_i, v_{i+1}) \in E$ for all $i$) 
with $v_n = v$. 
We define the {\em maximal path rank} of $\mcG$, denoted $\mr{mp}(\mcG)$, as
$\mr{mp}(\mcG) = \max(\mr{mp}(v) : v \in V)$.

We call a random variable {\em binary} if it can only take the value $0$ or $1$.
The following is a direct consequence of \cite[Corollary~A.1.14]{AlonSpencer} which in turn follows from the
Chernoff bound \cite{Chernoff}:

\begin{lem}\label{independent bernoulli trials}
Let $Z$ be the sum of $n$ independent binary random variables, each one with probability $p$ of having the value 1.
For every $\varepsilon > 0$ there is $c_\varepsilon > 0$, depending only on $\varepsilon$, such that the probability that
$|Z - pn| > \varepsilon p n$ is less than $2 e^{-c_\varepsilon p n}$.
\end{lem}

\begin{cor}\label{independent bernoulli trials, second version}
Let $p \in [0, 1]$ and let $\varepsilon > 0$.
Let $Z$ be the sum of $n$ independent binary random variables $Z_1, \ldots, Z_n$, where for each $i = 1, \ldots, n$ the
probability that $Z_i$ equals 1 belongs to the interval $[p-\varepsilon, p+\varepsilon]$.
Then there is $c_\varepsilon > 0$, depending only on $\varepsilon$, such that the probability that
$Z > (1 + \varepsilon)(p + \varepsilon) n$ or 
$Z < (1 - \varepsilon)(p - \varepsilon) n$ 
is less than $2 e^{-c_\varepsilon p n}$.
\end{cor}

\proof
Let $Z'_1, \ldots, Z'_n$ be  independent binary random variables where each $Z'_i$ takes value 1 with probability (exactly)
$p + \varepsilon$ (assuming $p + \varepsilon \leq 1$).
Let $Z' = Z'_1 + \ldots + Z'_n$.
By Lemma~\ref{independent bernoulli trials},
the probability that $Z' > (p + \varepsilon)n + \varepsilon(p + \varepsilon)n = (1 + \varepsilon)(p + \varepsilon)n$
is less than $2 e^{-a_\varepsilon p n}$  where $a_\varepsilon > 0$ depends only on $\varepsilon$.
Since, for each $i$, the probability that $Z_i = 1$ is less or equal to the probability that $Z'_1 = 1$ it follows that
the probability that $Z > (1 + \varepsilon)(p + \varepsilon) n$ is not larger than the probability that 
$Z' > (1 + \varepsilon)(p + \varepsilon) n$ which is less than $2 e^{-a_\varepsilon p n}$.
A similar argument shows that the probability that 
$Z < (1 - \varepsilon)(p - \varepsilon) n$ 
is less than $2 e^{-b_\varepsilon p n}$ for some $b_\varepsilon > 0$ depending only on $\varepsilon$.
\qed

\section{Connectives and aggregation functions}\label{Connectives and aggregation functions}

\noindent
The idea behind a $k$-ary connective is that it assigns a truth value to every $k$-tuple of truth values.
Recall from the previous section 
that $[0, 1]^{<\omega}$ is the set of all finite sequences $\bar{r}$ where each entry of $\bar{r}$
belongs to $[0, 1]$. 
So $\big([0, 1]^{<\omega}\big)^k$ is the set of all $k$-tuples $(\bar{r}_1, \ldots, \bar{r}_k)$ where,
for each $i = 1, \ldots, k$, $\bar{r}_i$ is a finite sequence of reals from $[0, 1]$. Note that $\bar{r}_i$
and $\bar{r}_j$ are allowed to have different length if $i \neq j$.
The role of a  $k$-ary aggregation function is to assign a truth value to every $k$-tuple  
$(\bar{r}_1, \ldots, \bar{r}_k) \in \big([0, 1]^{<\omega}\big)^k$.
In particular a unary (1-ary) aggregation
function takes only one sequence $\bar{r} \in [0, 1]^{<\omega}$ (of arbitrary finite length) as input.
When used in a logic, in Section~\ref{logics}, the role of
aggregation functions will 
be to ``aggregate over a domain'', similarly to (generalized) quantifiers
in the context of 0/1-valued logics.
Alternatively (as in e.g. \cite{Jae98a}), one can view a $k$-ary aggregation (or ``combination'') function as a mapping from
$k$-tuples of finite multisets (of reals in $[0, 1]$)  into $[0, 1]$.
However, the ``symmetry condition'' in Definition~\ref{definition of aggregation function} below implies that our notion of aggregation function is
exchangeable, in the context of this article, with the notion of an aggregation function as operating on multisets.

\begin{defi}\label{definition of connective} {\rm
A function $\msfC : [0, 1]^k \to [0, 1]$ where $k \in \mbbN^+$ will also be called a {\em connective}.
}\end{defi}

\begin{defi}\label{definition of aggregation function}{\rm
Let $F : \big([0, 1]^{<\omega}\big)^k \to [0, 1]$, so $F$ takes $k$ sequences (not necessarily of the same length) as input.
We call $F$ an {\em aggregation function} if $F$ is symmetric in the sense that if $\bar{r}_1, \ldots, \bar{r}_k  \in [0, 1]^{<\omega}$ 
and for each $i = 1, \ldots, k$, $\bar{\rho}_i$ is an arbitrary reordering of the entries of $\bar{r}_i$, then
$F(\bar{\rho}_1, \ldots, \bar{\rho}_k) = F(\bar{r}_1, \ldots, \bar{r}_k)$.
}\end{defi}

\begin{exa}\label{examples of aggregation functions}
 {\rm
(a) The aggregation functions listed below are common when analyzing data. 
For $\bar{r} = (r_1, \ldots, r_n) \in [0, 1]^{<\omega}$, define
\begin{enumerate}
\item $\max(\bar{r})$ to be the {\em maximum} of all $r_i$,
\item $\min(\bar{r})$ to be the {\em minimum} of all $r_i$,
\item $\mr{am}(\bar{r}) = (r_1 + \ldots + r_n)/n$, so `am' is the {\em arithmetic mean}.
\item $\mr{gm}(\bar{r}) = \big(\prod_{i=1}^n r_i\big)^{(1/n)}$, so `gm' is the {\em geometric mean}.
\item $\text{noisy-or}(\bar{r}) = 1 - \prod_{i=1}^n (1 - r_i)$.
\end{enumerate}
\medskip
(b) Another example of an aggregation function (now with arity 2) is the
pseudometric $\mu^u_1 : \big([0, 1]^{<\omega}\big)^2 \to [0, 1]$ of 
Definition~\ref{definition of the metric} which compares how similar (or close) two sequences 
$\bar{r}, \bar{\rho} \in [0, 1]^{<\omega}$ are if we disregard the ordering of the entries in each sequence.

\medskip
\noindent
(c) For another example, define $S(x) = (1 + e^{-x})^{-1}$ for all $x \in \mbbR$, 
let $w_1, \ldots, w_k \in [0, 1]$
be weights such that their sum equals 1, and define 
$G : \big([0, 1]^{<\omega}\big)^k \to [0, 1]$ by 
$G(\bar{r}_1, \ldots, \bar{r}_k) = S\big(\sum_{i=1}^k w_i \cdot \text{am}(\bar{r}_i)\big)$.
$G$ is used to define Bayesian networks in the context of 
{\em Domain-Size-Aware Relational Logistic Regression} models \cite{WRF23}.

\medskip
\noindent
(d) For $\alpha \in (0, 1)$ define $length_\alpha : [0, 1]^{<\omega} \to [0, 1]$ by 
$length_\alpha(\bar{r}) = 1/|\bar{r}|^\alpha$ for all $\bar{r} \in [0, 1]^{<\omega}$
(so $length_\alpha(\bar{r})$ depends only on the length of $\bar{r}$).
$length_\alpha$ can, for example, be used to define a $coPLA^+(\sigma)$-network 
where the signature $\sigma$ contains only a binary relation symbol such that this
network induces a probability distribution 
on directed graphs where each directed edge has probability $1/n^\alpha$, independently of
other edges,  where $n$ is the number of vertices. 

\medskip
\noindent
(e) More examples (e.g. conditional arithmetic means) are given in Section~5 of \cite{KW1}.

}\end{exa}

\noindent
Now we will isolate two classes of aggregation functions that are sufficiently benign that we can
prove results concerning the asymptotic expressivity of inference frameworks that use only such aggregation functions.
Later we will see that the two classes contain useful aggregation functions.
Before defining these two classes we need to define the notion of
convergence testing sequence of sequences from $[0, 1]^{<\omega}$.

Intuitively speaking, an infinite sequence (of finite sequences) $\bar{r}_n \in [0, 1]^{<\omega}$, $n \in \mbbN$,
is convergence testing if there are $k$ and $c_1, \ldots, c_k, \alpha_1, \ldots \alpha_k  \in  [0,1]$
such that, as $n \to \infty$, all entries in $\bar{r}_n$ congregate ever closer to the ``convergence points''
$c_1, \ldots, c_k$ and, for each $i$, the proportion of entries in $\bar{r}_n$ that are close to $c_i$
tends ever closer to $\alpha_i$.

\begin{defi}[Convergence testing sequence]\label{definition of convergence testing} {\rm 
A sequence $\bar{r}_n \in [0, 1]^{<\omega}$, $n \in \mbbN$,  is called {\em convergence testing} for parameters 
$c_1, \ldots, c_k \in [0,1]$ and $\alpha_1, \ldots \alpha_k  \in  [0,1]$ if the following hold, 
where $r_{n,i}$ denotes the $i$th entry of $\bar{r}_n$:
\begin{enumerate}
\item $|\bar{r}_n| < |\bar{r}_{n+1}|$ for all $n \in \mbbN$.
\item For every disjoint family of open (with respect to the induced topology on $[0, 1]$)
intervals $I_1, \ldots I_k \subseteq [0,1]$ 
 such that $c_i \in I_i$ for each $i$, 
there is an $ N \in \mbbN$ such that $\mathrm{rng}(\bar{r}_n) \subseteq \bigcup\limits_{j=1}^{k} I_j$ for all $n \geq N$, 
and for every $j \in \{1, \ldots, k \}$, 
\[
\lim\limits_{n \rightarrow \infty} \frac{\left| \{ i \leq |\bar{r}_n| : r_{n,i} \in I_j \} \right| }{|\bar{r}_n|} = \alpha_j.
\]
\end{enumerate}   

More generally, a sequence of $m$-tuples of sequences 
$(\bar{r}_{1, n}, \ldots, \bar{r}_{m, n}) \in  \big([0, 1]^{<\omega}\big)^m$, $n \in \mbbN$,  
is called {\em convergence testing} for parameters $c_{i,j} \in [0,1]$ and $\alpha_{i,j} \in [0,1]$, 
where $i \in \{1, \ldots, m\}$, $j \in \{ 1, \ldots, k_i \}$ and $k_1, \ldots k_m \in \mbbN^+$, if  for every fixed $i \in \{1, \ldots, m \}$
the sequence $\bar{r}_{i, n}$, $n \in \mbbN$, is convergence testing for $c_{i,1}, \ldots, c_{i, k_i}$, and $\alpha_{i,1}, \ldots, \alpha_{i, k_i}$.
}\end{defi}

\noindent
Next we define the notion of (strongly) admissible aggregation function.
The intuition behind admissibility and strong admissibility is that they are 
``(partial) continuity conditions'' suitable for aggregation functions. 
We would argue that strong admissibility is a more ``continuity-like'' condition
than admissibility because the aggregation functions max and min are admissible but not strongly admissible 
and one could argue that max and min, as aggregation functions, should not be considered to be continuous:
If $n$ is large then it is reasonable to view the sequences $(r_1, \ldots, r_n)$ and $(\rho_1, \ldots, \rho_n)$
as very similar (or ``close'' to each other) if $r_i = 0$ for all $i = 1, \ldots, n$, $\rho_i = 0$ for $i = 1, \ldots, n-1$ and $\rho_n = 1$, but
$\max(\bar{r}) = 0$ and $\max(\bar{\rho}) = 1$.

\begin{defi}[Admissibility and continuity] \label{alternative definition of admissible function} {\rm 
(i) An aggregation function $F : \big([0, 1]^{<\omega}\big)^m \to [0, 1]$ is called {\em strongly admissible},
or {\em continuous}, if the following two conditions hold:
\begin{enumerate}
\item For all $n_1, \ldots, n_m \in \mbbN^+$, $F$ is continuous on the set $[0, 1]^{n_1} \times \dots \times [0, 1]^{n_m}$.
\item For all convergence testing sequences of tuples
$(\bar{r}_{1, n}, \ldots, \bar{r}_{m, n}) \in  \big([0, 1]^{<\omega}\big)^m$, $n \in \mbbN$,
and $(\bar{\rho}_{1, n}, \ldots, \bar{\rho}_{m, n}) \in  \big([0, 1]^{<\omega}\big)^m$, $n \in \mbbN$,
with the same parameters $c_{i,j}, \alpha_{i,j} \in [0, 1]$, 
$\underset{n \rightarrow \infty}{\lim}  |F(\bar{r}_{1, n}, \ldots, \bar{r}_{m, n}) - F(\bar{\rho}_{1, n}, \ldots, \bar{\rho}_{m, n})| = 0$.
\end{enumerate}
(ii) An aggregation function $F : \big([0, 1]^{<\omega}\big)^m \to [0, 1]$ is called {\em admissible}
if condition~(1) above holds and condition~(2) above holds whenever the parameters $\alpha_{i, j}$ are {\em positive} for all $i$ and $j$.
}\end{defi}

\noindent
The next proposition and example show that a number of useful aggregation functions are indeed
(strongly) admissible.
Proposition~\ref{am and gm are strongly admissible} below
was proved in \cite{KW1}, because the proof in \cite{KW1} that the arithmetic and geometric means are admissible still works if one allows the parameters $\alpha_{i, j}$ to be 0, thus showing that the functions are strongly admissible.

\begin{prop}\label{am and gm are strongly admissible}
(i) The functions {\rm am} ({\em arithmetic mean}) and {\rm gm} ({\em geometric mean}) are strongly admissible
(in other words, continuous).\\
(ii) The functions {\rm max} and {\rm min} are admissible.
\end{prop}

\begin{exa}[More (strongly) admissible aggregation functions]\label{more about admissible functions} 
{\rm
The aggregation functions in parts (b), (c) and (d) in
Example~\ref{examples of aggregation functions} 
are strongly admissible.
In the case of (c) and (d) this follows easily from their definitions and the fact that arithmetic mean 
is strongly admissible. In the case of (b) it follows from the characterization of
strong admissibility given below by
Definition~\ref{definition of admissible function}
and Proposition~\ref{equivalence of definitions of admissibility}.
In Sections~5 and~6 of \cite{KW1} more examples of admissible aggregation functions are given, for example a
``conditional arithmetic mean''.
}\end{exa}

\noindent
It is not hard to see that max and min are not strongly admissible and that noisy-or is not even admissible.

The above given definition of (strong) admissibility is fairly straightforward and natural, as well as useful for proving that some functions
are (strongly) admissible.
But we do not see how it can be used directly in the proof
of Proposition~\ref{saturation implies elimination of strongly admissible functions}
which is used to prove one of our main results, 
Theorem~\ref{elimination of strongly admissible aggregation functions}.
Therefore we give a different characterization of (strong) admissibility 
(Definition~\ref{definition of admissible function}) 
below. For this we need to consider functional representations of sequences in $[0, 1]^{<\omega}$ and two pseudometrics
on $[0, 1]^{<\omega}$.

\begin{defi}[Functional representations of sequences]\label{definition of associated function} {\rm
Let $n \in \mbbN^+$ and let $\bar{r} = (r_1, \ldots, r_n) \in [0, 1]^n$. We will associate a function from $[0, 1]$ to $[0, 1]$ with $\bar{r}$ in
two different ways, one way where the order of the entries in $\bar{r}$ matters and one in which the order does not influence the associated function.
\begin{enumerate}
\item Define $\mff_{\bar{r}}$, which we call the {\em ordered functional representation of $\bar{r}$}, as follows: 
For every $a \in [0, 1/n)$, let $\mff_{\bar{r}}(a) = r_1$, 
for every $i = 1, \ldots, n-1$ and every $a \in [i/n, (i+1)/n)$, let $f(a) = r_{i+1}$ and finally let $f(1) = r_n$.

\item Define $\mfg_{\bar{r}}$, which we call the {\em unordered functional representation of $\bar{r}$}, as follows:
Let $\bar{\rho} = (\rho_1, \ldots, \rho_n)$ be a reordering of $\bar{r}$ such that, for all $i = 1, \ldots, n-1$, 
$\rho_i \leq \rho_{i+1}$ and let $\mfg_{\bar{r}} = \mff_{\bar{\rho}}$. 
\end{enumerate}
}\end{defi}

\begin{defi}[Pseudometrics on sequences]\label{definition of the metric} {\rm
\begin{enumerate}
\item First we recall the $L_1$ and $L_\infty$ norms: for every (bounded and integrable) $f : [0, 1] \to \mbbR$ they are defined as
\[
\| f \|_1 = \int_{[0,1]} | f(x) |dx \qquad \text{ and} \qquad
\| f \|_\infty = \sup\{|f(a)| : a \in [0, 1]\}.
\]
\item For $\bar{r}, \bar{\rho} \in [0, 1]^{<\omega}$ we define 
\begin{align*}
&\mu_1^u(\bar{r}, \bar{\rho}) =  \|\mfg_{\bar{r}} - \mfg_{\bar{\rho}}\|_1, \\
&\mu_\infty^o(\bar{r}, \bar{\rho}) =  \|\mff_{\bar{r}} - \mff_{\bar{\rho}}\|_\infty.
\end{align*}

\item For arbitrary $k > 1$ we can define a function on $\big([0, 1]^{<\omega}\big)^k$, also denoted 
$\mu_1^u$ and $\mu_\infty^o$
(to avoid making notation more complicated),
as follows: For all $(\bar{r}_1, \ldots, \bar{r}_k), (\bar{r}'_1, \ldots, \bar{r}'_k) \in \big([0, 1]^{<\omega}\big)^k$
let 
\[
\mu_1^u\big((\bar{r}_1, \ldots, \bar{r}_k), (\bar{r}'_1, \ldots, \bar{r}'_k)\big) = 
\max\big(\mu_\infty^u(\bar{r}_1, \bar{r}'_1), \ldots, \mu_1^u(\bar{r}_k, \bar{r}'_k)\big)
\]
and similarly for $\mu_\infty^o$.
\end{enumerate}
}\end{defi}

\noindent
From well-known results in analysis it follows that $\mu_1^u$ and  $\mu_\infty^o$ are symmetric and satisfy the triangle inequality so they are pseudometrics on $[0, 1]^{<\omega}$.
It is easy to see that none of them is a metric since it can happen that $\mu_1^u(\bar{r}, \bar{\rho}) = 0$ 
and $\bar{r} \neq \bar{\rho}$.
For example, if $\bar{r} = (0, 1/2, 1)$ and $\bar{\rho} = (0, 0, 1/2, 1/2, 1, 1)$
then $\mu_1^u(\bar{r}, \bar{\rho}) = 0$.
Note that for all $\bar{r}, \bar{\rho} \in [0, 1]^{<\omega}$, 
$\mu_1^u(\bar{r}, \bar{\rho}), \mu_\infty^o(\bar{r}, \bar{\rho}) \leq 1$.

\begin{defi}\label{definition of continuous aggregation function}
{\rm
Let $F : \big([0, 1]^{<\omega}\big)^k \to [0, 1]$ be an aggregation function and let 
$\mu$ be any of the the pseudometrics defined in Definition~\ref{definition of the metric}.
Also let $X \subseteq \big([0, 1]^{<\omega}\big)^k$.
We say that $F$ is {\em asymptotically uniformly continuous on $X$}
if for every $\varepsilon > 0$ there are $n$ and $\delta > 0$ such that 
if $(\bar{r}_1, \ldots, \bar{r}_k), (\bar{\rho}_1, \ldots, \bar{\rho}_k) \in X$,
$|\bar{r}_i|, |\bar{\rho}_i| \geq n$ for all $i$ and 
$\mu_1^u\big((\bar{r}_1, \ldots, \bar{r}_k), (\bar{\rho}_1, \ldots, \bar{\rho}_k)\big) < \delta$, then 
$\big|F(\bar{r}_1, \ldots, \bar{r}_k) - F(\bar{\rho}_1, \ldots, \bar{\rho}_k)\big| < \varepsilon$.
}\end{defi}

\begin{defi}[Alternative characterization of (strong) admissibility]\label{definition of admissible function} 
{\rm
An aggregation function $F : \big([0, 1]^{<\omega}\big)^m \to [0, 1]$ is called {\em strongly admissible sensu novo} 
if the following two conditions hold:
\begin{enumerate}
\item For all $k_1, \ldots, k_m \in \mbbN^+$ and all $c_{i, j}, \alpha_{i, j} \in [0, 1]$, 
for $i = 1, \ldots, m$ and $j = 1, \ldots, k_i$,
and all sufficiently small $\delta > 0$,
$F$ is asymptotically uniformly continuous on $X_1 \times \ldots \times X_m$ where, for each $i = 1, \ldots, m$,
\begin{align*}
X_i = &\big\{\bar{r} \in [0, 1]^{<\omega} : \rng(\bar{r}) \subseteq \{c_{i,1}, \ldots, c_{i,k_i}\} 
\text{ and, for each $j = 1, \ldots, k_i$,} \\
&\text{there are between $(\alpha_{i,j} - \delta)|\bar{r}|$ and $(\alpha_{i,j} + \delta)|\bar{r}|$ coordinates in $\bar{r}$ } \\
&\text{which equal $c_{i,j}$} \big\}.
\end{align*}

\item For all $k_1, \ldots, k_m \in \mbbN^+$ and all $c_{i, j}, \alpha_{i, j} \in [0, 1]$,
for $i = 1, \ldots, m$ and $j = 1, \ldots, k_i$,
and every $\varepsilon > 0$, there is $\delta > 0$ such that
if,  for $i = 1, \ldots, m$,  $\bar{r}_i, \bar{\rho}_i \in [0, 1]^{<\omega}$ and 
\begin{enumerate}
\item $|\bar{\rho}_i| = |\bar{r}_i|$, 
\item $\mu_\infty^o(\bar{r}_i, \bar{\rho}_i) < \delta$, 
\item $\rng(\bar{r}_i) \subseteq \{c_{i,1}, \ldots, c_{i,k_i}\}$, and
\item for each $j = 1, \ldots, k_i$, there are between $(\alpha_{i,j} - \delta)|\bar{r}_i|$ and 
$(\alpha_{i,j} + \delta)|\bar{r}_i|$ coordinates in $\bar{r}_i$ which equal $c_{i, j}$,
\end{enumerate}
then $|F(\bar{r}_1, \ldots, \bar{r}_m) - F(\bar{\rho}_1, \ldots, \bar{\rho}_m)| < \varepsilon$.
\end{enumerate}
An aggregation function $F : \big([0, 1]^{<\omega}\big)^m \to [0, 1]$ is called {\em admissible sensu novo} 
if the above conditions hold under the restriction that $\alpha_{i, j}$ is {\em positive} for all $i$ and $j$.
}\end{defi}

\begin{prop}\label{equivalence of definitions of admissibility}
(i) An aggregation function is strongly admissible sensu novo if and only if it is strongly admissible.\\
(ii) An aggregation function is admissible sensu novo if and only if it is admissible.
\end{prop}
 
\noindent
Proposition~\ref{equivalence of definitions of admissibility} is proved just like Proposition~6.5 in~\cite{KW1}, because
the only difference between strong admissibility (sensu novo) and admissibility (sensu novo) is that in the
former notions (but not the latter) we allow the parameters $\alpha_{i, j}$ to be zero. This does not affect
the proof of  Proposition~6.5 in~\cite{KW1}.

\section{Logics}\label{logics}

\noindent
In this section we define the general notion of logic that we will use as well as the various concrete logics that will be studied. We are pragmatic and minimalistic and will define a logic (with values in the unit interval)
to be something that has a few key properties necessary for making sense of definitions and results
that follow. 
(We are not aware of any commonly accepted notion of a many-valued logic in general.)
In the context of 0/1-valued logics a commonly used definition of
a logic appears in \cite[Definition~1.1.1, p. 27]{Ebb85} and this definition is stronger than the one we give below,
i.e. every logic in that sense is a logic in our sense.
A difference between \cite[Definition~1.1.1, p. 27]{Ebb85} and our notion of a logic is 
(even when restricting to 0/1-valued logics)
that we allow formulas of a logic to have free variables. 
In the rest of this section (as in the whole article) let $\sigma$ be a finite  relational signature.

\begin{defi}\label{definition of logic}
(i) By a {\em logic (for $\sigma$)} we mean a set $L$ of objects, called {\em formulas}, such that the following hold:
\begin{enumerate}
\item For every $\varphi \in L$ a finite set $Fv(\varphi)$ of so-called {\em free variables} of $\varphi$ is associated to $\varphi$.
If we write $\varphi(\bar{x})$ where $\varphi \in L$ then we mean that $Fv(\varphi) \subseteq \rng(\bar{x})$ and when using
this notation we assume that there are no repetitions in the sequence $\bar{x}$ (although we occasionally repeat this assumption).
\item To every triple $(\varphi(\bar{x}), \mcA, \bar{a})$ such that
$\varphi(\bar{x}) \in L$, $\mcA$ is a finite
$\sigma$-structure and $\bar{a} \in A^{|\bar{x}|}$ a number $\alpha \in [0, 1]$ is associated.
We write
$\mcA(\varphi(\bar{a})) = \alpha$ to express that $\alpha$ is the number, or {\em value}, associated to the triple
$(\varphi(\bar{x}), \mcA, \bar{a})$. 
(We allow $\bar{x}$ and $\bar{a}$ to be empty which is the case if the formula has no free variables.)
\end{enumerate}
(ii) We let the expressions  `$\mcA \models \varphi(\bar{a})$'  and 
`$\mcA \not\models \varphi(\bar{a})$' mean the same as
$\mcA(\varphi(\bar{a})) = 1$ and $\mcA(\varphi(\bar{a})) = 0$, respectively.\\
(iii) Suppose that $L$ is a logic, $\varphi(\bar{x}, \bar{y}) \in L$
(where $\rng(\bar{x}) \cap \rng(\bar{y}) = \es$), $\mcA$ is a finite $\sigma$-structure and $\bar{a} \in A^{|\bar{x}|}$.
Then $\varphi(\bar{a}, \mcA) = \{\bar{b} \in A^{|\bar{y}|} : \mcA(\varphi(\bar{a}, \bar{b}) = 1\}$.\\
(iv) Let $L$ and $L'$ be logics. Then let us write $L \leq L'$ if for every $\varphi(\bar{x}) \in L$ there is
$\varphi'(\bar{x}) \in L'$ such that for every finite $\sigma$-structure $\mcA$ and every $\bar{a} \in A^{|\bar{x}|}$,
  $\mcA(\varphi(\bar{a})) = \mcA(\varphi'(\bar{a}))$.
\end{defi}

\begin{defi}\label{definition of equivalent}
Suppose that $L$ is a logic for $\sigma$.
We say that $\varphi(\bar{x}) \in L$ and $\psi(\bar{x}) \in L$ are {\em equivalent} if,
for every finite $\sigma$-structure $\mcA$ and every $\bar{a} \in A^{|\bar{x}|}$, $\mcA(\varphi(\bar{a})) = \mcA(\psi(\bar{a}))$.
\end{defi}

\subsection{Some notation and concepts concerning first-order logic}

\noindent
Some basic concepts and notation regarding first-order logic will be used in the sequel and are defined below.

\begin{defi}\label{definition of first-order formulas}{\rm
(i) We let $FO(\sigma)$, respectively $qfFO(\sigma)$, 
denote the set of all first-order formulas, respectively quantifier-free first-order formulas, 
that can be constructed by using the signature $\sigma$.\\
(ii) Constructions of the form `$x = y$' and `$R(x_1, \dots, x_r)$', where $x, y$, $x_1, \ldots, x_r$ are variables, 
and $R \in \sigma$ has arity $r$, 
are called {\em atomic first-order formulas (over $\sigma$)}. 
By a {\em first-order literal (over $\sigma$)} we mean a first-order atomic formula 
(over $\sigma$) or a negation of such one.\\
(iii) If $\varphi(\bar{x}) \in L$, $\mcA$ is a $\sigma$-structure and $\bar{a} \in A^{|\bar{x}|}$,
then the notation `$\mcA \models \varphi(\bar{a})$' has the usual meaning of first-order logic,
and we let `$\mcA(\varphi(\bar{a})) = 1$' have the same meaning as `$\mcA \models \varphi(\bar{a})$'.
}\end{defi}

\begin{defi}[Atomic $\sigma$-types]\label{definition of atomic type} {\rm
(i) A consistent set $p$ of first-order literals over $\sigma$
is called an {\em atomic $\sigma$-type}.
If an atomic $\sigma$-type is denoted by $p(\bar{x})$ it is understood that every variable that occurs in a formula in $p(\bar{x})$
occurs in the sequence $\bar{x}$. \\
(ii) An atomic $\sigma$-type $p(\bar{x})$ is called {\em complete} if for every first-order 
atomic formula $\varphi(\bar{x})$ over $\sigma$,
either $\varphi(\bar{x})$ or $\neg\varphi(\bar{x})$ belongs to $p(\bar{x})$.\\
(iii) If $p(\bar{x})$ is an atomic $\sigma$-type and $\rng(\bar{y}) \subseteq \rng(\bar{x})$, then
$p(\bar{x}) \uhrc \bar{y}$ (or $p \uhrc \bar{y}$) denotes the set of all formulas $\varphi \in p(\bar{x})$
such that every variable of $\varphi$ occurs in $\bar{y}$.\\
(iv) If $p(\bar{x})$ is an atomic $\sigma$-type and $\sigma' \subset \sigma$, then 
$p(\bar{x}) \uhrc \sigma' = p(\bar{x}) \cap FO(\sigma')$.\\
(v) If $p(\bar{x})$ is an atomic $\sigma$-type, where $\bar{x} = (x_1, \ldots, x_k)$,
 then the {\em identity fragment} of $p$ is the set of all formulas in $p(\bar{x})$
of the form $x_i = x_j$ or $\neg(x_i = x_j)$ (abbreviated $x_i \neq x_j$). 
}\end{defi}

\noindent
When convenient we will identify, notationally, an atomic $\sigma$-type $p(\bar{x})$ with the formula obtained by taking the 
conjunction of all formulas in $p(\bar{x})$.
With this convention, if $\mcA$ is a $\sigma$-structure and $\bar{a} \in A^{|\bar{x}|}$ the notation
$\mcA \models p(\bar{a})$ makes sense and means, with model theoretic language, that $\bar{a}$ {\em realizes} $p(\bar{x})$ 
(in the structure $\mcA$).
Note that if $\sigma = \es$, then an atomic $\sigma$-type $p(\bar{x})$ will only contain literals of the form
$z = y$ or $z \neq y$ where $z, y \in \rng(\bar{x})$.

\subsection{Probabilistic logics with aggregation functions}

\noindent
We now define a quite general logic, $PLA^+$
(where $PLA$ stands for {\em probability logic with aggregation functions}), with truth values in $[0, 1]$.
Most of the logics that we will study more closely will be sublogics of $PLA^+$.
It follows directly from the definitions that the {\em probability logic} of Jaeger in \cite{Jae98a} is a
sublogic of $PLA^+$.

\begin{defi}[Syntax of $PLA^+(\sigma)$]\label{syntax of PLA+} {\rm 
The formulas of the logic $PLA^+(\sigma)$ 
are constructed as follows.\footnote{
The `+' in $PLA^+$ is there because the syntax of $PLA^+(\sigma)$ allows,
unlike $PLA(\sigma)$ defined in \cite{KW1}, that $p^=$ in item~(5) is not complete, 
and because item~(4) is more general than 
the corresponding part (about connectives) for $PLA(\sigma)$ in \cite{KW1}.}
\begin{enumerate}
\item  For each $c \in [0, 1]$, $c \in PLA^+(\sigma)$ (i.e. $c$ is a formula) and $Fv(c) = \es$. We also let $\bot$ and $\top$
denote $0$ and $1$, respectively.

\item For all variables $x$ and $y$, `$x = y$' belongs to $PLA^+(\sigma)$ and $Fv(x = y) = \{x, y\}$.

\item For every $R \in \sigma$, say of arity $r$, and any choice of variables $x_1, \ldots, x_r$, $R(x_1, \ldots, x_r)$ belongs to 
$PLA^+(\sigma)$ and  $Fv(R(x_1, \ldots, x_r)) = \{x_1, \ldots, x_r\}$.

\item If $n \in \mbbN^+$, $\varphi_1(\bar{x}), \ldots, \varphi_n(\bar{x})_n \in PLA^+(\sigma)$ and
$\msfC : [0, 1]^n \to [0, 1]$ is a continuous connective, then 
$\msfC(\varphi_1, \ldots, \varphi_n)$ is a formula of $PLA^+(\sigma)$ and
its set of free variables is $Fv(\varphi_1) \cup \ldots \cup Fv(\varphi_n)$.

\item If $k \in \mbbN^+$, $\varphi_1(\bar{x}, \bar{y}), \ldots, \varphi_k(\bar{x}, \bar{y}) \in PLA^+(\sigma)$,
$p^=(\bar{x}, \bar{y})$ is an atomic $\es$-type 
(i.e. a possibly empty description of the equalities and nonequalities between the variables in the sequence $\bar{x}\bar{y}$),
where $\bar{x}$ and $\bar{y}$ are sequences of distinct variables such that $\rng(\bar{x}) \cap \rng(\bar{y}) = \es$
and $F : \big( [0, 1]^{<\omega} \big)^k \to [0, 1]$ is an aggregation function,
then 
\[
F(\varphi_1(\bar{x}, \bar{y}), \ldots, \varphi_k(\bar{x}, \bar{y}) : \bar{y} : p^=(\bar{x}, \bar{y}))
\]
is a formula of $PLA^+(\sigma)$ and its set of free variables is
\[
\big( \bigcup_{i=1}^k Fv(\varphi_i)\big) \setminus \rng(\bar{y}),
\] 
so this construction binds the variables in $\bar{y}$.
\end{enumerate}
}\end{defi}

\begin{defi}[Semantics of $PLA^+(\sigma)$]\label{semantics of PLA+} {\rm
For each $\sigma$-structure $\mcA$, each formula $\varphi(\bar{x}) \in PLA^+(\sigma)$ and every $\bar{a} \in A^{|\bar{x}|}$,
we define a real number, denoted
$\mcA(\varphi(\bar{a}))$, in the interval $[0, 1]$, called the 
{\em value of $\varphi(\bar{a})$ in $\mcA$}, as follows
(where if $\varphi$ has no free variable we just omit $\bar{x}$ and $\bar{a}$):

\begin{enumerate}
\item For every $c \in [0, 1]$ and every $\sigma$-structure $\mcA$, $\mcA(c) = c$.

\item For every $\sigma$-structure $\mcA$ and all $a, b \in A$, $\mcA(a = b) = 1$ if $\mcA \models a = b$ 
and otherwise $\mcA(a = b) = 0$.

\item For every $R \in \sigma$, of arity $r$ say, every finite $\sigma$-structure $\mcA$ and all $\bar{a} \in A^r$,
$\mcA(R(\bar{a})) = 1$ if $\mcA \models R(\bar{a})$ and otherwise $\mcA(R(\bar{a})) = 0$.

\item If $n \in \mbbN^+$, $\varphi_1(\bar{x}), \ldots, \varphi_n(\bar{x}) \in PLA^+(\sigma)$ and 
$\msfC : [0, 1]^n \to [0, 1]$ is a continuous connective, then for every finite $\sigma$-structure $\mcA$ and every $\bar{a} \in A^{|\bar{x}|}$,
\[
\mcA\big(\msfC(\varphi_1(\bar{a}), \ldots, \varphi_n(\bar{a}))\big) \ = \ 
\msfC\big(\mcA(\varphi_1(\bar{a})), \ldots, \mcA(\varphi_n(\bar{a}))\big).
\]

\item If $k \in \mbbN^+$, $\bar{x}$ and $\bar{y}$ are sequences of distinct variables such that 
$\rng(\bar{x}) \cap \rng(\bar{y}) = \es$, 
$\varphi_1(\bar{x}, \bar{y}), \ldots, \varphi_k(\bar{x}, \bar{y}) \in PLA^+(\sigma)$,
$p^=(\bar{x}, \bar{y})$ is an atomic $\es$-type,
$F : \big([0, 1]^{<\omega}\big)^k \to [0, 1]$ is an aggregation function, 
$\mcA$ is a finite $\sigma$-structure and $\bar{a} \in A^{|\bar{x}|}$, then
\[
\mcA\big(F(\varphi_1(\bar{a}, \bar{y}), \ldots, \varphi_k(\bar{a}, \bar{y}) : \bar{y} : p^=(\bar{a}, \bar{y}))\big) = 
F(\bar{r}_1, \ldots, \bar{r}_k)
\]
if there is some $\bar{b} \in A^{|\bar{y}|}$ such that $p^=(\bar{a}, \bar{b})$ holds and, 
for $i = 1, \ldots, k$, 
\[
\bar{r}_i = \big(\mcA(\varphi_i(\bar{a}, \bar{b})) : \bar{b} \in A^{|\bar{y}|} 
\text{ and $p^=(\bar{a}, \bar{b})$ holds} \big),
\]
and otherwise 
$\mcA\big(F(\varphi_1(\bar{a}, \bar{y}), \ldots, \varphi_k(\bar{a}, \bar{y}) : \bar{y} : p^=(\bar{a}, \bar{y}))\big) = 0$.
\end{enumerate}
}\end{defi}

\noindent
With Definition~\ref{definition of connective} of a connective we can, 
by using the semantics of Lukasiewicz logic 
(see for example \cite[Section~11.2]{Ber}), define continuous connectives
which, when restricted to $\{0, 1\}$, have the usual meanings of $\neg$, $\wedge$, $\vee$, and
$\rightarrow$.

\begin{defi}[Some special continuous connectives]\label{special connectives}  {\rm
\begin{enumerate}
\item Let $\neg : [0, 1] \to [0, 1]$ be defined by $\neg(x) = 1 - x$.
\item Let $\wedge : [0, 1]^2 \to [0, 1]$ be defined by $\wedge(x, y) = \min(x, y)$.
\item Let $\vee : [0, 1]^2 \to [0, 1]$ be defined by $\vee(x, y) = \max(x, y)$.
\item Let $\rightarrow : [0, 1]^2 \to [0, 1]$ be defined by $\rightarrow(x, y) = \min(1, \ 1 - x + y)$.
\item Let $\text{wm} : [0, 1]^3 \to [0, 1]$ (where wm stands for {\em weighted mean})
be defined by $\text{wm}(x, y, z) = x\cdot y + (1 - x)\cdot z$.
\end{enumerate}
}\end{defi}

\noindent
We now give examples of the expressivity of $PLA^+(\sigma)$.
Note that in Examples~\ref{similarity measure} and~\ref{similarity profile}
only strongly admissible aggregation functions are used.
In Example~\ref{SimRank} only admissible aggregation functions are used (but the details are in \cite{KW1}).

\begin{exa}[Similarity measure]\label{similarity measure}  {\rm
Let $E_1, \ldots, E_k \in \sigma$ be binary relation symbols.
A measure of the {\em similarity of two elements $x$ and $y$}, with respect to
$E_1, \ldots, E_k$, is given by considering the fraction of 
elements which have the same connections to $x$ and $y$. This can be expressed in $PLA(\sigma)$ by
the formula
\begin{align*}
  \mathrm{am}\bigg( \bigwedge_{i=1}^k \big( ( E_i(z,x) \leftrightarrow E_i(z,y) ) \wedge 
( E_i(x,z) \leftrightarrow E_i(y,z) ) \big): 
 z:  y \neq z \wedge x \neq z \bigg)
\end{align*}
which we call $\psi(x, y)$. 
By nesting aggregation functions we can express other relations, properties and statements.
  ``The similarity to $x$ of the most similar other element'' is given by the formula
$
  \mathrm{max}( \psi(x,y) : y : x \neq y ).
$
  ``The average similarity of $x$ to other elements'' is given by
$
  \mathrm{am}( \psi(x,y) : y : x \neq y  ).
$
  ``The lowest similarity score between any two elements'' is expressed by
$
  \mathrm{min}(( \mathrm{min}( \psi(x,y):y:x \neq y )) : x : \ ).
$
In this example we considered atomic reations $E_i$, but it is possible to replace
$E_i$ with an arbitrary $PLA^+(\sigma)$-formula with two free variables.

}\end{exa}

\begin{exa}[Similarity profile]\label{similarity profile}  {\rm
Suppose that $E_1, \ldots, E_k \in \sigma$, $\psi(x, y)$ is the formula from 
Example~\ref{similarity measure}, 
$\mcA$ is a finite $\sigma$-structure and $a \in A$.
Then $\bar{p} = (\mcA(\psi(a, c)) : c \in A, c \neq a)$
is a sequence containing the similarity scores of $(a, c)$ as $c$ ranges over all other elements in $A$.
Let us call $\bar{p}$ the ``similarity profile'' of $a$ to other elements.
Let now $b \in A$ and $\bar{q} = (\mcA(\psi(b, c)) : c \in A)$.
Recall the pseudometric $\mu^u_1$ on $[0, 1]^{<\omega}$ from 
Definition~\ref{definition of the metric}, 
which is a strongly admissible aggregation function
$\mu^u_1 : \big([0, 1]^{<\omega}\big)^2 \to [0, 1]$, and
$\mu^u_1(\bar{p}, \bar{q})$ measures how close the similarity profile of $a$ is to the
similarity profile of $b$; the smaller $\mu^u_1(\bar{p}, \bar{q})$ is, the closer are
the similarity profiles of $a$ and $b$.
So the formula $\neg\mu^u_1(\psi(x, z), \psi(y, z) : z )$ 
gives a higher value, or ``similarity profile score'', if the similarity profiles of $x$ and $y$ are closer.
Intuitively speaking, if the similarity profile score is close to 1 then $x$ and $y$ are two possibly quite unrelated
entities with ``nearly isomorphic" connections (with respect to $E_1, \ldots, E_k$) to the rest of the world.
}\end{exa}

\begin{exa}[SimRank and PageRank]\label{SimRank}  {\rm
In \cite{KW1} it is demonstrated that every stage of SimRank \cite{JW}
can be expressed by a $PLA(\sigma)$-formula 
(defined in Definition~\ref{restrictions of PLA+} below)
that uses only admissible aggregation functions.
One can also show (which is simpler) that every stage of PageRank \cite{BP} can be expressed by a 
$PLA(\sigma)$-formula with only admissible aggregation functions.
}\end{exa}

\begin{defi}[Aggregation-free and basic probability formulas]\label{definition of basic probability formula}
{\rm \phantom{x} \\
(i) A formula of $PLA^+(\sigma)$ in which no aggregation function appears is called {\em aggregation-free}.\\
(ii) If $n \in \mbbN^+$, $\alpha_1, \ldots, \alpha_n \in [0, 1]$ and $\psi_1(\bar{x}), \ldots, \psi_n(\bar{x}) \in PLA(\sigma)$ 
are such that each $\psi_i$ 
is a conjunction of first-order literals, then the formula $\bigwedge_{i=1}^n \big(\psi_i(\bar{x}) \rightarrow \alpha_i\big)$
is called a {\em basic probability formula}.
}\end{defi}

\begin{rem}\label{remark on basic probability sentences} {\rm
A basic probability formula which is also a sentence, that is, a formula without free variables, 
has the form $\bigwedge_{i=1}^n(\top \to c_i)$ where $c_i \in [0, 1]$ (and recall that $\top = 1$).
The formula $\bigwedge_{i=1}^n(\top \to c_i)$ is equivalent to $c$ where $c = \min\{c_1, \ldots, c_n\}$,
so every basic probability sentence is equivalent to a sentence of the form $c$ for some $c \in [0, 1]$.
}\end{rem}

\noindent
The next lemma is proved in the same way as the corresponding result in \cite[Lemma~3.10]{KW1} where $PLA(\sigma)$-formulas are considered.

\begin{lem}\label{quantifier-free formulas are equivalent to bpf}
If $\varphi(\bar{x}) \in PLA^+(\sigma)$ is aggregation-free then $\varphi(\bar{x})$ is equivalent to a basic probability formula.
\end{lem}

\noindent
We now define different sublogics of $PLA^+(\sigma)$ which we will study, both as query languages
and as languages for defining networks that induce probability distributions.
These sublogics will be obtained from $PLA^+(\sigma)$ by restricting the kind of aggregation functions, or connectives, that are
allowed in formulas to such ones considered in Section~\ref{Connectives and aggregation functions}.

\begin{defi}[Sublogics of $PLA^+(\sigma)$]\label{restrictions of PLA+} {\rm 
\begin{enumerate}
\item $PLA(\sigma)$ is defined like $PLA^+(\sigma)$ except that in~(5) in Definition~\ref{syntax of PLA+}
we require that $p^=(\bar{x}, \bar{y})$ is a {\em complete} atomic $\es$-type
and~(4) in Definition~\ref{syntax of PLA+} is restricted to only apply to the continuous connectives
$\neg$, $\wedge$, $\vee$, $\to$ and $\text{wm}$.

\item $aPLA^+(\sigma)$, respectively $aPLA(\sigma)$, is the subset of $PLA^+(\sigma)$,
respectively $PLA(\sigma)$,
where only {\em admissible} aggregation functions $F$ are allowed in part~(5) of
Definition~\ref{syntax of PLA+}.

\item $coPLA^+(\sigma)$, respectively $coPLA(\sigma)$, is the subset of $PLA^+(\sigma)$,
respectively $PLA(\sigma)$,
where only {\em strongly admissible} (or {\em continuous})
aggregation functions $F$ are allowed in parts~(5) of
Definition~\ref{syntax of PLA+}. 

\item $\text{\em afPLA}(\sigma)$  is the set of all aggregation-free formulas in $PLA^+(\sigma)$.
\end{enumerate}
}\end{defi}

\noindent
Note that every aggregation-free formula in $PLA^+(\sigma)$ is, by 
Lemma~\ref{quantifier-free formulas are equivalent to bpf},
equivalent to a basic probability formula, which is a $PLA(\sigma)$ formula.
Observe also that $\text{\em afPLA}(\sigma) \subseteq coPLA(\sigma) \subseteq coPLA^+(\sigma) \subseteq PLA^+(\sigma)$,
$coPLA(\sigma) \subseteq aPLA(\sigma) \subseteq PLA(\sigma)$ and
$coPLA(\sigma^+)$ $\subseteq$ $aPLA^+(\sigma)$ $\subseteq$ $PLA^+(\sigma)$.

The following was proved in \cite[Lemma~3.11]{KW1} for $PLA(\sigma)$ and the (simple) proof is easily generalized to $PLA^+(\sigma)$.

\begin{lem}[Truth value invariance under isomorphisms]\label{isomorphisms preserve the value}   
Let $\mcA$ and $\mcB$ be isomorphic $\sigma$-structures and let $f$ denote an isomorphism from $\mcA$ to $\mcB$.
If $\varphi(\bar{x}) \in PLA^+(\sigma)$ and $\bar{a} \in A^{|\bar{x}|}$, then 
$\mcA(\varphi(\bar{a})) = \mcB(\varphi(f(\bar{a})))$.
\end{lem}

\subsection{Conditional probability logic}

\noindent
Besides $PLA^+$ and its sublogics we will also consider {\em conditional probability logic ($CPL$)} \cite{Kop20}
as a logic for queries and for defining networks. 
$CPL$ is a two-valued logic which extends first-order logic and with which one can express that the 
relative frequency of tuples that satisfy (some condition) $\varphi_1(\bar{x})$,
conditioned on tuples satisfying $\varphi_2(\bar{x})$ at least as large as
the relative frequency of tuples that satisfy $\varphi_3(\bar{x})$, conditioned on
tuples satisfying $\varphi_4(\bar{x})$.
If a probability distribution is given on the set of structures with a given finite domain,
then we can use $CPL$ to ask what the probability is that the frequency of 
one event (possibly conditioned on another event) is larger than the frequency of another event (possibly conditioned on yet another event).
When $CPL$ is used in the definition of a {\em lifted Bayesian network} in the sense of 
\cite{Kop20}, it expresses ``threshold conditions'' when the probability of a relation changes from 
one value to another.
Concrete examples of its expressive power are found in
Example~3.5 and remarks~3.4 and~3.6 in \cite{Kop20}.

\begin{defi}[Syntax of $CPL(\sigma)$]\label{definition of CPL}  {\rm
Suppose that $\sigma$ is a finite relational signature.
Then the set of {\em conditional probability formulas over $\sigma$}, denoted $CPL(\sigma)$, is defined inductively as follows:
\begin{enumerate}
\item Every atomic $\sigma$-formula belongs to $CPL(\sigma)$ (where `atomic' has the same meaning as in first-order logic with equality).

\item If $\varphi, \psi \in CPL(\sigma)$ then $(\neg\varphi), (\varphi\wedge\psi), (\varphi\vee\psi), (\varphi\rightarrow\psi), 
(\varphi\leftrightarrow\psi), (\exists x \varphi) \in CPL(\sigma)$ where $x$ is a variable.
(As usual, in practice we do not necessarily write out all parentheses.)
We consider $\forall x \varphi$ to be an abbreviation of $\neg \exists x \neg \varphi$.

\item If $r \geq 0$ is a real number, $\varphi, \psi, \theta, \tau \in CPL(\sigma)$ and $\bar{y}$ is a sequence of distinct variables, then
\begin{align*}
&\Big( r + \| \varphi \ |  \ \psi \|_{\bar{y}} \ \geq \ 
\| \theta \ |  \ \tau \|_{\bar{y}} \Big) \in CPL(\sigma)
\ \ \text{ and} \\
&\Big( \| \varphi \ |  \ \psi \|_{\bar{y}} \ \geq \ 
\| \theta \ |  \ \tau \|_{\bar{y}} + r \Big) \in CPL(\sigma).
\end{align*}
In both these new formulas all variables of $\varphi, \psi, \theta$ and $\tau$ that appear in the sequence $\bar{y}$ become {\em bound}.
So this construction can be seen as a sort of quantification, which may become more clear by the provided semantics below.
\end{enumerate}
}\end{defi}

\noindent
A formula $\varphi \in CPL(\sigma)$ is called {\em quantifier-free} if it contains no quantifier, that is, if it is constructed from atomic formulas
using only the connectives $\neg, \wedge, \vee, \rightarrow, \leftrightarrow$.

\begin{defi}[Semantics of $CPL(\sigma)$]\label{semantics of CPL}  {\rm
\begin{enumerate}
\item The interpretations of $\neg, \wedge, \vee, \rightarrow, \leftrightarrow$ and $\exists$ are as in first-order logic.

\item Suppose that $\mcA$ is a {\em finite} $\sigma$-structure and let $\varphi(\bar{x}, \bar{y}), \psi(\bar{x}, \bar{y}), 
\theta(\bar{x}, \bar{y}), \tau(\bar{x}, \bar{y}) \in CPL(\sigma)$.
Let $\bar{a} \in A^{|\bar{x}|}$.
\begin{enumerate}
\item We define $\varphi(\bar{a}, \mcA) = \big\{\bar{b} \in A^{|\bar{y}|} : \mcA \models \varphi(\bar{a}, \bar{b}) \big\}$.

\item The expression 
\[
\mcA \  \models \ 
\Big( r + \| \varphi(\bar{a}, \bar{y}) \ | \ \psi(\bar{a}, \bar{y}) \|_{\bar{y}} \ \geq \ 
\| \theta(\bar{a}, \bar{y}) \ | \ \tau(\bar{a}, \bar{y}) \|_{\bar{y}} \Big)
\]
means that $\psi(\bar{a}, \mcA) \neq \es$, $\tau(\bar{a}, \mcA) \neq \es$ and
\[
r + \frac{\big| \varphi(\bar{a}, \mcA) \cap \psi(\bar{a}, \mcA) \big|}{\big| \psi(\bar{a}, \mcA) \big|} \ \geq \ 
\frac{\big| \theta(\bar{a}, \mcA) \cap \tau(\bar{a}, \mcA) \big|}{\big| \tau(\bar{a}, \mcA) \big|}
\]
and in this case we say that 
$\Big( r + \| \varphi(\bar{a}, \bar{y}) \ | \ \psi(\bar{a}, \bar{y}) \|_{\bar{y}} \ \geq \ 
\| \theta(\bar{a}, \bar{y}) \ | \ \tau(\bar{a}, \bar{y}) \|_{\bar{y}} \Big)$
is true (or holds) in $\mcA$.
If $\psi(\bar{a}, \mcA) = \es$ or $\tau(\bar{a}, \mcA) = \es$ or
\[
r + \frac{\big| \varphi(\bar{a}, \mcA) \cap \psi(\bar{a}, \mcA) \big|}{\big| \psi(\bar{a}, \mcA) \big|} \ < \ 
\frac{\big| \theta(\bar{a}, \mcA) \cap \tau(\bar{a}, \mcA) \big|}{\big| \tau(\bar{a}, \mcA) \big|}
\]
then we write
\[
\mcA \  \not\models \ 
\Big( r + \| \varphi(\bar{a}, \bar{y}) \ | \ \psi(\bar{a}, \bar{y}) \|_{\bar{y}} \ \geq \ 
 \| \theta(\bar{a}, \bar{y}) \ | \ \tau(\bar{a}, \bar{y}) \|_{\bar{y}} \Big)
\]
and say that 
$\Big( r + \| \varphi(\bar{a}, \bar{y}) \ | \ \psi(\bar{a}, \bar{y}) \|_{\bar{y}} \ \geq \ 
\| \theta(\bar{a}, \bar{y}) \ | \ \tau(\bar{a}, \bar{y}) \|_{\bar{y}} \Big)$
is false in $\mcA$.

\item The meaning of 
\[
\mcA \  \models \ 
\Big( \| \varphi(\bar{a}, \bar{y}) \ | \ \psi(\bar{a}, \bar{y}) \|_{\bar{y}} \ \geq \ 
\| \theta(\bar{a}, \bar{y}) \ | \ \tau(\bar{a}, \bar{y}) \|_{\bar{y}} + r \Big)
\]
is defined similarly.
\end{enumerate}
\end{enumerate}
}\end{defi}

\section{Probabilistic graphical models, sequences of probability distributions, 
and asymptotic elimination of aggregation functions}\label{PGM}

\noindent
In this section we  define the (parametrized) probabilistic graphical models that will be used
for defining probability distributions on $\mbW_n$, the set of $\sigma$-strutures, for some finite relational
signature $\sigma$, with domain $[n]$.
In particular we define the notion of $L(\sigma)$-network where $L(\sigma)$ is an
arbitrary logic for $\sigma$. The notion of $L(\sigma)$-network is general enough to encompass all
directed probabilistic graphical models that we are aware of 
(e.g. relational Bayesian networks, \cite{Jae98a}, lifted Bayesian networks \cite{Kop20}, 
relational logistic regression models \cite{MBGS, PBKKN}) 
by choosing an appropriate logic $L(\sigma)$).
In most results mentioned in this article we 
consider $L(\sigma)$-networks where $L(\sigma)$ is a sublogic of $PLA^+(\sigma)$ obtained
by putting restrictions on the kind of aggregation functions that may be used.
This is because we want to understand the role of aggregation functions in 
the interplay between those logics used to define probability distributions and (possibly other)
logics used to define queries.
It follows immediately from the definitions that every relational Bayesian network (for $\sigma$) in \cite{Jae98a} 
is  a  $PLA^+(\sigma)$-network, modulo some notational and terminological differences.

In Section~\ref{L-Networks}
we state our first main result,
Theorem~\ref{elimination of strongly admissible aggregation functions},
which is about $PLA^+(\sigma)$-networks and queries that use only strongly admissible (or continuous) aggregation functions,
and its corollaries.
As discussed in Section~\ref{Introduction},
Theorem~\ref{elimination of strongly admissible aggregation functions}
and Corollary~\ref{corollary to main results} 
differ from other results with similar general aims in the field of statistical relational AI, 
e.g. \cite{Jae98a, MBGS, PBKKN, Kop20, KW1},
because they apply to {\em different} sequences of probability distributions or to {\em different} queries.

Throughout this section (as in the rest of the article) we assume that $\sigma$ is a finite relational signature and that $\mbW_n$
denotes the set of all $\sigma$-structures with domain $[n]$.
We begin by defining some notions that are relevant to all  probabilistic graphical models and logics that we consider.
Then we define $L(\sigma)$-networks, lifted Bayesian networks and then state the main results
of \cite{Kop20, KW1} which will be used later.

\subsection{Sequences of probability distributions and asymptotic equivalence}

\noindent
Let $L$ be a logic for $\sigma$.

\begin{defi}\label{definition of asymptotic probability distribution}{\rm
By a {\em sequence of probability distributions} we mean a sequence
$(\mbbP_n : n \in \mbbN^+)$ such that for every $n$, $\mbbP_n$ is a probability distribution on $\mbW_n$.
}\end{defi}

\begin{defi}\label{notation P-n(varphi)}{\rm
Let $\mbbP_n$ is a probability distribution on $\mbW_n$, let $\varphi(\bar{x}) \in L$,
and let $\bar{a} \in [n]^{|\bar{x}|}$. Then 
\[
\mbbP_n(\varphi(\bar{a})) = \mbbP_n\big(\{\mcA \in \mbW_n : \mcA(\varphi(\bar{a})) = 1\}\big).
\]
If $p(\bar{x})$ is a finite set of formulas (e.g. an atomic $\sigma$-type) then we let $\mbbP_n(p(\bar{a})) = \mbbP_n(\varphi(\bar{a}))$
where $\varphi(\bar{x})$ is the conjunction of all formulas in $p(\bar{x})$.
}\end{defi}

\begin{defi}[Asymptotic total variation equivalence]\label{definitions of asymptotic equivalences} {\rm
Two sequences of probability distributions $(\mbbP_n : n \in \mbbN^+)$ and $(\mbbP'_n : n \in \mbbN^+)$ 
are called {\em asymptotically total variation equivalent}, denoted 
$(\mbbP_n : n \in \mbbN^+) \sim_{tv} (\mbbP'_n : n \in \mbbN^+)$, 
if there is a function $\delta : \mbbN^+ \to \mbbR$ such that $\lim_{n\to\infty} \delta(n) = 0$ and 
for all sufficiently large $n$ and every $\mbX \subseteq \mbW_n$, $|\mbbP_n(\mbX) - \mbbP'_n(\mbX)| \leq \delta(n)$.
}\end{defi}

\begin{defi}[Asymptotic equivalence of formulas]\label{definition of asymptotically equivalent formulas}  {\rm
Let $\varphi(\bar{x}), \psi(\bar{x}) \in  L$.\\
(i) Let $\mbbP = (\mbbP_n : n \in \mbbN^+)$ be a sequence of probability distributions.
We say that $\varphi(\bar{x})$ and $\psi(\bar{x})$ are {\em asymptotically equivalent with respect to $\mbbP$} 
if for all $\varepsilon > 0$
\[
\mbbP_n\Big(\big\{\mcA \in \mbW_n : \text{ there is $\bar{a} \in A^{|\bar{x}|}$ such that 
$|\mcA(\varphi(\bar{a})) - \mcA(\psi(\bar{a}))| > \varepsilon$}\big\} \Big) \to 0
\]
as $n \to \infty$.\\
(ii) Suppose in addition that $L$ is a logic such that for every $\varphi(\bar{x}) \in L$, every finite $\sigma$-structure $\mcA$
and every $\bar{a} \in A^{|\bar{a}|}$, the value $\mcA(\varphi(\bar{a}))$ is either 0 or 1.
Then we say that $\varphi(\bar{x}) \in L$ and $\psi(\bar{x}) \in L$ are {\em almost surely equivalent with respect to $\mbbP$} if 
\[
\lim_{n\to\infty} \mbbP_n\big(\big\{\mcA \in \mbW_n : \text{ for all } \bar{a} \in [n]^{|\bar{x}|}, 
\mcA(\varphi(\bar{a})) = \mcA(\psi(\bar{a})) \}\big) \  = \  1.
\]
}\end{defi}

\noindent
For logics with only the truth values 0 and 1, the notions of asymptotic equivalence and almost sure equivalence are equivalent
as stated by the following lemma, the straightforward proof of which is left to the reader.

\begin{lem}\label{asymptotic equivalence is equivalent to almost sure equivalence}
Let $\mbbP = (\mbbP_n : n \in \mbbN^+)$ be a sequence of probability distributions and
suppose that $L$ is a logic such that for every $\varphi(\bar{x}) \in L$, every finite $\sigma$-structure $\mcA$
and every $\bar{a} \in A^{|\bar{a}|}$, the value $\mcA(\varphi(\bar{a}))$ is either 0 or 1.
Let $\varphi(\bar{x}), \psi(\bar{x}) \in L$.
Then $\varphi(\bar{x})$ and  $\psi(\bar{x})$ are asymptotically equivalent with respect to 
$\mbbP$ if and only if they are almost surely equivalent with respect to $\mbbP$.
\end{lem}

\subsection{\texorpdfstring{$L(\sigma)$-Networks}{L(sigma)-Networks}}\label{L-Networks}

Now we define the notion of $L(\sigma)$-network, where $L(\sigma)$ is a logic for $\sigma$, and how
it induces a probability distribution on $\mbW_n$. 

\begin{defi}[$L(\sigma)$-network]\label{definition of L-network}  {\rm
Suppose that $L(\sigma)$ is a logic for $\sigma$ and suppose that, for every $\sigma' \subset \sigma$, 
$L(\sigma')$ is a logic for $\sigma'$ and whenever $\sigma'' \subset \sigma' \subset \sigma$, then
$L(\sigma'') \subset L(\sigma') \subset L(\sigma)$.\\
(i) A {\em $L(\sigma)$-network} is determined by the following two components:
\begin{enumerate}
\item A DAG $\mbbG$ with vertex set $\sigma$.

\item To each relation symbol $R \in \sigma$ a formula $\theta_R(\bar{x}) \in L(\mr{par}(R))$ 
(where $\mr{par}(R)$ is the set of parents of $R$ in the DAG)
is associated where $|\bar{x}|$ equals the arity of $R$.
We call $\theta_R$ the {\em probability formula associated to $R$} by the network.
\end{enumerate}
(ii) For technical reasons it will be convenient to consider, for the empty signature $\sigma = \es$, a (unique) $L(\sigma)$-network,
denoted $\mbbG^\es$, such that its underlying DAG has empty vertex set and consequently no
probability formula.\\
(iii) Let $\mbbG$ denote an $L(\sigma)$-network, let $\sigma' \subseteq \sigma$, and suppose that for
every $R \in \sigma'$, $\mr{par}(R) \subseteq \sigma'$. 
Then the $L(\sigma')$-network specified by the induced subgraph of the underlying DAG of $\mbbG$ with vertex set $\sigma'$
and the probability formulas $\theta_R$ for all $R \in \sigma'$ will be called the
{\em $L(\sigma')$-subnetwork of $\mbbG$ induced by $\sigma'$}.
}\end{defi}

\noindent
{\em We use the convention to denote an $L(\sigma)$-network by the same symbol (e.g. $\mbbG$) as its underlying DAG.}

It follows immediately from its definition that $PLA^+(\sigma)$ and all sublogics of it described in
Definition~\ref{restrictions of PLA+}
satisfy the conditions in the beginning of the above definition, so 
it makes sense to talk about for example a $PLA^+(\sigma)$-network, a $PLA(\sigma)$-network, a $aPLA(\sigma)$-network,
a $coPLA^+(\sigma)$-network, and an $afPLA(\sigma)$-network.

Now we define how an $L(\sigma)$-network defines a probability distribution on $\mbW_n$.

\begin{defi}[The sequence of probability distributions induced by an $L(\sigma)$-network]\label{the probability distribution induced by an L-network}
 {\rm
Let $L(\sigma)$ be a logic for $\sigma$ which satisfies 
the conditions in the beginning of
Definition~\ref{definition of L-network}
and let $\mbbG$ be an $L(\sigma)$-network.\\
(i) If $\sigma$ is empty then $\mbbP_n$, the probability distribution on $\mbW_n$ induced by $\mbbG$, 
is the unique probability distribution on (the singleton set) $\mbW_n$.\\
(ii) Now suppose that $\sigma$ is nonempty and suppose tat
for each $R \in \sigma$,  its arity is denoted by $k_R$ and the
probability formula
corresponding to $R$ is denoted by $\theta_R(\bar{x})$ where $|\bar{x}| = k_R$.
Suppose that the underlying DAG of $\mbbG$ has mp-rank $\rho$.
For each $0 \leq r \leq \rho$ let $\mbbG_r$ be the subnetwork which is
induced by $\sigma_r = \{R \in \sigma : \mr{mp}(R) \leq r\}$ and note that $\mbbG_\rho = \mbbG$.
Also let $\mbbG_{-1} = \mbbG^\es$ and let $\mbbP^{-1}_n$ be the unique probability distribution on $\mbW^{-1}_n = \mbW^\es_n$.
By induction on $r$ we define, for every $r = 0, 1, \ldots, \rho$, a probability distribution $\mbbP^r_n$ on the set $\mbW^r_n$ of all
$\sigma_r$-structures with domain $[n]$ as follows:
For every $\mcA \in \mbW^r_n$, let $\mcA' = \mcA \uhrc \sigma_{r-1}$ and
\[
\mbbP^r_n(\mcA) \ = \ \mbbP^{r-1}_n(\mcA') 
\prod_{R \in \sigma_r \setminus \sigma_{r-1}} \ \prod_{\bar{a} \in R^\mcA} 
\mcA' \big(\theta_R(\bar{a})\big) \ 
\prod_{\bar{a} \in [n]^{k_R} \ \setminus \ R^\mcA} \big( 1 - \mcA' \big(\theta_R(\bar{a})\big) \big).
\]
Finally we let $\mbbP_n = \mbbP^\rho_n$ and note that $\mbW_n = \mbW^\rho_n$, so 
$\mbbP_n$ is a probability distribution on $\mbW_n$ and we call
$(\mbbP_n : n \in \mbbN^+)$
{\em the sequence of probability distributions induced by $\mbbG$}.
}\end{defi}

\noindent
From the above definition it follows immediately that if a probability distribution on $\mbW_n$ can be 
induced by a relational Bayesian network (for $\sigma$) in the sense of \cite{Jae98a} then it can be induced by a
$PLA^+(\sigma)$-network.
The following lemma is a straightforward consequence of Definition~\ref{definition of L-network}
and it uses the notation and assumptions of this definition.

\begin{lem}\label{lemma on conditional probability for R}
If $0 \leq r \leq \rho$, $R \in \sigma_r$ has arity $k$, $\theta_R(\bar{x}) \in L(\mr{par}(R))$ is the formula associated to $R$ according to 
Definition~\ref{definition of L-network},
$n \in \mbbN^+$, $\bar{a} \in [n]^k$ and $\mcA \in \mbW^{r-1}_n$, then
\[
\mbbP^r_n\big( \{ \mcB \in \mbW^r_n : \mcB \models R(\bar{a}) \} \ | \ 
\{ \mcB \in \mbW_n : \mcB \uhrc \sigma_{r-1} = \mcA \}\big) \  = \ 
\mcA(\theta_R(\bar{a})).
\]
\end{lem}

\begin{defi}\label{asymptotically euivalent with respect to a network}{\rm
Let $\mbbG$ be an $L(\sigma)$-network.
We say that $\varphi(\bar{x})$ and $\psi(\bar{x})$ are {\em asymptotically equivalent with respect to to $\mbbG$},
denoted $\varphi(\bar{x}) \sim_{\mbbG} \psi(\bar{x})$, if $\varphi(\bar{x})$ and $\psi(\bar{x})$ are asymptotically 
equivalent with respect to the sequence of probability distributions induced by $\mbbG$.
}\end{defi}

\begin{exa}\label{example of PLA-networks} {\rm
We give examples of (conditional) probabilities that can be modelled with a $PLA^+(\sigma)$-network.
Let $\sigma$ be a finite relational signature and 
suppose that $\mbbG$ is a DAG with vertex set $\sigma$.
Let $R \in \sigma$. We show how the probability of $R$, conditioned on $\mr{par}(R)$, 
may be expressed. In all examples, we need to define a 
formula $\theta_R(\bar{x}) \in PLA^+(\sigma)$
as in Definition~\ref{definition of L-network}
which assigns the probability that $R(\bar{x})$ holds, conditioned on the values of the relations in $\mr{par}(R)$.

If $R$ $\mr{par}(R) = \es$ (i.e. if $R$ has no parents in the DAG) 
then we can let $\theta_R(\bar{x})$ be some constant $c \in [0, 1]$, meaning
that the probability of $R(\bar{x})$ is $c$. Or we can (for example) let $\alpha \in (0, 1)$
and then let $\theta_R(\bar{x})$ be 
$length_\alpha( y = y : y : )$ (where $y \notin \rng(\bar{x})$). 
This expresses that the probability of $R(\bar{x})$ is $n^{-\alpha}$ where $n$
is the cardinality of the domain.

For all remaining examples suppose that $\mr{par}(R) \neq \es$.
If we want to express that the probability of $R(\bar{x})$ depends (only) on the complete
atomic $\mr{par}(R)$-type that $\bar{x}$ satisifies, then we can let $\theta_R(\bar{x})$ be
$\bigwedge_i (p_i(\bar{x}) \to \gamma_i)$ where $p_i(\bar{x})$ ranges over all complete atomic $\mr{par}(R)$-types
in the variables $\bar{x}$. This expresses that if $p_i(\bar{x})$ holds then the probability of $R(\bar{x})$ is 
the value of the formula $\gamma_i$ (without free variables).
Here $\gamma_i$ may (for example) be a constant $c_i \in [0, 1]$ expressing that if $p_i(\bar{x})$ holds,
then the probability of $R(\bar{x})$ is $c_i$, or it may (for example)
be a $PLA^+(\sigma)$-formula of the form $length_\alpha( y = y : y : )$, expressing that if $p_i(\bar{x})$ holds,
then the probability of $R(\bar{x})$ is $n^{-\alpha}$ where $n$ is the cardinality of the domain.

Let $\varphi_1(x, y), \ldots, \varphi_k(x, y) \in PLA^+(\mr{par}(R))$.
If $R$ is binary then we can let $\theta_R(x, y)$ be the formula $\psi(x, y)$ from 
Example~\ref{similarity measure}
with $E_i$ replaced by $\varphi_i$, 
which expresses that the probability of $R(x, y)$ equals the similarity of $x$ and $y$ with respect to this
similarity measure.

If $R$ is unary then we can instead let 
$\theta_R(x)$ be the formula
$\mr{am}(\psi(x, y) : y : x \neq y)$
which expresses that the probability of $R(x)$ equals the average similiarity of $x$ to other elements
with respect to $\varphi_1(x, y), \ldots, \varphi_k(x, y)$.

Suppose again that $R$ is binary. Then we can let $\theta_R(x, y)$ be
\[
1 - \mu^u_1\big( \psi(x, z), \psi(y, z) : z : x \neq z, y \neq z \big)
\]
which 
(recalling Example~\ref{similarity profile})
expresses that the probability of $R(x, y)$ equals the ``proximity'' of the similarity profiles 
of $x$ and $y$.
Alternatively, if $E \in \mr{par}(R)$ then $\theta_R(x, y)$ can, for any $l \in \mbbN$, be
the $PLA^+(\mr{par}(R))$-formula which expresses the $l$th stage (or approximation) of SimRank with respect
to $E$. This means that the probability of $R(x, y)$ equals the 
$l$th  stage of the approximation of SimRank of $x$ and $y$.
}\end{exa}

\noindent
Now we have the concepts that allow us to state our first main result.
Recall Definition~\ref{restrictions of PLA+}
of $coPLA(\sigma)$.

\begin{thm}[Asymptotic elimination of continuous aggregation functions]\label{elimination of strongly admissible aggregation functions} 
Let $\sigma$ be a finite relational signature, let 
$\mbbG$ be a $coPLA^+(\sigma)$-network and 
let $(\mbbP_n : n \in \mbbN^+)$ be the sequence of probability distributions induced by $\mbbG$.
\begin{enumerate}
\item[(i)] If $\varphi(\bar{x}) \in coPLA^+(\sigma)$ 
then $\varphi(\bar{x})$ is asymptotically equivalent to a basic probability formula with respect to $\mbbG$.

\item[(ii)] For every atomic $\sigma$-type $p(\bar{x})$, every $m \in \mbbN^+$ and every $\bar{a} \in [m]^{|\bar{x}|}$,
$\lim_{n\to\infty} \mbbP_n(p(\bar{a}))$ exists and depends only on $p$ and $\mbbG$.
\end{enumerate}
\end{thm}

\begin{rem}[Expressivity of $coPLA^+$] {\rm
Recall that $coPLA^+$ may only use continuous aggregation functions.
The aggregation functions am and $\mu_1^u$ are continuous, as stated before.
Therefore the formulas from examples~\ref{similarity measure} and~\ref{similarity profile} that express
``the similarity of $x$ and $y$'', respectively ``the similarity profile score och $x$ and $y$''
are in $coPLA^+$.

But max and min are not continuous and cannot be expressed in some indirect way by $coPLA^+$,
for otherwise Corollary~\ref{corollary to main results} would contradict a nonconvergence result in \cite{SS}
as discussed in the introduction. It follows that the first-order quantifiers cannot be expressed by $coPLA^+$,
so it does not subsume first-order logic.

However in some contexts one may be more interested in the proportion (rather than existence) of elements
that satisfy some (two-valued) formula. For this we can use the continuous aggregation function am which returns this proportion.
If we want to express that the proportion of elements that satisfy a formula is at least as large as the proportion that
satisfies another formula, then we could, tentatively, use the connective $\leq(x, y)$ which returns 1 if $x \leq y$ and 0
otherwise. 
But since this connective is not continuous it cannot be used in $PLA^+$, so instead we consider a continuous
approximation of $\leq$, say $\leq_\varepsilon(x, y)$ = 1 if $x \leq y$, 0 if $x \geq y + \varepsilon$,
and $(y + \varepsilon - x)/\varepsilon$ if $y < x < y + \varepsilon$, where $\varepsilon > 0$ is small.
With $\leq_\varepsilon$ and am we can, within $coPLA^+$, 
express approximations of statements about proportions.
The idea of considering continuous approximations of connectives or aggregation functions can
of course be used more generally to get $coPLA^+$-formulas that estimate discrete/discontinuous properties.
}\end{rem}

\noindent
The proof of Theorem~\ref{elimination of strongly admissible aggregation functions}
consists of two parts. One part is to show that if, with high probability, structures in $\mbW_n$
satisfy certain saturation conditions (defined in Section~\ref{asymptotic elimination of aggregation functions})
then strongly admissible aggregation functions can be asymptotically eliminated.
This is stated by Proposition~\ref{saturation implies elimination of strongly admissible functions}
and proved in Section~\ref{asymptotic elimination of aggregation functions}.
The other part is to prove that with probability tending to 1 as the domain size tends to infinity,
a random structure from $\mbW_n$ satisfies these saturation conditions. 
This is proved in Section~\ref{proof of main theorem}.
The following corollary is quickly derived from Theorem~\ref{elimination of strongly admissible aggregation functions}.

\begin{cor}[Convergence of probability]\label{corollary to main results} 
Let $\sigma$ be a finite relational signature, let
$\mbbG$ be a $coPLA^+(\sigma)$-network, and
let $(\mbbP_n : n \in \mbbN^+)$ be the sequence of probability distributions induced by $\mbbG$.
If $\varphi(\bar{x}) \in coPLA^+(\sigma)$ 
then there are $c_1, \ldots, c_k \in [0, 1]$, depending only on $\varphi$ and $\mbbG$, such that
for every $m \in \mbbN^+$, every $\bar{a} \in [m]^{|\bar{x}|}$ and every $\varepsilon > 0$,
\begin{align*}
&\lim_{n\to\infty} \mbbP_n\big(\big\{ \mcA \in \mbW_n : 
\mcA(\varphi(\bar{a})) \in \bigcup_{i=1}^k[c_i - \varepsilon, c_i + \varepsilon] \big\} \big) \ = \ 1\\ 
&\text{and for all $i = 1, \ldots, k$ } \\
&\mbbP_n\big(\big\{\mcA \in \mbW_n : |\mcA(\varphi(\bar{a})) - c_i| < \varepsilon \big\}\big) 
\text{ converges as  $n\to\infty$} \\
&\text{to a number which depends only on $\varphi$, $c_i$ and $\mbbG$.}
\end{align*}
\end{cor}

\proof
Let $\mbbG$ and $\varphi(\bar{x})$ be as assumed.
By Part~(i) of
Theorem~\ref{elimination of strongly admissible aggregation functions},
there is a basic probability formula $\psi(\bar{x})$ which is asymptotically equivalent to $\varphi(\bar{x})$
with respect to $\mbbG$.
Then $\psi(\bar{x})$ has the form $\bigwedge_{i=1}^k (\psi_i(\bar{x}) \to c_i)$ where, for each $i$,
$c_i \in [0, 1]$ and $\psi_i(\bar{x})$ is a
conjunction of first-order literals. 
Without loss of generality we can assume that each $\psi_i$ is the conjunction of all 
formulas in a complete atomic $\sigma$-type and that every complete atomic $\sigma$-type in the variables $\bar{x}$ 
is represented by some $\psi_i$.
Note that for every $\mcA \in \mbW_n$ and every $\bar{a} \in [n]^{|\bar{x}|}$ we have 
$\mcA(\psi(\bar{a})) \in \{c_1, \ldots, c_k\}$ and $\mcA(\psi(\bar{a})) = c_i$ if $\mcA \models \psi_i(\bar{a})$.
Let $c \in \{c_1, \ldots, c_k\}$ and suppose that $i_1, \ldots, i_t$ enumerates all $i$ such that $c_i = c$.
Then 
\[
\mbbP_n\big(\{\mcA \in \mbW_n : \mcA(\psi(\bar{a})) = c\}\big) = \mbbP_n\big(\bigvee_{j=1}^t \psi_{i_j}(\bar{a})\big)
= \sum_{j=1}^t \mbbP_n\big(\psi_{i_j}(\bar{a})\big).
\]
By Part~(ii) of
Theorem~\ref{elimination of strongly admissible aggregation functions}, 
it follows that the above probability converges, as $n\to\infty$,
to a number which depends only on $\psi$ and $\mbbG$.
Since $\varphi(\bar{x}) \sim_\mbbG \psi(\bar{x})$ the conclusions of the corollary follow.
\qed

\subsection{Lifted Bayesian networks}\label{Lifted Bayesian networks}

\noindent
The notions and results of this subsection are not needed until 
Section~\ref{inference frameworks},
but we introduce them here since 
Section~\ref{PGM}
has introduced the other probabilistic graphical model considered here, the $L(\sigma)$-network (e.g. $PLA^+(\sigma)$-network).
In one of our main results, Theorem~\ref{partial order of inference frameworks},
we  will consider (sequences of) probability distributions induced by lifted Bayesian networks in 
the sense of \cite{Kop20}. 
It is not hard (by combining $CPL$, constants in $[0, 1]$, and the ``weighted mean'' connective in
Definition~\ref{special connectives}) to construct a logic, say $CPL^*(\sigma)$, such that every 
lifted Bayesian network for $\sigma$ corresponds to an $CPL^*(\sigma)$-network.
In order to avoid the somewhat cumbersome translation of results in \cite{Kop20} about lifted Bayesian networks to
results about $CPL^*(\sigma)$-networks we do not however do this, but instead 
we formulate the results in this section in their original form which uses
the notion of a lifted Bayesian network.
The intuition behind the next definition is that if $R \in \sigma$ and the condition expressed by 
the $CPL$-formula $\chi_{R, i}(\bar{x})$
holds, then the probability of $R(\bar{x})$ is (the number) $\mu(R \ | \ \chi_{R, i})$.

\begin{defi}[Lifted Bayesian network]\label{definition of BN}  {\rm
Let $\sigma$ be a finite relational signature.
A {\em lifted Bayesian network for $\sigma$} is determined by the following components:
\begin{enumerate}
\item[(a)] An acyclic directed graph (DAG) $\mbbG$ with vertex set $\sigma$.

\item[(b)] For each $R \in \sigma$, a number $\nu_R \in \mbbN^+$, formulas $\chi_{R, i}(\bar{x}) \in CPL(\mr{par}(R))$,
for $i = 1, \ldots, \nu_R$, where $|\bar{x}|$ equals the arity of $R$, such that
$\forall \bar{x} \big( \bigvee_{i = 1}^{\nu_R} \chi_{R, i}(\bar{x})\big)$ is valid (i.e. true in all $\mr{par}(R)$-structures) and if
$i \neq j$ then $\exists \bar{x} \big(\chi_{R, i}(\bar{x}) \wedge \chi_{R, j}(\bar{x})\big)$
is unsatisfiable.
Each $\chi_{R, i}$ will be called an {\em aggregation formula (of $\mbbG$)}.

\item[(c)] For each $R \in \sigma$ and each $1 \leq i \leq \nu_R$, 
a number denoted $\mu(R \ | \ \chi_{R, i})$ (or $\mu(R(\bar{x}) \ | \ \chi_{R, i}(\bar{x}))$)
in the interval $[0, 1]$.
\end{enumerate}
}\end{defi}

\noindent
Observe that Definition~\ref{definition of BN} makes sense if $\sigma$ is empty. In this case
the underlying DAG has empty vertex set (and edge set) and no numbers or formulas as
in parts~(b) and~(c) of the definition need to be specified.
A lifted Bayesian network $\mbbG$ for $\sigma$ induces a probability distribution on $\mbW_n$
in a way explained in \cite[Definition~3.11]{Kop20}, but as we will not need the details we omit the definition here.
We refer to \cite[Example~5.3]{KW1} for an example of a lifted Bayesian network.

Now we state previous results about lifted Bayesian networks, $CPL$ and $PLA$ 
in \cite{Kop20} and \cite{KW1} that will be 
used in proving Theorem~\ref{partial order of inference frameworks}
which describes the relative asymptotic expressivity of the inference frameworks considered in this article.
These results use the notion of a (non)critical $CPL$-formula. 
The intuition behind this notion is that a $CPL(\sigma)$-formula is {\em critical} with respect to a lifted Bayesian network 
$\mbbG$ for $\sigma$ if it contains a subformula
of the form
\begin{align*}
\Big( r + \| \varphi \ |  \ \psi \|_{\bar{y}} \ \geq \ 
\| \theta \ |  \ \tau \|_{\bar{y}} \Big) 
\ \ \text{ or} \ \ 
\Big( \| \varphi \ |  \ \psi \|_{\bar{y}} \ \geq \ 
\| \theta \ |  \ \tau \|_{\bar{y}} + r \Big).
\end{align*}
where $r = \alpha - \beta$, $\alpha = \lim_{n\to\infty}\frac{\mbbP_n(\varphi_1(\bar{a}))}{\mbbP_n(\varphi_2(\bar{a}))}$,
$\beta =  \lim_{n\to\infty}\frac{\mbbP_n(\psi_1(\bar{a}))}{\mbbP_n(\psi_2(\bar{a}))}$ and 
$\varphi(\bar{x})$ and $\psi(\bar{x})$ are quantifier-free first-order formulas, the length of $\bar{x}$ is
bounded in terms of the length of the subformula, 
$\varphi_1(\bar{x})$ implies $\varphi_2(\bar{x})$, and $\psi_1(\bar{x})$ implies $\psi_2(\bar{x})$.
Otherwise the $CPL(\sigma)$-formula is {\em noncritical} with respect to $\mbbG$.
Observe that {\em every first-order formula is noncritical with respect to every lifted Bayesian network}, since
a first-order formula does not contain a subformula of the form considered above
The numbers $\alpha$ and $\beta$ above depend only on the underlying DAG of $\mbbG$ and
the numbers $\mu(R \ | \ \chi_{R, i})$ in the definition of a lifted Bayesian network.
Therefore it makes sense to say that an aggregation formula of $\mbbG$ is (or is not) noncritical with respect to $\mbbG$.
The exact definition of noncritical formula in \cite{Kop20} is quite technical and appears in 
\cite[Definitions~4.29 and~4.30]{Kop20}.
A simplified and stronger definition of noncritical $CPL(\sigma)$-formula
is given in \cite[Definition~6.7]{KW1}.

\begin{thmC}[{\cite[Theorems 3.14 -- 3.16]{Kop20}}]\label{main result on quantifier elimination} 
Let $\mbbG$ be a lifted Bayesian network for $\sigma$, suppose that all aggregation functions of $\mbbG$
are noncritical with respect to $\mbbG$, and
let $\mbbP = (\mbbP_n : n \in \mbbN^+)$ be the sequence of probability distributions induced by $\mbbG$.
\begin{enumerate}
\item[(i)] If $\varphi(\bar{x}) \in CPL(\sigma)$ is noncritical with respect to $\mbbG$, then $\varphi(\bar{x})$
is almost surely equivalent to a quantifier-free formula (with respect to $\mbbP$). 
Moreover, if $m \in \mbbN^+$ and $\bar{a} \in [m]^{|\bar{x}|}$, then $\lim_{n\to\infty} \mbbP_n(\varphi(\bar{a}))$
exists. In particular, if $\varphi$ has no free variable (i.e. is a sentence), then $\lim_{n\to\infty} \mbbP_n(\varphi)$ 
is either~0 or~1.

\item[(ii)] For every aggregation formula $\chi_{R, i}$ of $\mbbG$ there is a
quantifier-free formula $\chi'_{R, i} \in CPL(\sigma)$ such that every relation symbol of $\chi'_{R, i}$ occurs
in $\chi_{R, i}$ and such that if $\mbbG'$ is the lifted Bayesian network for $\sigma$
\begin{enumerate}
\item with the same underlying DAG as $\mbbG$, and

\item where, for every $R \in \sigma$, the aggregation formula $\chi_{R, i}$ is replaced by $\chi'_{R, i}$
and $\mu(R \ | \ \chi'_ {R, i}) = \mu(R \ | \ \chi_{R, i})$,
\end{enumerate}
then $(\mbbP'_n : n \in \mbbN^+) \sim_{tv} (\mbbP_n : n \in \mbbN^+)$ where 
$(\mbbP'_n : n \in \mbbN^+)$ is the sequence of probability distributions induced by $\mbbG'$.
\end{enumerate}
\end{thmC}

\begin{rem}{\rm
The conclusion of  Theorem~3.16 in \cite{Kop20} is somewhat weaker than part~(ii)
of Theorem~\ref{main result on quantifier elimination} above since the former does not refer to 
asymptotic total variation equivalence, but the proof of \cite[Theorem~3.16]{Kop20}
shows, using the notation above, that $(\mbbP'_n : n \in \mbbN^+) \sim_{tv} (\mbbP_n : n \in \mbbN^+)$
(as seen from the proof of \cite[Corollary~4.42~(c)]{Kop20}).
}\end{rem}

\begin{thmC}[{\cite[Theorem~6.8]{KW1}}]\label{elimination of aggregation functions for a lifted BN} 
Let $\mbbG$ be a lifted Bayesian network for $\sigma$ and suppose that all aggregation functions of $\mbbG$
are noncritical with respect to $\mbbG$.
If $\varphi(\bar{x}) \in PLA(\sigma)$ and all aggregation functions in $\varphi$ are admissible, 
then $\varphi(\bar{x})$ is asymptotically equivalent to a basic probability formula with respect to $\mbbG$.
\end{thmC}

\section{Asymptotic elimination of strongly admissible aggregation functions}\label{asymptotic elimination of aggregation functions}

\noindent
In this section we prove a result, 
Proposition~\ref{saturation implies elimination of strongly admissible functions}, 
which is one component of the proof of 
Theorem~\ref{elimination of strongly admissible aggregation functions}
and of independent interest
since it shows that if a saturation condition 
(given by Definition~\ref{definition of saturation} and
Assumption~\ref{assumption about saturation})
holds almost surely,
then strongly admissible aggregation functions can be eliminated from $PLA^+(\sigma)$-formulas.

Throughout this section, for each $n \in \mbbN^+$, $\mbbP_n$ is a probability distribution on $\mbW_n$, 
the set of all $\sigma$-structures with domain $[n]$,
where $\sigma$ is a finite relational signature.
When saying that two formulas are asymptotically equivalent then it is with respect to $\mbbP = (\mbbP_n : n \in \mbbN^+)$.
When denoting an atomic $\sigma$-type by notation like $p(\bar{x}, \bar{y})$, or a formula by $\varphi(\bar{x}, \bar{y})$, we assume that
$\bar{x}$ and $\bar{y}$ are sequences of different variables and
$\rng(\bar{x}) \cap \rng(\bar{y}) = \es$ although this assumption may be repeated.

Let $p(\bar{x}, \bar{y})$ be an atomic $\es$-type, let $\mcA$ be a finite $\sigma$-structure
 and let $\bar{a} \in A^{|\bar{x}|}$. 
The equalities and inequalities in $p(\bar{x}, \bar{y})$ specifies how many ``degrees of freedom'' we have 
for choosing $\bar{b} \in A^{|\bar{y}|}$ such that $p(\bar{a}, \bar{b})$ holds.
Since the ``degrees of freedom'' for $\bar{y}$ when $\bar{x}$ has been assigned some fixed values will
play a role in the proof that will follow
we now define the $\bar{y}$-dimension of $p(\bar{x}, \bar{y})$ which captures this idea.

\begin{defi}\label{definition of y-dimension}{\rm
Let $p(\bar{x}, \bar{y})$ be an atomic $\sigma$-type. The {\em $\bar{y}$-dimension of $p(\bar{x}, \bar{y})$},
denoted $\dim_{\bar{y}}(p)$, is the maximal $d \in \mbbN$ such that there are a $\sigma$-structure $\mcA$,
$\bar{a} \in A^{|\bar{x}|}$ and $\bar{b} \in A^{|\bar{y}|}$ such that $\mcA \models p(\bar{a}, \bar{b})$
and $|\rng(\bar{b}) \setminus \rng(\bar{a})| \geq d$.
}\end{defi}

\noindent
Observe that if $p^=(\bar{x}, \bar{y})$ is an atomic $\es$-type, $d = \dim_{\bar{y}}(p)$, $\mcA \in \mbW_n$,
and $\bar{a} \in [n]^{|\bar{x}|}$ satisfies $p^= \uhrc \bar{x}$, 
then (for large enough $n$) $|p^=(\bar{a}, \mcA)| = (n - |\rng(\bar{a})|)^d \sim n^d$.
The intuitive content of Assumption~\ref{assumption about saturation} below
is that for every complete atomic $\sigma$-type $p(\bar{x}, \bar{y})$ and $q(\bar{x}) = p \uhrc \bar{x}$,
there is $\alpha \in [0, 1]$ such that with high probability (for large $n$), if $q(\bar{a})$ holds
then the proportion of $\bar{b}$ such that $p(\bar{a}, \bar{b})$ holds is close to $\alpha$.
The next definition will be used to make this idea more precise.

\begin{defi}[Saturation and unsaturation]\label{definition of saturation} {\rm
Let $\bar{x}$ and $\bar{y}$ be sequences of different variables such that 
$\rng(\bar{x}) \cap \rng(\bar{y}) = \es$ and let
$p(\bar{x}, \bar{y})$ and $q(\bar{x})$ be atomic $\sigma$-types such that $q \subseteq p$.
Let also $0 \leq \alpha \leq 1$ and $d = \dim_{\bar{y}}(p)$.
\begin{enumerate}
\item A finite $\sigma$-structure $\mcA$
is called {\em $(p, q, \alpha)$-saturated} if, whenever $\bar{a} \in A^{|\bar{x}|}$ and $\mcA \models q(\bar{a})$, then
$\big| \{\bar{b} \in A^{|\bar{y}|} : \mcA \models p(\bar{a}, \bar{b})\} \big| \geq \alpha |A|^d$.

\item A finite $\sigma$-structure $\mcA$
is called {\em $(p, q, \alpha)$-unsaturated} if, whenever $\bar{a} \in A^{|\bar{x}|}$ and $\mcA \models q(\bar{a})$, then
$\big| \{\bar{b} \in A^{|\bar{y}|} : \mcA \models p(\bar{a}, \bar{b})\} \big| \leq \alpha |A|^d$.
\end{enumerate}
}\end{defi}

\begin{asm}\label{assumption about saturation}{\rm
For all $m, n \in \mbbN^+$ and $\delta > 0$ there is $\mbY^{m, \delta}_n \subseteq \mbW_n$
such that
\begin{enumerate}
\item $\lim_{n\to\infty}\mbbP_n(\mbY^{m, \delta}_n) = 1$, and

\item for every complete atomic $\sigma$-type $p(\bar{x}, \bar{y})$ such that $|\bar{x}| + |\bar{y}| \leq m$,
if $q(\bar{x}) = p \uhrc \bar{x}$ and $\dim_{\bar{y}}(p) > 0$, 
then there is $\alpha_{p, q} \in [0, 1]$ depending only on $p$, $q$ and $\mbbP$, such that 
every $\mcA \in \mbY^{m, \delta}_n$ 
is $(p, q, \alpha_{p, q} - \delta)$-saturated and $(p, q, \alpha_{p, q} + \delta)$-unsaturated.
\end{enumerate}
}\end{asm}

\begin{prop}\label{saturation implies elimination of strongly admissible functions}
If Assumption~\ref{assumption about saturation} holds and $\varphi(\bar{x}) \in PLA^+(\sigma)$ contains only
strongly admissible aggregation functions, then $\varphi(\bar{x})$ is asymptotically equivalent to a basic probability formula.
\end{prop}

\noindent
The rest of this section is devoted to proving 
Proposition~\ref{saturation implies elimination of strongly admissible functions}
and its proof is concluded by 
Corollary~\ref{elimination of aggregation functions if saturation conditions hold}.
The proofs follow the pattern of the proofs in \cite[Section 7]{KW1}, 
but there are subtle differences throughout.
The reason is partly that, unlike $PLA(\sigma)$, $PLA^+(\sigma)$ allows constructions
as in part~(5) of Definition~\ref{syntax of PLA+}
where $p^=$ is not necessarity a {\em complete} atomic $\sigma$-type,
and partly that we allow the numbers $\alpha_{p, q}$ in Assumption~\ref{assumption about saturation}
to be 0 and we did not need to bother with this ``convergence to 0 case'' in \cite{KW1}.

\begin{rem}[Eliminating aggregation functions of higher arities]\label{remark on higher arities} {\rm
The results below up to Proposition~\ref{elimination of one aggregation function}
are stated and proved only for (unary) admissible aggregations functions $F : [0, 1]^{<\omega} \to [0, 1]$ but the results hold
also for admissible aggregation functions $F: \big([0, 1]^{<\omega}\big)^k \to [0, 1]$, where $k > 1$,
and basic probability formulas $\psi_i(\bar{x}, \bar{y})$, $i = 1, \ldots, k$ 
(in place of $\psi(\bar{x}, \bar{y})$). 
The proofs in the general case work out in the same way but the notation becomes messier since the assumptions and notation introduced 
in Assumption~\ref{assumptions in results for eliminating an aggregation function} below
for $\psi(\bar{x}, \bar{y})$ need to be considered for all $\psi_i(\bar{x}, \bar{y})$.
}\end{rem}

\begin{lem}\label{reduction to a simpler basic formula}
Suppose that $\psi(\bar{x}, \bar{y}) = \bigwedge_{i=1}^t (\psi_i(\bar{x}, \bar{y}) \to c_i)$ is a basic formula where each $\psi_i$ is
a conjunction of literals.
Let be $p^=(\bar{x}, \bar{y})$ be an atomic $\es$-type and 
let $F : [0, 1]^{<\omega} \to [0, 1]$ be an aggregation function.
\begin{enumerate}
\item[(i)] Suppose that, for some $1 \leq s \leq t$,
$\psi_i(\bar{x}, \bar{y}) \wedge p^=(\bar{x}, \bar{y})$ is consistent if $1 \leq i \leq s$ and inconsistent if $i > s$.
Then for every finite $\sigma$-structure $\mcA$ and every $\bar{a} \in A^{|\bar{x}|}$ that
satisfies $p^= \uhrc \bar{x}$,
\[
\mcA\big(F\big(\psi(\bar{a}, \bar{y}) : \bar{y} : p^=(\bar{a}, \bar{y})\big)\big) \ = \ 
\mcA\big(F\big(\bigwedge_{i=1}^s (\psi_i(\bar{a}, \bar{y}) \to c_i) : \bar{y} : p^=(\bar{a}, \bar{y})\big)\big).
\]
\item[(ii)] If $F$ is admissible and
$\psi_i(\bar{x}, \bar{y}) \wedge p^=(\bar{x}, \bar{y})$ is inconsistent for all $i = 1, \ldots, t$,
then for all finite $\sigma$-structures $\mcA$ and all 
$\bar{a} \in A^{|\bar{x}|}$ that satisfy $p^= \uhrc \bar{x}$,
\[
\mcA\big(F\big(\psi(\bar{a}, \bar{y}) : \bar{y} : p^=(\bar{a}, \bar{y})\big)\big) \ = \ F(\bar{r})
\]
where $\bar{r}$ is the sequence of length 1 the unique entry of which is 1.
\end{enumerate}
\end{lem}

\proof
(i) Suppose that the assumptions of~(i) hold.
If $i > s$ then $\psi_i(\bar{x}, \bar{y}) \wedge p^=(\bar{x}, \bar{y})$ is inconsistent, so
for every finite $\sigma$-structure $\mcA$, every $\bar{a} \in A^{|\bar{x}|}$ and every $\bar{b} \in A^{|\bar{y}|}$,
if $\mcA \models p^=(\bar{a}, \bar{b})$ then $\mcA(\psi_i(\bar{a}, \bar{b})) = 0$ and hence
$\mcA(\psi_i(\bar{a}, \bar{b}) \to c_i) = 1$.
Now we get 
\begin{align*}
&\mcA\big(\bigwedge_{i=1}^t (\psi_i(\bar{a}, \bar{b}) \to c_i)\big) = 
\min\{ \mcA(\psi_i(\bar{a}, \bar{b}) \to c_i) : i = 1, \ldots, t\} = \\
&\min\{ \mcA(\psi_i(\bar{a}, \bar{b}) \to c_i) : i = 1, \ldots, s\} =
\mcA\big(\bigwedge_{i=1}^s(\psi_i(\bar{a}, \bar{b}) \to c_i)\big).
\end{align*}

(ii) Suppose that $\psi_i(\bar{x}, \bar{y}) \wedge p^=(\bar{x}, \bar{y})$ is inconsistent for all $i = 1, \ldots, t$.
Then, for every finite $\sigma$-structure $\mcA$, every $\bar{a} \in A^{|\bar{x}|}$ and every $\bar{b} \in A^{|\bar{y}|}$,
if $\mcA \models p^=(\bar{a}, \bar{b})$ then
we have  $\mcA(\psi_i(\bar{a}, \bar{b}) \to c_i) = 1$ for all $i$, so 
$\mcA(\psi(\bar{a},\bar{b})) = 1$. 
Hence, the sequence $\bar{r} = (\mcA(\psi(\bar{a}, \bar{y})) : \bar{y} : p^=(\bar{a}, \bar{y}))$ 
is constantly 1. 
Let $\bar{r}'$ be the sequence of length 1 the only entry of which is 1.
Then $\mu_1^u(\bar{r}, \bar{r}') = 0$. Since $F$ is strongly admissible it is strongly admissible sensu novo
(according to Proposition~\ref{equivalence of definitions of admissibility})
and by condition~(1) in the definition of admissibility sensu novo 
(Definition~\ref{definition of admissible function})
we get $F(\bar{r}) = F(\bar{r}')$.
\qed

\begin{asm}\label{assumptions in results for eliminating an aggregation function} {\rm
Until Proposition~\ref{elimination of one aggregation function}
we make, without loss of generality, the following assumptions:
Let $\kappa \in \mbbN^+$ and let $\bar{x}$ and $\bar{y}$ be sequences of distinct variables such that 
$\rng(\bar{x}) \cap \rng(\bar{y}) = \es$ and $|\bar{x}| + |\bar{y}| \leq \kappa \in \mbbN^+$.
Let $\psi(\bar{x}, \bar{y})$ be a basic probability formula.
Then $\psi(\bar{x}, \bar{y})$ is equivalent to  a basic probability formula of the form
\[
\bigwedge_{i=1}^s\bigwedge_{j=1}^{t_i} \big(p_{i,j}(\bar{x}, \bar{y}) \rightarrow c_{i,j}\big),
\]
where each $p_{i,j}(\bar{x}, \bar{y})$ is a complete atomic $\sigma$-type and 
(since $c_{i, j}$ may be zero) every complete atomic $\sigma$-type in the variables $\bar{x}\bar{y}$
equals $p_{i, j}$ for some $1 \leq i \leq s$ and $1 \leq j \leq t_i$.
Furthermore,
we may assume (by reordering if necessary) 
that for all $i = 1, \ldots, s$ and all $1 \leq j, j' \leq t_i$, $p_{i, j} \uhrc \bar{x} = p_{i, j'} \uhrc \bar{x}$.
Let $q_i(\bar{x}) = p_{i,1} \uhrc \bar{x}$ for each $i$.
{\em Without loss of generality we may therefore assume that $\psi(\bar{x}, \bar{y})$ has the above described form.}
}\end{asm}

\begin{lem}\label{crucial step in convergence of probability}
Suppose that $F : [0, 1]^{<\omega} \to [0, 1]$ is a strongly  admissible aggregation function and let
$p^=(\bar{x}, \bar{y})$ be an atomic $\es$-type.
Fix an index $1 \leq i \leq s$.
There is $d_i \in [0, 1]$, depending only on $\psi$, $p^=$ and $F$, such that
for every $\varepsilon > 0$ there is $\delta > 0$ such that for all sufficiently large $n$, all $\mcA \in \mbY_n^{\kappa, \delta}$, 
and all $\bar{a} \in [n]^{|\bar{x}|}$,
if $\mcA \models q_i(\bar{a})$, then
\[
\big| \mcA\big(F\big( \psi(\bar{a}, \bar{y}) : \bar{y} : p^=(\bar{a}, \bar{y})\big)\big) - d_i \big| < \varepsilon.
\]
\end{lem}

\proof
Fix $1 \leq i \leq s$.
If $p_{i, j}(\bar{x}, \bar{y}) \wedge p^=(\bar{x}, \bar{y})$ is inconsistent for all $j = 1, \ldots, t_i$, then,
by Lemma~\ref{reduction to a simpler basic formula}~(ii),
the conclusion is immediate.

So now suppose that  there is at least one $j$ such that $p_{i, j}(\bar{x}, \bar{y}) \wedge p^=(\bar{x}, \bar{y})$ is consistent.
By  Lemma~\ref{reduction to a simpler basic formula}~(i),
we may without loss of generality modify $\psi$ and assume that $p_{i, j}(\bar{x}, \bar{y}) \wedge p^=(\bar{x}, \bar{y})$ 
is consistent, hence $p^= \subseteq p_{i, j}$, for all $j = 1, \ldots, t_i$,
and that $p_{i, j}$,  $j = 1, \ldots, t_i$, enumerates all complete atomic $\sigma$-types in the variables 
$\bar{x}\bar{y}$ which extend $q_i$ and $p^=$.

Let $l = \dim_{\bar{y}}(p^=)$ and, for $j = 1, \ldots, t_i$, let 
$l_j = \dim_{\bar{y}}(p_{i, j})$.
Suppose first that $l = 0$.
Then $p^= \wedge q_i(\bar{x})$ has a unique extension to a complete atomic $\sigma$-type with 
variables $\bar{x}, \bar{y}$, so $t_i = 1$. Also, if $\mcA \models p_{i, j}(\bar{a}, \bar{b})$, then
$\rng(\bar{b}) \subseteq \rng(\bar{a})$.
It follows that for every finite $\sigma$-structure $\mcA$ and $\bar{a} \in A^{|\bar{a}|}$ such that $q_i(\bar{a})$ holds
there is a unique $\bar{b} \in A^{|\bar{y}|}$ such that $\mcA \models p_{i, j}(\bar{a}, \bar{b})$ and hence
the sequence $(\mcA(\psi(\bar{a}, \bar{y}) : \bar{y} : p^=(\bar{a}, \bar{y}))$ has a single coordinate
which is $c_{i, 1}$. Therefore we can let $d_i = F(\bar{r})$ where $\bar{r}$ is the sequence of length 1 where the only
coordinate is $c_{i, 1}$.

Now suppose that $l > 0$.
If $l_j < l$ then let $\alpha_j = 0$.
If $l_j = l$ then let $\alpha_j \in [0, 1]$ be the number associated to $p_{i, j}$ and $q_i$ by
Assumption~\ref{assumption about saturation}.

Let $\delta > 0$,
$\mcA_1 \in \mbY_{n_1}^{\kappa, \delta}$, $\mcA_2 \in \mbY_{n_2}^{\kappa, \delta}$, $\bar{a}_1 \in[n_1]^{|\bar{x}|}$,
$\bar{a}_2 \in [n_2]^{|\bar{x}|}$, $\mcA_1 \models q_i(\bar{a}_1)$, $\mcA_2 \models q_i(\bar{a}_2)$,
and, for $k = 1, 2$, let
\begin{equation}\label{definition of r-k}
\bar{r}_k = \big(\mcA_k\big(\psi(\bar{a}_k, \bar{b})\big) : 
\bar{b} \in [n_k]^{|\bar{y}|} \text{ and $\mcA \models p^=(\bar{a}, \bar{b})$}\big).
\end{equation}
It follows directly from the definition of $\bar{r}_k$ and assumptions about $\psi$ that 
$\rng(\bar{r}_1), \rng(\bar{r}_2) \subseteq \{c_{i, 1}, \ldots, c_{i, t_i}\}$.
To prove the lemma it now suffices to show that for any $\varepsilon > 0$ and all large enough $n_1$ and $n_2$,
$|F(\bar{r}_1) - F(\bar{r}_2)| < \varepsilon$.
Since we assume that $F$ is strongly admissible it follows from 
Proposition~\ref{equivalence of definitions of admissibility} that $F$ is strongly admissible sensu novo.
From Condition~(1) of the definition of strong admissibility sensu novo
(Definition~\ref{definition of admissible function})
it follows that it now suffices to show that there is a constant $C > 0$ which depends only on $\psi$ and $p^=$ such that
if $n_1$ and $n_2$ are sufficiently large, then 
$\mu_1^u(\bar{r}_1, \bar{r}_2) < \delta C$.

Let $k \in \{1, 2\}$ and note that $|\bar{r}_k| \sim (n_k)^l$.
By Assumption~\ref{assumption about saturation}, if $n_k$ is large enough the following holds for each $j = 1, \ldots, t_i$:
\begin{itemize}
\item If $l_j < l$ then $|p_{i, j}(\bar{a}_k, \mcA_k)| \leq (n_k)^{l_j}$ where $\frac{(n_k)^{l_j}}{(n_k)^l} \to 0$
as $n_k \to \infty$, so $|p_{i, j}(\bar{a}_k, \mcA_k)| \leq \delta (n_k)^l = (\alpha_j + \delta)(n_k)^l$, as 
$\alpha_j = 0$ in this case.

\item If $l_j = l$ then $(\alpha_j - \delta)(n_k)^l \leq |p_{i, j}(\bar{a}_k, \mcA_k)| \leq (\alpha_j + \delta)(n_k)^l$.
\end{itemize}
Let $c \in [0, 1]$ and suppose that there are exactly $m$ indices $j = j_1, \ldots, j_m$ such that $c_{i, j} = c$.
Every $\bar{b} \in p_{i, j}(\bar{a}_k, \mcA_k)$ contributes to 
a coordinate $c_{i, j}$ in the sequence $\bar{r}_k$.
Therefore the number $c$ will occur between 
\[
(\alpha_{j_1} + \ldots + \alpha_{j_m} - m\delta)(n_k)^l  \ \  \text{ and } \ \
(\alpha_{j_1} + \ldots + \alpha_{j_m} + m\delta) (n_k)^l
\]
times in $\bar{r}_k$.
This implies that $\mu_1^u(\bar{r}_1, \bar{r}_2) \leq \delta C$ for a constant
$C$ that depends only on $t_i$ which in turn depends only on $\psi$ and $p^=$.
This concludes the proof.
\qed

\begin{cor}\label{crucial step in eliminating F}
Let
$F : [0, 1]^{<\omega} \to [0, 1]$ be a strongly admissible aggregation function and let
$p^=(\bar{x}, \bar{y})$ be an atomic $\es$-type.
Then there is a basic probability formula $\theta(\bar{x})$ such that for every $\varepsilon > 0$
there is $\delta > 0$ such that for all sufficiently large $n$, all $\mcA \in \mbY_n^{\kappa, \delta}$, 
and all $\bar{a} \in [n]^{|\bar{x}|}$,
\[
\big| \mcA\big(F\big(\psi(\bar{a}, \bar{y}) : \bar{y} : p^=(\bar{a}, \bar{y})\big)\big) - 
\mcA\big(\theta(\bar{a})\big) \big| < \varepsilon.
\]
\end{cor}

\proof
Recall that from 
Assumption~\ref{assumptions in results for eliminating an aggregation function}
$q_i(\bar{x}) = p_{i, j} \uhrc \bar{x}$ for all $i$ (and all $j = 1, \ldots, t_i$).
By  Lemma~\ref{reduction to a simpler basic formula}~(i),
we may without loss of generality modify $\psi$ and assume that $p_{i, j}(\bar{x}, \bar{y}) \wedge p^=(\bar{x}, \bar{y})$ 
is consistent for all $j = 1, \ldots, t_i$,
and that $p_{i, j}$,  $j = 1, \ldots, t_i$, enumerates all complete atomic $\sigma$-types in the variables 
$\bar{x}\bar{y}$ which extend $q_i$ and $p^=$.

For every $i = 1, \ldots, s$, let $d_i \in [0, 1]$ be as in 
Lemma~\ref{crucial step in convergence of probability}.
Let $q'_1(\bar{x}), \ldots, q'_m(\bar{x})$ enumerate all complete atomic $\es$-types in 
the variables $\bar{x}$ which are different from $p^= \uhrc \bar{x}$.
We show that if $\theta(\bar{x})$ is the formula
$\bigwedge_{i=1}^s (q_i(\bar{x}) \to d_i) \ \wedge \ 
\bigwedge_{j=1}^m (q'_j(\bar{x}) \to 0)$
then the lemma holds.
Let $\varepsilon > 0$.
Let $\mcA \in  \mbY_n^{\kappa, \delta}$ and $\bar{a} \in [n]^{|\bar{x}|}$.
If $\bar{a}$ does not satisfy $p^= \uhrc \bar{x}$,
then it satisfies $q'_j(\bar{x})$ for some $j$ and (no matter what $\delta$ is)
\[
\mcA\big(F\big(\psi(\bar{a}, \bar{y}) : \bar{y} : p^=(\bar{a}, \bar{y})\big)\big) \ = \ 
0 \ = \ \mcA\Big(\bigwedge_{i=1}^s (q_i(\bar{a}) \to d_i) \ \wedge \ 
\bigwedge_{j=1}^m (q'_j(\bar{a}) \to 0) \Big).
\]
Now suppose that $\bar{a}$ satisfies $p^= \uhrc \bar{x}$ and hence it satisfies $q_i(\bar{x})$ for some $i$.
Then 
\begin{equation}\label{the value of theta}
\mcA\Big( \bigwedge_{i=1}^s (q_i(\bar{a}) \to d_i) \ \wedge \ 
\bigwedge_{j=1}^m (q'_j(\bar{a}) \to 0) \Big) \ = \  d_i.
\end{equation}
From Lemma~\ref{crucial step in convergence of probability}
we have
that if $\delta > 0$ is small enough, then for every $i = 1, \ldots, s$, 
all sufficiently large $n$, all $\mcA \in \mbY_n^{\kappa, \delta}$, 
and all $\bar{a} \in [n]^{|\bar{x}|}$,
if $\mcA \models q_i(\bar{a})$, then
\begin{equation}\label{repeating the conclusion of the lemma}
\big| \mcA\big(F\big( \psi(\bar{a}, \bar{y}) : \bar{y} : p^=(\bar{a}, \bar{y})\big)\big) - d_i \big| < \varepsilon.
\end{equation}
The corollary now follows from~(\ref{the value of theta}) and~(\ref{repeating the conclusion of the lemma}).
\qed

\begin{prop}\label{elimination of one aggregation function}
Suppose that $\varphi(\bar{x}, \bar{y}), \psi(\bar{x}, \bar{y}) \in PLA^+(\sigma)$ are asymptotically equivalent and that
$\psi(\bar{x}, \bar{y})$ is a basic probability formula.
Let $p^=(\bar{x}, \bar{y})$ be an atomic $\es$-type. 
If $F : [0, 1]^{<\omega} \to [0, 1]$ is a strongly admissible aggregation function,
then $F\big(\varphi(\bar{x}, \bar{y}) : \bar{y} : p^=(\bar{x}, \bar{y})\big)$ 
is asymptotically equivalent to a basic probability formula.
\end{prop}

\proof
Suppose that $\varphi(\bar{x}, \bar{y}), \psi(\bar{x}, \bar{y}) \in PLA(\sigma)$ are asymptotically equivalent and that
$\psi(\bar{x}, \bar{y})$ is a basic probability formula.
Without loss of generality we may assume that $\psi$ has the form described
in Assumption~\ref{assumptions in results for eliminating an aggregation function}.
Let $\kappa = |\bar{x}| + |\bar{y}|$ and $\varepsilon > 0$.
By Corollary~\ref{crucial step in eliminating F}
there is a basic probability formula $\theta(\bar{x})$ such that 
for all small enough $\delta > 0$ and large enough $n$, if $\mcA \in \mbY_n^{\kappa, \delta}$ 
and $\bar{a} \in [n]^{|\bar{x}|}$, then
\begin{equation}\label{the conclusion of a previous corollary}
\big| \mcA\big(F\big(\psi(\bar{a}, \bar{y}) : \bar{y} : p^=(\bar{a}, \bar{y})\big)\big) \ - \ 
\mcA\big(\theta(\bar{a})\big) \big| < \varepsilon/2.
\end{equation}

For $\delta > 0$ and  $n \in \mbbN^+$ let 
\begin{align*}
\mbX_n^\delta = \big\{ \mcA \in \mbW_n : &\text{ for all $\bar{a} \in [n]^{|\bar{x}|}$ and all $\bar{b} \in [n]^{|\bar{y}|}$
such that $p^=(\bar{a}, \bar{b})$ holds},\\
&\big|\mcA(\varphi(\bar{a}, \bar{b})) - \mcA(\psi(\bar{a}, \bar{b}))\big| < \delta \big\}.
\end{align*}
Since $\varphi(\bar{x}, \bar{y})$ and $\psi(\bar{x}, \bar{y})$ are asymptotically equivalent we have
$\lim_{n\to\infty}\mbbP_n(\mbX_n^\delta) =~1$.
By Assumption~\ref{assumption about saturation}
we have $\lim_{n\to\infty}\mbbP_n(\mbY_n^{\kappa, \delta}) = 1$ and hence
$\lim_{n\to\infty}\mbbP_n(\mbX_n^\delta \cap \mbY_n^{\kappa, \delta}) =~1$.

It now suffices to prove that if $\delta > 0$ is small enough, then for all sufficiently large $n$,
all $\mcA \in \mbX_n^\delta \cap \mbY_n^{\kappa, \delta}$ and all $\bar{a} \in [n]^{|\bar{x}|}$,
\begin{equation}\label{to be proved in eliminating F in the general case}
\big| \mcA\big(F\big(\varphi(\bar{a}, \bar{y}) : \bar{y} : p^=(\bar{a}, \bar{y})\big)\big) - 
\mcA\big(\theta(\bar{a})\big) \big| < \varepsilon.
\end{equation}
The statement~(\ref{to be proved in eliminating F in the general case}) 
follows from~(\ref{the conclusion of a previous corollary})
and the following (to be proved)
\begin{equation}\label{difference between F(varphi) and F(psi)}
\big| \mcA\big(F\big(\varphi(\bar{a}, \bar{y}) : \bar{y} : p^=(\bar{a}, \bar{y})\big)\big) - 
\mcA\big(F\big(\psi(\bar{a}, \bar{y}) : \bar{y} : p^=(\bar{a}, \bar{y})\big)\big) < \varepsilon/2.
\end{equation}
Hence it remains to prove that if $\delta > 0$ is small enough 
then~(\ref{difference between F(varphi) and F(psi)}) holds
for all sufficiently large $n$,
all $\mcA \in \mbX_n^\delta \cap \mbY_n^{\kappa, \delta}$ and all $\bar{a} \in [n]^{|\bar{x}|}$.

Let $\mcA \in \mbX_n^\delta \cap \mbY_n^{\kappa, \delta}$ and $\bar{a} \in [n]^{|\bar{x}|}$.
If $\bar{a}$ does not satisfy $p^= \uhrc \bar{x}$ then 
\[
\mcA\big(F\big(\varphi(\bar{a}, \bar{y}) : \bar{y} : p^=(\bar{a}, \bar{y})\big)\big) \ = \ 0 \ = \ 
\mcA\big(F\big(\psi(\bar{a}, \bar{y}) : \bar{y} : p^=(\bar{a}, \bar{y})\big)\big).
\]
Now suppose that $\bar{a}$ satisfies $p^= \uhrc \bar{x}$ and hence $\bar{a}$
satisfies $q_i(\bar{x})$ 
(as in Assumption~\ref{assumptions in results for eliminating an aggregation function})
for some $i$. Then the following two sequences
are nonempty:
\begin{align*}
&\bar{r} = \big(\mcA\big(\varphi(\bar{a}, \bar{b})\big) : 
\bar{b} \in [n]^{|\bar{y}|} \text{ and $p^=(\bar{a}, \bar{b})$ holds} \big), \\
&\bar{\rho} = \big(\mcA\big(\psi(\bar{a}, \bar{b})\big) : 
\bar{b} \in [n]^{|\bar{y}|} \text{ and $p^=(\bar{a}, \bar{b})$ holds} \big).
\end{align*}
First suppose that every $p_{i, j}(\bar{x}, \bar{y})$ 
(as in Assumption~\ref{assumptions in results for eliminating an aggregation function})
is {\em in}consistent with $p^=$. 
Then all entries in $\bar{\rho}$ are equal to 1.
Since $\mcA \in \mbX_n^\delta$ we get $\mu_\infty^o(\bar{r}, \bar{\rho}) < \delta$.
Since $F$ is strongly admissible, hence strongly admissible sensu novo, if follows 
from Condition~(2) of the definition of strong admissibility sensu novo,
that if $\delta$ is small enough, then $|F(\bar{r}) - F(\bar{\rho})| < \varepsilon/2$
and~(\ref{difference between F(varphi) and F(psi)}) follows immediately from this.

Now suppose that at least one $p_{i, j}$ is consistent with $p^=$.
By Lemma~\ref{reduction to a simpler basic formula} we may, without loss of generality, assume that 
every $p_{i, j}$ is consistent with $p^=$.
Then we can argue in the same way as we argued in the proof of 
Lemma~\ref{crucial step in convergence of probability} and conclude 
that there are $\alpha_j, c_{i, j} \in [0, 1]$, corresponding to $p_{i, j}$, for $j = 1, \ldots, t_i$, such that 
if $c \in [0, 1]$ and $j_1, \ldots, j_m$ enumerates all $c_{i, j}$ such that $c_{i, j} = c$, then
$c$ appears between
\[
(\alpha_{j_1} + \ldots + \alpha_{j_m} - m\delta) n^l \ \ \text{ and } \ \
(\alpha_{j_1} + \ldots + \alpha_{j_m} + m\delta) n^l
\]
times in $\bar{\rho}$.
As $\mcA \in  \mbX_n^\delta$ we get $\mu_\infty^o(\bar{r}, \bar{\rho}) < \delta$.
Since $F$ is strongly admissible, hence strongly admissible sensu novo, it follows 
from Condition~(2) of the definition of admissibility sensu novo,
that if $\delta$ is small enough, then $|F(\bar{r}) - F(\bar{\rho})| < \varepsilon/2$
which implies that~(\ref{difference between F(varphi) and F(psi)}) holds.
\qed

\begin{cor}\label{elimination of aggregation functions if saturation conditions hold}
Let $\varphi(\bar{x}) \in PLA^+(\sigma)$ and suppose that
all aggregation functions in $\varphi$ are strongly admissible.
Then $\varphi(\bar{x})$ is asymptotically equivalent to a basic probability formula.
\end{cor}

\proof
We use induction on the complexity of formulas.
If $\varphi(\bar{x})$ is of one of the forms described in parts~(1) and~(2) of 
the definition of $PLA^+(\sigma)$
(Definition~\ref{syntax of PLA+}), then $\varphi(\bar{x})$ is aggregation-free and then
$\varphi(\bar{x})$ is asymptotically equivalent to a basic probability formula by virtue of 
Lemma~\ref{quantifier-free formulas are equivalent to bpf}, since equivalence implies asymptotic equivalence.

Now suppose that $\msfC : [0, 1]^k \to [0, 1]$ is a continuous connective and 
$\varphi_1(\bar{x}), \ldots, \varphi_k(\bar{x}) \in PLA^+(\sigma)$, so
$\msfC(\varphi_1(\bar{x}), \ldots, \varphi_k(\bar{x})) \in PLA^+(\sigma)) \in PLA^+(\sigma)$.
If each $\varphi_i(\bar{x})$ is asymptotically equivalent to a basic probability formula $\varphi'_i(\bar{x})$, 
then it follows from the continuity of $\sf{C}$ that 
$\msfC(\varphi_1(\bar{x}), \ldots, \varphi_k(\bar{x}))$ is asymptotically equivalent to
$\msfC(\varphi'_1(\bar{x}), \ldots, \varphi'_k(\bar{x}))$.
Since $\msfC(\varphi'_1(\bar{x}), \ldots, \varphi'_k(\bar{x}))$ is aggregation-free it follows from 
Lemma~\ref{quantifier-free formulas are equivalent to bpf}
that it is equivalent to a basic probability formula $\psi(\bar{x})$.
But then $\msfC(\varphi_1(\bar{x}), \ldots, \varphi_k(\bar{x}))$  is asymptotically equivalent to $\psi(\bar{x})$.

Now suppose that $F : \big([0, 1]^{<\omega}\big)^k \to [0,1]$
is a strongly admissible aggregation function, $p^=(\bar{x}, \bar{y})$ is an atomic $\sigma$-type,
$\varphi_1(\bar{x}, \bar{y}), \ldots, \varphi_k(\bar{x}, \bar{y}) \in PLA^+(\sigma)$
and that each $\varphi_i(\bar{x}, \bar{y})$ is asymptotically equivalent to a 
basic probability formula $\varphi'_i(\bar{x}, \bar{y})$.
Then Proposition~\ref{elimination of one aggregation function}
combined with 
Remark~\ref{remark on higher arities}
implies that 
$F(\varphi_1(\bar{x}, \bar{y}), \ldots, \varphi_k(\bar{x}, \bar{y}) : \bar{y} : p^=(\bar{x}, \bar{y}))$
is asymptotically equivalent to a basic probability formula.
\qed

\section{Saturation and convergence of atomic types: finishing the proof of 
Theorem~\ref{elimination of strongly admissible aggregation functions}}\label{proof of main theorem}

\noindent
Theorem~\ref{elimination of strongly admissible aggregation functions}
follows from Proposition~\ref{saturation implies elimination of strongly admissible functions}
and the proofs in this section.
We consider a finite relational signature $\sigma$ and a $PLA^+(\sigma)$-network $\mbbG$ 
whose every probability formula uses only {\em strongly admissible} aggregation functions (if it uses any at all).
We want to prove that the claims of 
Theorem~\ref{elimination of strongly admissible aggregation functions}
hold.
For this we will use induction on the maximal path rank, or mp-rank, of the underlying DAG of $\mbbG$, also denoted $\mbbG$.
In order to make the inductive step work out we have to prove
(in the base case and in the inductive step) a few claims none of which explicitly states that
the claims of Theorem~\ref{elimination of strongly admissible aggregation functions} hold.
But these few claims 
(labelled (1)--(5) in Assumption~\ref{inductive assumptions}) in conjunction with 
Proposition~\ref{saturation implies elimination of strongly admissible functions}
imply Theorem~\ref{elimination of strongly admissible aggregation functions}.
Perhaps a bit counter-intuitively the base case of the induction will {\em not} be the case when $\mbbG$ has mp-rank 0
(i.e. when $\mbbG$ has no edges).
Instead the base case will be when $\sigma = \es$, in other words when the DAG has no vertices (and hence no edges) 
and in this case we have the convention that the empty DAG has mp-rank `$-1$'.
The base case, for an empty signature, is stated by
Lemma~\ref{the case of empty sigma'}. 
To make the notation consistent with the notation used in the inductive step formulated in
Assumption~\ref{inductive assumptions}, 
we denote the empty signature of the base case (Lemma~\ref{the case of empty sigma'}) by $\sigma'$.

\begin{lem}[The base case]\label{the case of empty sigma'}
Suppose that $\sigma' = \es$ and, for all $n \in \mbbN^+$, let $\mbW'_n$ be the set of all $\sigma'$-structures
with domain $[n]$ (so $\mbW'_n$ is a singleton set), and let $\mbbP'_n$ be the unique probability distribution
on $\mbW'_n$.
\begin{enumerate}
\item[(a)] Let $k \in \mbbN^+$, $\varepsilon' > 0$,
and let $\delta' : \mbbN^+ \to \mbbR^{\geq 0}$ be any function such that $\lim_{n\to\infty} \delta'(n) = 0$, 
Then there are
$\mbY'_n \subseteq \mbW'_n$, for $n \in \mbbN^+$,  such that the following hold:
\begin{enumerate}
\item[(1)] $\lim_{n\to\infty} \delta'(n) = 0$.

\item[(2)] $\mbbP'_n(\mbY'_n) \geq 1 - \delta'(n)$ for all sufficiently large $n$.

\item[(3)] For every complete atomic $\sigma'$-type $p'(\bar{x})$ with $|\bar{x}| \leq k$
there is a number which we denote 
$\msfP'(p'(\bar{x}))$, or just $\msfP'(p')$, such that for all sufficiently large $n$ and all $\bar{a} \in [n]$ 
which realize the identity fragment of $p'$,
\[
\big| \mbbP'_n\big(\{\mcA' \in \mbW'_n : \mcA' \models p'(\bar{a})\}\big) \ -  \ \msfP'(p'(\bar{x})) \big| \ \leq \ \varepsilon'.
\]

\item[(4)] For every complete atomic $\sigma'$-type $p'(\bar{x}, \bar{y})$ with $|\bar{x}\bar{y}| \leq k$ and
$0 < \dim_{\bar{y}}(p'(\bar{x}, y)) = d$,
if $q'(\bar{x}) = p' \uhrc \bar{x}$, then there is $\alpha \in [0, 1]$ such that, for all sufficiently large $n$, every
$\mcA' \in \mbY'_n$ is $(p', q', \alpha - \varepsilon')$-saturated and $(p', q', \alpha + \varepsilon')$-unsaturated.
\end{enumerate}

\item[(b)] If $\varphi(\bar{x}) \in PLA^+(\es)$ and every aggregation function in $\varphi$ is strongly admissible, 
  then $\varphi(\bar{x})$ is asymptotically equivalent, with respect to $(\mbbP'_n : n \in \mbbN^+)$, to a basic probability formula.
  \end{enumerate}
\end{lem}

\proof
(a) Suppose that $\sigma' = \es$ and let $k \in \mbbN^+$ and $\varepsilon' > 0$ be given.
Also let $\delta' : \mbbN^+ \to \mbbR^{\geq 0}$ be any function such that $\lim_{n\to\infty} \delta'(n) = 0$, 
so~(1) holds.
For every complete atomic $\sigma'$-type $p'(\bar{x})$ let $\msfP'(p'(\bar{x})) = 1$.
Observe that, for every $n$, if $\bar{a} \in [n]$ and $\bar{a}$ realizes the identity fragment of $p'(\bar{x})$,
then $\bar{a}$ realizes $p'(\bar{x})$ in the unique $\mcA'$ of $\mbW'_n$.
Hence, for trivial reasons we have~(3).

For every $n$ let $\mbY'_n$ be the set of all $\mcA' \in \mbW'_n$ such that for every
complete atomic $\sigma'$-type $p'(\bar{x}, \bar{y})$ with $|\bar{x}\bar{y}| \leq k$ and
$0 < \dim_{\bar{y}}(p'(\bar{x}, \bar{y})) = |\bar{y}|$,
if $q(\bar{x}) = p \uhrc \bar{x}$, then for all sufficiently large $n$, every
$\mcA' \in \mbY'_n$ is $(p', q', 1 - \varepsilon')$-saturated and $(p', q', 1 + \varepsilon')$-unsaturated.
Suppose that $p'(\bar{x}, \bar{y})$ is a complete atomic $\sigma'$-type with $|\bar{x}\bar{y}| \leq k$ and
$0 < \dim_{\bar{y}}(p'(\bar{x}, \bar{y})) = |\bar{y}|$.
Let $q'(\bar{x}) = p' \uhrc \bar{x}$ and suppose that $\mcA' \models q'(\bar{a})$ where $\mcA' \in \mbW'_n$.
Then $\mcA' \models p'(\bar{a}, \bar{b})$ for every $\bar{b} \in [n]^{|\bar{y}|}$ consisting of different elements no one of which 
occurs in $\bar{a}$. There are $n^{|\bar{y}|} - Cn^{|\bar{y}|-1}$ such $\bar{b}$ for some constant $C$.
So if $n^{|\bar{y}|} - Cn^{|\bar{y}|-1} \geq (1 - \varepsilon')n^{|\bar{y}|}$ then
$\mcA'$ is $(p', q', 1 - \varepsilon')$-saturated.
For trivial reasons, $\mcA'$ is also $(p', q', 1 + \varepsilon')$-unsaturated.
Hence we have proved~(4).
The above argument shows that for all large enough $n$ the unique member of $\mbW'_n$ belongs to $\mbY'_n$, so 
it follows that~(2) holds.

(b) Note that we have proved that~(1) --~(4) hold for all choices of $k \in \mbbN^+$ and $\varepsilon' > 0$.
Therefore Assumption~\ref{assumption about saturation} holds and consequently, by
Proposition~\ref{saturation implies elimination of strongly admissible functions},
if $\varphi(\bar{x}) \in PLA^+(\es)$ and every aggregation function in $\varphi$ is strongly admissible, then
$\varphi(\bar{x})$ is asymptotically equivalent to a basic probability formula.
\qed
\\

\noindent
{\em Note that Lemma~\ref{the case of empty sigma'} implies the statement of 
Theorem~\ref{elimination of strongly admissible aggregation functions} 
in the case when $\sigma = \es$.}
{\em For the rest of this section we make the following assumptions}:
$\sigma$ is a nonempty finite relational signature.
$\mbbG$ is a $PLA^+(\sigma)$-network with mp-rank $\rho \geq 0$.
For every $R \in \sigma$, the corresponding probability formula $\theta_R$ contains
only strongly admissible aggregation functions.
$\sigma'$ is the set of $R \in \sigma$ such that the mp-rank of $R$ is less than $\rho$ and $\mbbG'$ is the 
$PLA^+(\sigma')$-subnetwork induced by $\sigma'$. 
$\mbW'_n$ be the set of all $\sigma'$-structures with domain $[n]$ and $\mbW_n$ 
is the set of $\sigma$-structures with domain $[n]$.
$\mbbP'_n$ and $\mbbP_n$ are the probability distributions induced by $\mbbG'$ and $\mbbG$ on 
$\mbW'_n$ and $\mbW_n$, respectively.
Now we assume the following (which, by Lemma~\ref{the case of empty sigma'}, holds if $\sigma' = \es$):

\begin{asm}[Induction hypothesis]\label{inductive assumptions} {\rm 
For all $k \in \mbbN^+$ and all $\varepsilon' > 0$ there are $\delta' : \mbbN^+ \to \mbbR^{\geq 0}$ and 
$\mbY'_n \subseteq \mbW'_n$, for $n \in \mbbN^+$, such that the following hold:
\begin{itemize}[align=left]
\item[(1)] $\lim_{n\to\infty} \delta'(n) = 0$.

\item[(2)] $\mbbP'_n(\mbY'_n) \geq 1 - \delta'(n)$ for all sufficiently large $n$.

\item[(3)] For every complete atomic $\sigma'$-type $p'(\bar{x})$ with $|\bar{x}| \leq k$
there is a number which we denote 
$\msfP'(p'(\bar{x}))$, or just $\msfP'(p')$, such that for all sufficiently large $n$ and all $\bar{a} \in [n]$ 
which realize the identity fragment of $p'$,
\[
\big| \mbbP'_n\big(\{\mcA' \in \mbW'_n : \mcA' \models p'(\bar{a})\}\big) \ -  \ \msfP'(p'(\bar{x})) \big| \ \leq \ \varepsilon'.
\]

\item[(4)] For every complete atomic $\sigma'$-type $p'(\bar{x}, \bar{y})$ with $|\bar{x}\bar{y}| \leq k$ and
$0 < \dim_{\bar{y}}(p'(\bar{x}, \bar{y})) = d$,
if $q'(\bar{x}) = p' \uhrc \bar{x}$, then there is $\alpha \in [0, 1]$ such that, for all sufficiently large $n$, every
$\mcA' \in \mbY'_n$ is $(p', q', \alpha - \varepsilon')$-saturated and $(p', q', \alpha + \varepsilon')$-unsaturated.

\item[(5)] For every $R \in \sigma \setminus \sigma'$ there is a basic probability formula $\chi_R(\bar{x}) \in PLA^+(\sigma')$ such that 
$\chi_R(\bar{x}) \sim_{\mbbG'} \theta_R(\bar{x})$, where $\theta_R \in PLA^+(\sigma')$ is the probability formula corresponding to $R$ in $\mbbG$,
and for all sufficiently large $n$, all $\mcA' \in \mbY'_n$ and all $\bar{a} \in [n]^{|\bar{x}|}$,
\[
\big| \mcA'(\theta_R(\bar{a})) - \mcA'(\chi_R(\bar{a})) \big| \leq \varepsilon'.
\]
\end{itemize}
}\end{asm}

\noindent
{\bf Outline of the inductive step and how conditions~(1) --~(5) imply Theorem~\ref{elimination of strongly admissible aggregation functions}.}
We fix some arbitrary $k \in \mbbN^+$ and $\varepsilon' > 0$.
The we carry out an argument, finished in
Proposition~\ref{finishing the induction step}, which shows that 
we can find $\varepsilon > 0$, which can be made as small as we like if $\varepsilon'$ is chosen small enough,
$\delta : \mbbN^+ \to \mbbR^{\geq 0}$, and 
$\mbY_n \subseteq \mbW_n$, for $n \in \mbbN^+$, such that if 
$\sigma'$, $\varepsilon'$, $\delta'$, $\mbW'_n$, $\mbY'_n$ and  $\mbbP'_n$ are replaced by
$\sigma$, $\varepsilon$, $\delta$, $\mbW_n$, $\mbY_n$ and  $\mbbP_n$, respectively,
then parts~(1) -- (4) of Assumption~\ref{inductive assumptions} hold.
It follows, as stated in Corollary~\ref{inductive assumption holds for all k and epsilon}, that
{\em for every $k$ and every $\varepsilon > 0$}, there are
$\delta : \mbbN^+ \to \mbbR^{\geq 0}$ and $\mbY_n \subseteq \mbW_n$, for $n \in \mbbN^+$, such that if 
$\sigma'$, $\varepsilon'$, $\delta'_n$, $\mbW'_n$, $\mbY'_n$ and  $\mbbP'_n$ are replaced by
$\sigma$, $\varepsilon$, $\delta_n$, $\mbW_n$, $\mbY_n$ and  $\mbbP_n$, respectively,
then parts~(1) -- (4) of Assumption~\ref{inductive assumptions} hold.
Hence Assumption~\ref{assumption about saturation} holds and therefore 
Proposition~\ref{saturation implies elimination of strongly admissible functions}
implies that if $\varphi(\bar{x}) \in PLA^+(\sigma)$ and all aggregation functions in $\varphi$ are strongly admissible,
then $\varphi(\bar{x})$ is asymptotically equivalent to a basic probability formula.
This is stated in Corollary~\ref{concluding elimination of aggregation functions}.

It remains to show that Part~(5) of Assumption~\ref{inductive assumptions} holds for every $R \in \sigma^* \setminus \sigma$
if $\sigma \subset \sigma^*$ and the corresponding probability formula $\theta_R \in PLA^+(\sigma)$
(of a $PLA(\sigma^*)$-network $\mbbG^*$ having $\mbbG$ as a subnetwork)
contains only strongly admissible aggregation functions.
So suppose 
that $\sigma^* \supset \sigma$ and $\mbbG^*$ is a $PLA^+(\sigma^*)$-network such that
$\mbbG$ is a subnetwork of $\mbbG^*$ and
for every $R \in \sigma^* \setminus \sigma$ the corresponding probability formula $\theta_R \in PLA^+(\sigma)$
contains only strongly admissible aggregation functions.
Then, 
by Corollary~\ref{concluding elimination of aggregation functions}, 
for every $R \in \sigma^* \setminus \sigma$ there is a basic probability formula $\chi_R$ such that $\theta_R \sim_\mbbG \chi_R$.
Since $\sigma^*$ is finite, it follows that for any $\varepsilon > 0$ there 
are $\mbY^*_n \subseteq \mbW_n$ for all $n \in \mbbN^+$ such that
$\lim_{n\to\infty}\mbbP_n(\mbY^*_n) = 1$ and
the following holds for all $n$:
If $R \in \sigma^* \setminus \sigma$, $\mcA \in \mbY^*_n$ and $\bar{a} \in [n]^r$, where $r$ is the arity of $R$,
then 
\begin{equation}\label{small difference between theta-R and chi-R}
|\mcA(\theta_R(\bar{a})) - \mcA(\chi_R(\bar{a}))| \leq \varepsilon.
\end{equation}
We can now replace $\mbY_n$ by $\mbY_n \cap \mbY^*_n$, but in order to not introduce new notation we still call
this new set $\mbY_n$. By modifying $\delta_n$ slightly if necessary the conditions~(1)~--~(4) still hold
if 
$\sigma'$, $\varepsilon'$, $\delta'_n$, $\mbW'_n$, $\mbY'_n$ and  $\mbbP'_n$ are replaced by
$\sigma$, $\varepsilon$, $\delta_n$, $\mbW_n$, $\mbY_n$ and  $\mbbP_n$, respectively.
In addition, condition~(5) now holds if $\sigma'$, $\sigma$, $\mbbG'$, $\mbbG$ and $\mbY'_n$ are replaced by
$\sigma$, $\sigma^*$, $\mbbG$, $\mbbG^*$ and $\mbY_n$, respectively.

This reduces the proof of the inductive step to
Proposition~\ref{finishing the induction step},
Corollary~\ref{inductive assumption holds for all k and epsilon} and
Corollary~\ref{concluding elimination of aggregation functions}.
Thus it remains to prove these results.
But before doing this, we explain how 
Theorem~\ref{elimination of strongly admissible aggregation functions}
follows from the conditions (1)~--~(5) of 
Assumption~\ref{assumption about saturation}
with $\sigma'$, $\varepsilon'$, $\delta'_n$, $\mbW'_n$, $\mbY'_n$ and  $\mbbP'_n$ replaced by
$\sigma$, $\varepsilon$, $\delta_n$, $\mbW_n$, $\mbY_n$ and  $\mbbP_n$, respectively.
Part~(ii) of Theorem~\ref{elimination of strongly admissible aggregation functions} follows from
condition~(3) with the mentioned replacements
as stated by Corollary~\ref{inductive assumption holds for all k and epsilon}.
Condition~(4) (with the mentioned replacements) implies that 
Assumption~\ref{assumption about saturation}
holds and therefore
Proposition~\ref{saturation implies elimination of strongly admissible functions}
implies Part~(i) of 
Theorem~\ref{elimination of strongly admissible aggregation functions}.
The statement which is relevant for Part~(i) of 
Theorem~\ref{elimination of strongly admissible aggregation functions}
appears clearly in 
Corollary~\ref{concluding elimination of aggregation functions}.

{\em From now until Proposition~\ref{finishing the induction step} we fix $k \in \mbbN^+$ and $\varepsilon' > 0$.}
In the proofs that follow we will consider relativizations of $\mbbP_n$ to some subsets of $\mbW_n$ according to the next definition.

\begin{defi}{\rm
(i) If $\mbY' \subseteq \mbW'_n$ then we define
\begin{align*}
&\mbW^{\mbY'} = \{\mcA \in \mbW_n : \mcA \uhrc \sigma' \in \mbY'\} \quad \text{ and if $\mcA \in \mbW^{\mbY'}$ and
$\mcA \uhrc \sigma' = \mcA'$, then} \\
&\mbbP^{\mbY'}(\mcA) \ = \
\frac{\mbbP'_n(\mcA')}{\mbbP'_n(\mbY')} 
\prod_{R \in \sigma \setminus \sigma'} \ \prod_{\bar{a} \in R^\mcA} 
\mcA' \big(\theta_R(\bar{a})\big) \ 
\prod_{\bar{a} \in [n]^{k_R} \ \setminus \ R^\mcA} \big( 1 - \mcA' \big(\theta_R(\bar{a})\big) \big)
\end{align*}
where $k_R$ is the arity of $R$. \\ 
(ii) If $\mcA' \in \mbW'_n$, then we let
\begin{align*}
&\mbW^{\mcA'} = \mbW^{\{\mcA'\}} \quad \text{ and, for every $\mcA \in \mbW^{\mcA'}$,} \\
&\mbbP^{\mcA'}(\mcA) = \mbbP^{\{\mcA'\}}(\mcA) = 
\prod_{R \in \sigma \setminus \sigma'} \ \prod_{\bar{a} \in R^\mcA} 
\mcA' \big(\theta_R(\bar{a})\big) \ 
\prod_{\bar{a} \in [n]^{k_R} \ \setminus \ R^\mcA} \big( 1 - \mcA' \big(\theta_R(\bar{a})\big) \big)
\end{align*}
}\end{defi}

\noindent
Then $\mbbP^{\mbY'}$ and $\mbbP^{\mcA'}$ are probability distributions on $\mbW^{\mbY'}$ and $\mbW^{\mcA'}$, respectively.
Note also that if $\mbY' \subseteq \mbW'_n$, $\mcA' \in \mbY'$ and $\mcA \in \mbW^{\mcA'}$, then
\begin{equation}\label{first basic equality about restricted distributions}
\mbbP^{\mbY'}(\mcA) = \frac{\mbbP'_n(\mcA')}{\mbbP'_n(\mbY')}\mbbP^{\mcA'}(\mcA),
\end{equation}
and in particular, taking $\mbY' = \mbW'_n$, we have, for every $\mcA \in \mbW_n$,
\begin{equation}\label{second basic equality about restricted distributions}
\mbbP_n(\mcA) = \mbbP'_n(\mcA \uhrc \sigma')\mbbP^{\mcA \uhr \sigma'}(\mcA).
\end{equation}

\noindent
We now state a few basic lemmas which will be useful.

\begin{lem}\label{P and P' agree when constraints speak only about W'}
For every $n$, if $\mbY' \subseteq \mbW'_n$ then $\mbbP_n(\mbW^{\mbY'}) = \mbbP'_n(\mbY')$.
\end{lem}

\proof
By using~(\ref{second basic equality about restricted distributions}) in the first line below we get
\begin{align*}
&\mbbP_n(\mbW^{\mbY'}) \ = \ \sum_{\mcA' \in \mbY'} \sum_{\mcA \in \mbW^{\mcA'}} \mbbP_n(\mcA) \ = \ 
\sum_{\mcA' \in \mbY'} \sum_{\mcA \in \mbW^{\mcA'}} \mbbP'_n(\mcA') \mbbP^{\mcA'}(\mcA) \ = \\ 
&\sum_{\mcA' \in \mbY'}  \mbbP'_n(\mcA')  \sum_{\mcA \in \mbW^{\mcA'}} \mbbP^{\mcA'}(\mcA) \ = \ 
\sum_{\mcA' \in \mbY'}  \mbbP'_n(\mcA') \ = \ \mbbP'_n(\mbY'). \qedhere
\end{align*}

\begin{lem}\label{P(X given Y) equals P_Y(X cut Y)}
For every $n$, \\
(i) if $\mbX \subseteq \mbW_n$ and $\mcA' \in \mbW'_n$, 
then $\mbbP_n(\mbX \ | \ \mbW^{\mcA'}) = \mbbP^{\mcA'}(\mbX \cap \mbW^{\mcA'})$, and\\
(ii) if $\mbX \subseteq \mbW_n$ and $\mbY' \subseteq \mbW'_n$, 
then $\mbbP_n(\mbX \ | \ \mbW^{\mbY'}) = \mbbP^{\mbY'}(\mbX \cap \mbW^{\mbY'})$.
\end{lem}

\proof
Let $\mbX \subseteq \mbW_n$.

(i) Let $\mcA' \in \mbW'_n$. Using Lemma~\ref{P and P' agree when constraints speak only about W'} in the first line below
and~(\ref{second basic equality about restricted distributions}) in the second line below, we get
\begin{align*}
&\mbbP_n(\mbX \ | \ \mbW^{\mcA'}) \ = \
\frac{\mbbP_n(\mbX \cap \mbW^{\mcA'})}{\mbbP_n(\mbW^{\mcA'})} \ = \ 
\frac{\mbbP_n(\mbX \cap \mbW^{\mcA'})}{\mbbP'_n(\mcA')} \ = \\ 
&\frac{\mbbP'_n(\mcA') \sum_{\mcA \in \mbX \cap \mbW^{\mcA'}} \mbbP^{\mcA'}(\mcA)}{\mbbP'_n(\mcA')} \ = \ 
\mbbP^{\mcA'}(\mbX \cap \mbW^{\mcA'}).
\end{align*}

(ii) Let $\mbY' \subseteq \mbW'_n$. Using that $\mbX \cap \mbW^{\mbY'}$ is the disjoint union of all
$\mbX \cap \mbW^{\mcA'}$ such that $\mcA' \in \mbY'$, 
Lemma~\ref{P and P' agree when constraints speak only about W'},
Part~(i) of this lemma 
and~(\ref{first basic equality about restricted distributions}), we get
\begin{align*}
&\mbbP_n(\mbX \ | \ \mbW^{\mbY'}) \ = \ 
\frac{\mbbP_n(\mbX \cap \mbW^{\mbY'})}{\mbbP_n(\mbW^{\mbY'})} \ = \ 
  \sum_{\mcA' \in \mbY'} \frac{\mbbP_n(\mbX \cap \mbW^{\mcA'})}{\mbbP_n(\mbW^{\mbY'})} \ = \\ 
&\sum_{\mcA' \in \mbY'} \frac{\mbbP_n(\mbW^{\mcA'})}{\mbbP_n(\mbW^{\mbY'})} \ \mbbP_n(\mbX \ | \ \mbW^{\mcA'}) \ = \ 
\sum_{\mcA' \in \mbY'} \frac{\mbbP'_n(\mcA')}{\mbbP'_n(\mbY')} \mbbP^{\mcA'}(\mbX \cap \mbW^{\mcA'}) \ = \\ 
&\sum_{\mcA' \in \mbY'} \mbbP^{\mbY'}(\mbX \cap \mbW^{\mcA'}) \  = \ 
\mbbP^{\mbY'}(\mbX \cap \mbW^{\mbY'}). \qedhere
\end{align*}

\begin{rem}[Properties of $\mbbP^{\mcA'}$]\label{remark on P restricted to A'}{\rm 
Fix any $n$ and any $\mcA' \in \mbW'_n$.\\
(i) If $\mcA \in \mbW^{\mcA'}$ then (by the definitions of $\mbbP_n$ and $\mbbP^{\mcA'}$)
\[
\mbbP^{\mcA'}(\mcA) = 
\mbbP_n\big(\mcA \ | \ \mcA \uhrc \sigma' = \mcA' \big).
\]
It follows that if $p'(\bar{x})$ is a complete atomic $\sigma'$-type, $\mcA' \models p'(\bar{a})$
and $p(\bar{x}) \supset p'(\bar{x})$ is an atomic $\sigma$-type, then
\[
\mbbP^{\mcA'}\big(\{\mcA \in W^{\mcA'} : \mcA \models p(\bar{a})\}\big) \  = \ 
\mbbP_n\big(\{\mcA \in \mbW_n: \mcA \models p(\bar{a}) \ | \ \{\mcA \in \mbW_n : \mcA \uhrc \sigma' = \mcA'\}\big).
\]
(ii) For every $\alpha \in \{0, 1\}$, every $R \in \sigma \setminus \sigma'$ and every $\bar{a} \in [n]^r$, where $r$ is the arity of $R$,
let
$\mbE^\alpha_{R, \bar{a}} = \{\mcA \in \mbW^{\mcA'} : \mcA \models R^\alpha(\bar{a})\}$
where $R^0$ and $R^1$ denote $\neg R$ and $R$, respectively.
It follows from the definition of $\mbbP^{\mcA'}$ that
\begin{enumerate}
\item[(a)] for every $R \in \sigma \setminus \sigma'$ and every $\bar{a} \in [n]^r$, where $r$ is the arity of $R$,
$\mbbP^{\mcA'}(\mbE^1_{R, \bar{a}}) = \mcA'(\theta_R(\bar{a}))$, 
$\mbbP^{\mcA'}(\mbE^0_{R, \bar{a}}) = 1 - \mcA'(\theta_R(\bar{a}))$,
and

\item[(b)] if $\alpha_1, \ldots, \alpha_m \in \{0, 1\}$, $R_1, \ldots, R_m \in \sigma \setminus \sigma'$
and $\bar{a}_1, \ldots, \bar{a}_m$ are tuples where $|\bar{a}_i|$ equals the arity of $R_i$ for each $i$,
and for all  $1 \leq i < j \leq m$,  $R_i \neq R_j$ or $\bar{a}_i \neq \bar{a}_j$, then
the events $\mbE^{\alpha_1}_{R_1, \bar{a}_1}, \ldots, \mbE^{\alpha_m}_{R_m, \bar{a}_m}$
are independent.
\end{enumerate}
}\end{rem}

\begin{lem}\label{uniformity of probabilities under qf}
Suppose that $p'(\bar{x})$ is a complete atomic $\sigma'$-type and that $p(\bar{x}) \supseteq p'(\bar{x})$
is a (possibly partial) atomic $\sigma$-type.
There is a number which we denote $\msfP(p(\bar{x}) \ | \ p'(\bar{x}))$, or just $\msfP(p \ | \ p')$,
such that for all sufficiently large $n$, all $\bar{a} \in [n]^{|\bar{x}|}$ and all $\mcA' \in \mbY'_n$
such that $\mcA' \models p'(\bar{a})$,
\[
\big| \mbbP^{\mcA'}\big(\{\mcA \in \mbW^{\mcA'} : \mcA \models p(\bar{a})\} \big) \ - \ \msfP(p(\bar{x}) \ | \ p'(\bar{x})) \big| \ 
\leq \ \varepsilon_p
\]
where $\varepsilon_p > 0$ depends only on $\varepsilon$ and on $p$ and $\varepsilon_p$ can be made arbitrarily small by taking
$\varepsilon'$ sufficiently small.
\end{lem}

\proof
Let $\mcA' \in \mbY'_n$.
Suppose that $\bar{x} = (x_1, \ldots, x_m)$ and $\bar{a} = (a_1, \ldots, a_m) \in [n]^m$.
It follows from Remark~\ref{remark on P restricted to A'}
and Lemma~\ref{lemma on conditional probability for R}
that
\begin{align*}
&\mbbP^{\mcA'}\big(\{\mcA \in W^{\mcA'} : \mcA \models p(\bar{a})\}\big) \  = \\ 
&\mbbP_n\big(\{\mcA \in \mbW_n: \mcA \models p(\bar{a}) \ | \ \{\mcA \in \mbW_n : \mcA \uhrc \sigma' = \mcA'\}\big) \ = \\
&\prod_{\substack{R \in \sigma \setminus \sigma' \text{ and} \\ R(x_{i_1}, \ldots, x_{i_r}) \in p(\bar{x})}}
\mcA'(\theta_R(a_{i_1}, \ldots, a_{i_r}))
\prod_{\substack{R \in \sigma \setminus \sigma' \text{ and} \\ \neg R(x_{i_1}, \ldots, x_{i_r}) \in p(\bar{x})}}
\big(1 - \mcA'(\theta_R(a_{i_1}, \ldots, a_{i_r}))\big).
\end{align*}
Assumption~\ref{inductive assumptions}~(5) now says that for every $R \in \sigma \setminus \sigma'$ there is
a basic probability formula $\chi_R$ which is asymptotically equivalent to $\theta_R$, with respect to $\mbbG'$
(where $\theta_R$ is the probability formula associated to $R$ by $\mbbG$), and if $r$ is the arity of $R$ then
the following holds for all $b_1, \ldots, b_r \in [n]$ (assuming that $n$ is large enough):
\[
\big| \mcA'(\theta_R(b_1, \ldots, b_r))  - \mcA'(\chi_R(b_1, \ldots, b_r)) \big| \leq \varepsilon'.
\]
Hence
\begin{align*}
&\Big| \mbbP^{\mcA'}\big(\{\mcA \in W^{\mcA'} : \mcA \models p(\bar{a})\}\big) \  - \\
&\prod_{\substack{R \in \sigma \setminus \sigma' \\ R(x_{i_1}, \ldots, x_{i_r}) \in p(\bar{x})}}
\mcA'(\chi_R(a_{i_1}, \ldots, a_{i_r}))
\prod_{\substack{R \in \sigma \setminus \sigma' \\ \neg R(x_{i_1}, \ldots, x_{i_r}) \in p(\bar{x})}}
\big(1 - \mcA'(\chi_R(a_{i_1}, \ldots, a_{i_r}))\big) \Big| \ \leq \ 
\varepsilon_p
\end{align*}
where $\varepsilon_p > 0$ depends only on $p$ and $\varepsilon'$.
Now we note that every number $\mcA'(\chi_R(a_{i_1}, \ldots, a_{i_r}))$ in the above expression depends only on $p'$.
The reason is that since $\chi_R$ is a basic probability formula the value $\mcA'(\chi_R(a_{i_1}, \ldots, a_{i_r}))$ depends only on 
the complete atomic $\sigma'$-type which is realized by $(a_{i_1}, \ldots, a_{i_r})$ in $\mcA'$ and we are assuming that
$\mcA' \models p'(\bar{a})$.
In other words, there is a constant, which we denote by $\msfP(p(\bar{x}) \ | \ p'(\bar{x}))$ such that 
for all sufficiently large $n$, all $\bar{a} \in [n]$ and all $\mcA' \in \mbY'_n$
such that $\mcA' \models p'(\bar{a})$,
\[
\big| \mbbP^{\mcA'}\big(\{\mcA \in \mbW^{\mcA'} : \mcA \models p(\bar{a})\} \big) \ - \ \msfP(p(\bar{x}) \ | \ p'(\bar{x})) \big| \ 
\leq \ \varepsilon_p. \qedhere
\]

\begin{defi}\label{definition of epsilon*}{\rm
Let $\varepsilon^*$ be the maximum of $\varepsilon'$ and of all $\varepsilon_p$ as in 
Lemma~\ref{uniformity of probabilities under qf} where the atomic $\sigma$-type $p(\bar{x}, \bar{y})$ is subject to the
constraint that $|\bar{x}\bar{y}| \leq k$.
}\end{defi}

\begin{lem}\label{P(p) is a product}
Suppose that $p'(\bar{x})$ is a complete atomic $\sigma'$-type and that $p(\bar{x}) \supseteq p'(\bar{x})$
is a (possibly partial) atomic $\sigma$-type.
Then for all sufficiently large $n$ and all $\bar{a} \in [n]^{|\bar{x}|}$ which realize the identity fragment of $p'(\bar{x})$ (and hence of $p$)
we have
\[
\big| \mbbP_n\big(\{\mcA \in \mbW_n : \mcA \models p(\bar{a})\} \ | \ \mbW^{\mbY'_n} \big) \ - \
\msfP(p(\bar{x}) \ | \ p'(\bar{x})) \cdot \msfP'(p'(\bar{x})) \big| \ \leq \ 7\varepsilon^*.
\]
\end{lem}

\proof
Let $\bar{a} \in [n]^{|\bar{x}|}$ realize the identity fragment of $p'(\bar{x})$. Furthermore,
\begin{itemize}
\item[] let $\mbX_n$ be the set of all $\mcA \in \mbW_n$ such that $\mcA \models p(\bar{a})$,
\item[] let $\mbX'_n$ be the set of all $\mcA' \in \mbW'_n$ such that $\mcA' \models p'(\bar{a})$, and
\item[] let $\mbZ'_n$ be the set of all $\mcA' \in \mbY'_n$ such that $\mcA' \models p'(\bar{a})$.
\end{itemize}
From parts~(2) and~(3) of Assumption~\ref{inductive assumptions} it easily follows that (for large enough $n$)
\begin{itemize}
\item[] $\mbbP'_n(\mbZ'_n) / \mbbP'_n(\mbY'_n)$ differs from $\mbbP'_n(\mbZ'_n)$ by at most $\varepsilon^*$,
\item[] $\mbbP'_n(\mbZ'_n)$ differs from $\mbbP'_n(\mbX'_n)$ by at most $\varepsilon^*$ and
\item[] $\mbbP'_n(\mbX'_n)$ differs from $\msfP'(p'(\bar{x}))$ by at most $\varepsilon^*$.
\end{itemize}
Consequently, 
\begin{equation}\label{a difference of at most 3varepsilon*}
\msfP'(p'(\bar{x})) - 3\varepsilon^* \leq \frac{\mbbP'_n(\mbZ'_n)}{\mbbP'_n(\mbY'_n)} \leq \msfP'(p'(\bar{x})) + 3\varepsilon^*.
\end{equation}
By Lemma~\ref{P(X given Y) equals P_Y(X cut Y)},
$\mbbP_n(\mbX_n \ | \ \mbW^{\mbY'_n}) = \mbbP^{\mbY'_n}(\mbX \cap \mbW^{\mbY'_n})$.
Then, using~(\ref{first basic equality about restricted distributions}), we have
\begin{align*}
&\mbbP^{\mbY'_n}\big(\mbX_n \cap \mbW^{\mbY'_n}\big) = 
\sum_{\mcA' \in \mbY'_n} \mbbP^{\mbY'_n}\big(\mbX_n \cap \mbW^{\mcA'}\big) =
\sum_{\mcA' \in \mbZ'_n} \mbbP^{\mbY'_n}\big(\mbX_n \cap \mbW^{\mcA'}\big) = \\
&\sum_{\mcA' \in \mbZ'_n} \ \sum_{\mcA \in \mbX_n \cap \mbW^{\mcA'}}\mbbP^{\mbY'_n}(\mcA) = 
\sum_{\mcA' \in \mbZ'_n} \ \sum_{\mcA \in \mbX_n \cap \mbW^{\mcA'}}
\frac{\mbbP'_n(\mcA')}{\mbbP'_n(\mbY'_n)} \mbbP^{\mcA'}(\mcA) = \\
&\sum_{\mcA' \in \mbZ'_n}\frac{\mbbP'_n(\mcA')}{\mbbP'_n(\mbY'_n)}\sum_{\mcA \in \mbX_n \cap \mbW^{\mcA'}}
\mbbP^{\mcA'}(\mcA) = 
\sum_{\mcA' \in \mbZ'_n}
\frac{\mbbP'_n(\mcA')}{\mbbP'_n(\mbY'_n)}\mbbP^{\mcA'}\big(\mbX_n \cap \mbW^{\mcA'}\big).
\end{align*}
By Lemma~\ref{uniformity of probabilities under qf} and~(\ref{a difference of at most 3varepsilon*}),
\begin{align*}
&\sum_{\mcA' \in \mbZ'_n}
\frac{\mbbP'_n(\mcA')}{\mbbP'_n(\mbY'_n)}\mbbP^{\mcA'}\big(\mbX_n \cap \mbW^{\mcA'}\big) \leq
\sum_{\mcA' \in \mbZ'_n}
\frac{\mbbP'_n(\mcA')}{\mbbP'_n(\mbY'_n)}\big(\msfP(p(\bar{x}) \ | \ p'(\bar{x})) + \varepsilon^*\big) = \\
&\frac{\mbbP'_n(\mbZ'_n)}{\mbbP'_n(\mbY'_n)}\big(\msfP(p(\bar{x}) \ | \ p'(\bar{x})) + \varepsilon^*\big) \leq
\big(\msfP'(p'(\bar{x})) + 3\varepsilon^*\big)\big(\msfP(p(\bar{x}) \ | \ p'(\bar{x})) + \varepsilon^*\big) \leq \\
&\msfP'(p'(\bar{x})) \cdot \msfP(p(\bar{x}) \ | \ p'(\bar{x})) + 7\varepsilon^*
\end{align*}
and in a similar way
\[
\sum_{\mcA' \in \mbZ'_n}
\frac{\mbbP'_n(\mcA')}{\mbbP'_n(\mbY'_n)}\mbbP^{\mcA'}\big(\mbX_n \cap \mbW^{\mcA'}\big) \geq
\msfP'(p'(\bar{x})) \cdot \msfP(p(\bar{x}) \ | \ p'(\bar{x})) - 7\varepsilon^*. \qedhere
\]

\begin{lem}\label{P(p) is a product, version 2}
Suppose that $p'(\bar{x})$ is a complete atomic $\sigma'$-type and that $p(\bar{x}) \supseteq p'(\bar{x})$
is an (possibly partial) atomic $\sigma$-type.
Then for all sufficiently large $n$ and all $\bar{a} \in [n]^{|\bar{x}|}$ which realize the identity fragment of $p'(\bar{x})$ we have
\[
\big| \mbbP_n\big(\{\mcA \in \mbW_n : \mcA \models p(\bar{a})\}\big) \ - \
\msfP(p(\bar{x}) \ | \ p'(\bar{x})) \cdot \msfP'(p'(\bar{x})) \big| \ < \ 9\varepsilon^*.
\]
\end{lem}

\proof
Let $\bar{a} \in [n]^{|\bar{x}|}$ realize the identity fragment of $p'(\bar{x})$ 
and we can assume that $n$ is large enough that $\delta'(n) \leq \varepsilon^*$.
Let $\mbX_n$ be the set of all $\mcA \in \mbW_n$ such that $\mcA \models p(\bar{a})$.
We have 
\[
\mbbP_n\big(\mbX_n\big) = \mbbP_n\big(\mbX_n \ | \ \mbW^{\mbY'_n}\big) \mbbP_n\big(\mbW^{\mbY'_n}\big)
+ 
\mbbP_n\big(\mbX_n \ | \ \mbW_n \setminus \mbW^{\mbY'_n}\big) \mbbP_n\big(\mbW_n \setminus \mbW^{\mbY'_n}\big).
\]
By the use of Lemma~\ref{P and P' agree when constraints speak only about W'} and by
Part~(2) of Assumption~\ref{inductive assumptions},
we also have
\[
\mbbP_n\big(\mbW_n \setminus \mbW^{\mbY'_n}\big) = 1 - \mbbP_n\big(\mbW^{\mbY'_n}\big) =
1 - \mbbP'_n(\mbY'_n) \leq \delta'(n).
\]
It follows that 
$\mbbP_n\big(\mbX_n \ | \ \mbW_n \setminus \mbW^{\mbY'_n}\big) \mbbP_n\big(\mbW_n \setminus \mbW^{\mbY'_n}\big)
\leq \delta'(n)$.
By Lemma~\ref{P and P' agree when constraints speak only about W'} 
and Part~(2) of Assumption~\ref{inductive assumptions},
$\mbbP_n\big(\mbW^{\mbY'_n}\big) = \mbbP'_n\big(\mbY'_n\big) \geq 1 - \delta'(n)$.
It now follows from 
Lemma~\ref{P(p) is a product}
that $\mbbP_n\big(\mbX_n\big)$ differs from 
$\msfP(p(\bar{x}) \ | \ p'(\bar{x})) \cdot \msfP'(p'(\bar{x}))$ 
by at most $7\varepsilon^* + 2\delta'(n) \leq 9\varepsilon^*$.
\qed

\begin{defi}\label{definition of P(p)}{\rm
For every (possibly partial) $\sigma$-type $p(\bar{x})$ such that 
$p'(\bar{x}) = p \uhrc \sigma'$ is a complete atomic $\sigma'$-type, we define
$
\msfP(p(\bar{x})) = 
\msfP'(p'(\bar{x})) \cdot \msfP(p(\bar{x}) \ | \ p'(\bar{x})).
$
}\end{defi}

\noindent
With this definition we can reformulate Lemma~\ref{P(p) is a product, version 2} as follows:

\begin{cor}\label{probability of p converges to P(p)}
If $p(\bar{x})$ is an (possibly partial) atomic $\sigma$-type such that $p \uhrc \sigma'$  is a complete atomic $\sigma'$-type,
then, for all sufficiently large $n$ and all $\bar{a} \in [n]^{|\bar{x}|}$ which realize the identity fragment of $p(\bar{x})$ we have
\[
\big| \mbbP_n\big(\{\mcA \in \mbW_n : \mcA \models p(\bar{a})\}\big) \ - \ \msfP(p(\bar{x})) \big| \ < \ 9\varepsilon^*.
\]
\end{cor}

\begin{lem}\label{the number of realizations in A'}
Suppose that $p(\bar{x}, y)$ and $q(\bar{x})$ are complete atomic $\sigma$-types such that $|\bar{x}y| \leq k$,
$\dim_{y}(p) = 1$ and $q \subseteq p$. 
Then there are $\gamma \in [0, 1]$ and $c > 0$ such that for all sufficiently large $n$ and all $\mcA' \in \mbY'_n$,
\begin{align*}
\mbbP^{\mcA'}\big(\{\mcA \in \mbW^{\mcA'} : &\text{ $\mcA$ is $(p, q, (\gamma - 5\varepsilon^*)$-saturated}\\
 &\text{and $(p, q, (\gamma + 5\varepsilon^*)$-unsaturated} \}\big) \ \geq \ 1 \ - \ e^{-cn}.
\end{align*}
\end{lem}

\proof
Let $n$ be large enough that Part~(4) of Assumption~\ref{inductive assumptions} holds.
Suppose that $p(\bar{x}, y)$ and $q(\bar{x})$ are complete atomic $\sigma$-types such that
$|\bar{x}y| \leq k$, $\dim_y(p) = 1$ and
$q \subseteq p$.
Let $p' = p \uhrc \sigma$ and $q' = q \uhrc \sigma'$.
Moreover, let $p^y(\bar{x}, y)$ include $p'(\bar{x}, y)$ and all $(\sigma \setminus \sigma')$-formulas
in $p(\bar{x}, y)$ which contain the variable $y$.

Let $\mcA' \in \mbY'_n$.
By Part~(4) of 
Assumption~\ref{inductive assumptions} there is $\alpha \in [0, 1]$ such that
$\mcA'$ is $(p', q', \alpha - \varepsilon')$-saturated and $(p', q', \alpha + \varepsilon')$-unsaturated.
By the same assumption, $\alpha$ does not depend on the particular choice of $\mcA' \in \mbY'_n$.
Let 
\[
\beta = \msfP(p^y(\bar{x}, y) \ | \ p'(\bar{x}, y)),
\]
where $\msfP(p^y(\bar{x}, y) \ | \ p'(\bar{x}, y))$ is like in 
Lemma~\ref{uniformity of probabilities under qf},
and let
\[
\gamma = \alpha \beta.
\]
For every $\bar{a} \in [n]^{|\bar{x}|}$ let
\[
B'_{\bar{a}} \ = \ \big\{b \in [n] : \mcA' \models p'(\bar{a}, b)\big\}.
\]
Since $\mcA'$ is  $(p', q', \alpha - \varepsilon')$-saturated and $(p', q', \alpha + \varepsilon')$-unsaturated we have
\[
(\alpha - \varepsilon')n \leq |B'_{\bar{a}}| \leq (\alpha + \varepsilon')n.
\]
For every $\bar{a} \in [n]^{|\bar{x}|}$ and every $\mcA \in \mbW^{\mcA'}$ let
\[
B_{\bar{a}, \mcA} \  = \ \big\{ b \in [n] : \mcA \models p^y(\bar{a}, b)\big\}
\]
and note that $B_{\bar{a}, \mcA} \subseteq B'_{\bar{a}}$ for every $\bar{a}$ and every $\mcA \in \mbW^{\mcA'}$.
It follows that if $\alpha = 0$ then the conclusion of the lemma follows with $\gamma = 0$ 
because $\varepsilon' \leq \varepsilon^*$. 
So for the rest of the proof we assume that $\alpha > 0$. 
By starting with a sufficiently small  $\varepsilon'$ we can assume that $\varepsilon^*$ is small enough so that
$\alpha, \beta > \varepsilon^*$.
Let 
\begin{align*}
\mbX_{\bar{a}} \ = \ \big\{ \mcA \in \mbW^{\mcA'} &: 
\text{ either $\mcA \not\models q(\bar{a})$ or} \\ 
&(1- \varepsilon^*)(\gamma - 2\varepsilon^*)n \leq |B_{\bar{a}, \mcA}| \leq 
(1 + \varepsilon^*)(\gamma + 2\varepsilon^*)n \big\}.
\end{align*}
Observe that if $\mcA \in \mbW^{\mcA'}$, $\mcA \models q(\bar{a})$ and $\mcA \models p^y(\bar{a}, b)$,
then $\mcA \models p(\bar{a}, b)$. 
Hence every $\mcA \in \bigcap_{\bar{a} \in [n]^{|\bar{x}|}} \mbX_{\bar{a}}$ is 
$(p, q, (1- \varepsilon^*)(\gamma - 2\varepsilon^*))$-saturated and $(p, q, (1+ \varepsilon^*)(\gamma + 2\varepsilon^*))$-unsaturated.

Fix any $\bar{a}$ such that $\mcA' \models q'(\bar{a})$ (and note that $\mcA \models q(\bar{a})$ implies $\mcA' \models q'(\bar{a})$).
By Remark~\ref{remark on P restricted to A'},
for all distinct $b, c \in B'_{\bar{a}}$, the events
\[
\mbE_b = \{\mcA \in \mbW^{\mcA'} : \mcA \models p^y(\bar{a}, b)\} \ \text{ and } \
\mbE_c = \{\mcA \in \mbW^{\mcA'} : \mcA \models p^y(\bar{a}, c)\}
\]
are independent.
Moreover,
by the choice of $\beta$ and the definition of $\varepsilon^*$
(Definition~\ref{definition of epsilon*}),
for each $b \in B'_{\bar{a}}$, 
\[
\beta - \varepsilon^* \leq \mbbP^{\mcA'}_n(\mbE_b)  \leq \beta + \varepsilon^*.
\]
Let $Z : \mbW^{\mcA'} \to \mbbN$ be the random variable defined by
\[
Z(\mcA) = \big| \{b \in B'_{\bar{a}} : \mcA \models p^y(\bar{a}, b) \}\big|.
\]

It follows from Corollary~\ref{independent bernoulli trials, second version} that
\[
\mbbP^{\mcA'}_n\big( Z > (1 + \varepsilon^*)(\beta + \varepsilon^*)|B'_{\bar{a}}|  \big) \  < \ 
2\exp\big( -c_{\varepsilon^*} (\beta + \varepsilon^*)|B'_{\bar{a}}| \big)
\]
and
\[
\mbbP^{\mcA'}_n\big( Z < (1 - \varepsilon^*)(\beta - \varepsilon^*)|B'_{\bar{a}}|  \big) \  < \ 
2\exp\big( -c_{\varepsilon^*} (\beta - \varepsilon^*)|B'_{\bar{a}}| \big)
\]
where the constant $c_{\varepsilon^*} > 0$ depends only on $\varepsilon^*$.
By using that 
$\alpha n \geq |B'_{\bar{a}}| - \varepsilon' n \geq |B'_{\bar{a}}| - \varepsilon^* n$ we see that
\begin{align*}
&(\gamma + 2\varepsilon^*)n \geq 
(\alpha \beta + 2\varepsilon^*)n \geq 
\beta \alpha n +  2\varepsilon^* n \geq
\beta(|B'_{\bar{a}}| - \varepsilon^*n) + 2\varepsilon^*n \geq \\
&\beta |B'_{\bar{a}}| - \beta \varepsilon^*n + 2\varepsilon^*n \geq 
\beta |B'_{\bar{a}}| + \varepsilon^*n \geq
\beta |B'_{\bar{a}}| + \varepsilon^*|B'_{\bar{a}}| =
(\beta + \varepsilon^*)|B'_{\bar{a}}|
\end{align*}
and by similar reasoning we get
\[
(\gamma - 2\varepsilon^*)n \leq (\beta - \varepsilon^*)|B'_{\bar{A}}|.
\]
It follows that if $Z > (1 + \varepsilon^*)(\gamma + 2\varepsilon^*)n$ then
$Z > (1 + \varepsilon^*)(\beta + \varepsilon^*)|B'_{\bar{a}}|$, and
if $Z < (1 - \varepsilon^*)(\gamma - 2\varepsilon^*)n$ then
$Z < (1 - \varepsilon^*)(\beta - \varepsilon^*)|B'_{\bar{a}}|$.
Hence we have
\begin{align*}
&\mbbP^{\mcA'}\big(\mbW^{\mcA'} \setminus \mbX_{\bar{a}}\big) \  < \ 
2\exp\big( -c_{\varepsilon^*} (\beta + \varepsilon^*)|B'_{\bar{a}}| \big) +
2\exp\big( -c_{\varepsilon^*} (\beta - \varepsilon^*)|B'_{\bar{a}}| \big) \ \leq \\
&2\exp\big( -c_{\varepsilon^*} (\beta + \varepsilon^*)(\alpha  - \varepsilon^*)n \big) +
2\exp\big( -c_{\varepsilon^*} (\beta - \varepsilon^*)(\alpha  - \varepsilon^*)n \big) \ \leq \
e^{-dn}
\end{align*}
for some constant $d > 0$ that depends only on $\varepsilon^*$, $p$ and $q$.
Since the argument works for all $\bar{a} \in [n]^{|\bar{x}|}$ it follows that
\[
\mbbP^{\mcA'}\bigg(\bigcap_{\bar{a} \in [n]^{|\bar{x}|}} \mbX_{\bar{a}}\bigg) \ \geq \
1 -  n^{|\bar{x}|} e^{-dn} \geq 1 - e^{-cn}
\]
for some constant $c > 0$.
Since 
\[
(1 \pm \varepsilon^*)(\gamma \pm 2\varepsilon^*) = 
\gamma \pm 2\varepsilon^* \pm \varepsilon^*\gamma \pm 2(\varepsilon^*)^2
\]
and (if $0 < \varepsilon^* < 1$) $|2\varepsilon^* \pm \varepsilon^*\gamma \pm 2(\varepsilon^*)^2| \leq 5\varepsilon^*$
we get the conclusion of the lemma. 
\qed
\\

\noindent
The next lemma generalizes the previous one to types $p(\bar{x}, \bar{y})$ where the length of
$\bar{y}$ is greater than one.

\begin{lem}\label{the number of realizations in A' for y-bar}
Suppose that $p(\bar{x}, \bar{y})$ and $q(\bar{x})$ are complete atomic $\sigma$-types such that $|\bar{x}\bar{y}| \leq k$,
$\dim_{\bar{y}}(p) = |\bar{y}|$ and $q \subseteq p$.
Then there are $\gamma \in [0, 1]$ and $c  > 0$ such that for all large enough $n$ and all $\mcA' \in \mbY'_n$,
\begin{align*}
\mbbP^{\mcA'}\big(\{\mcA \in \mbW^{\mcA'} : &\text{ $\mcA$ is $(p, q, \alpha - 11^{|\bar{y}| -1}\cdot 5\varepsilon^*)$-saturated}\\
 &\text{and $(p, q, \alpha + 11^{|\bar{y}| +1}\cdot 5\varepsilon^*)$-unsaturated} \}\big) \ \geq \ 
1 - e^{-cn}.
\end{align*}
\end{lem}

\proof
We prove the lemma by induction on $m = |\bar{y}|$. The base case $m = 1$ is given by
Lemma~\ref{the number of realizations in A'} (since then $11^{m-1}\cdot 5\varepsilon^* = 5\varepsilon^*$).
Let $p(\bar{x}, \bar{y})$ and $q(\bar{x})$ be as assumed in the lemma where $\bar{y} = (y_1, \ldots, y_{m+1})$.
Let $p_m(\bar{x}, y_1, \ldots, y_m)$ be the restriction of $p$ to formulas with variables among $\bar{x}, y_1, \ldots, y_m$.
By induction hypothesis, there are $\alpha \in [0, 1]$ and  $a > 0$ such that 
for all sufficiently large $n$ and all $\mcA' \in \mbY'_n$,
\begin{align*}
\mbbP^{\mcA'}\big(\{\mcA \in \mbW^{\mcA'} : &\text{ $\mcA$ is $(p_m, q, \alpha - 11^{m-1}\cdot 5\varepsilon^*)$-saturated}\\
 &\text{and $(p_m, q, \alpha+ 11^{m-1}\cdot 5\varepsilon^*)$-unsaturated} \}\big) \ \geq \ 
1 - e^{-an}.
\end{align*}
By Lemma~\ref{the number of realizations in A'}, there are $\beta \in [0, 1]$ and $b > 0$ such that 
for all sufficiently large $n$ and all $\mcA' \in \mbY'_n$,
\begin{align*}
\mbbP^{\mcA'}\big(\{\mcA \in \mbW^{\mcA'} : &\text{ $\mcA$ is $(p, p_m, (\beta - 5\varepsilon^*)$-saturated}\\
 &\text{and $(p, p_m, (\beta + 5\varepsilon^*)$-unsaturated} \}\big) \ \geq \ 1 \ - \ e^{-bn}.
\end{align*}
Suppose that $\mcA \in \mbW^{\mcA'}$ is 
$(p_m, q, \alpha - 11^{m-1}\cdot 5\varepsilon^*)$-saturated
and 
$(p, p_m, \beta - 5\varepsilon^*)$-saturated.
Then one straightforwardly finds that $\mcA$ is 
\[
(p, q, ( \alpha - 11^{m-1}\cdot 5\varepsilon^*)(\beta - 5\varepsilon^*))\text{-saturated}
\]
and by calculations we get the following quite crude estimate
\[
(\alpha - 11^{m-1}\cdot 5\varepsilon^*)(\beta - 5\varepsilon^*) \geq \alpha\beta - 11^m \cdot 5\varepsilon^*.
\]
Hence $\mcA$ is $(p, q, \alpha\beta - 11^m \cdot 5\varepsilon^*)$-saturated.
In a similar way it follows that if
$\mcA \in \mbW^{\mcA'}$ is 
$(p_m, q, \alpha + 11^{m-1}\cdot 5\varepsilon^*)$-unsaturated
and 
$(p, p_m, \beta + 5\varepsilon^*)$-unsaturated, then $\mcA$ is 
$(p, q,  \alpha\beta + 11^m \cdot 5\varepsilon^*)$-unsaturated.
Let $\gamma = \alpha \beta$.
Since there is $c > 0$ such that
$e^{-an} + e^{-bn} \leq e^{-cn}$ for all large enough $n$
we get the desired estimate of the probability in the statement of the lemma.
\qed

\begin{cor}\label{all members of Y-n are sufficiently saturated}
Let $p(\bar{x}, \bar{y})$ and $q(\bar{x})$ are complete atomic $\sigma$-types such that
$|\bar{x}\bar{y}| \leq k$, $d = \dim_{\bar{y}}(p) > 0$, $q \subseteq p$.
Then there are $\gamma \in [0, 1]$, depending only on $p$, $q$ and $\mbbG$, and $c > 0$ such that
for all sufficiently large $n$ and all $\mcA' \in \mbY'_n$,
\begin{align*}
\mbbP^{\mcA'}\big(\{\mcA \in \mbW^{\mcA'} : &\text{ $\mcA$ is $(p, q, \alpha - 11^{d -1}\cdot 5\varepsilon^*)$-saturated}\\
 &\text{and $(p, q, \alpha + 11^{d +1}\cdot 5\varepsilon^*)$-unsaturated} \}\big) \ \geq \ 
1 - e^{-cn}.
\end{align*}
\end{cor}

\proof
Suppose that $p(\bar{x}, \bar{y})$, where $\bar{y} = (y_1, \ldots, y_m)$ 
is a complete $\sigma$-type and let $q(\bar{x}) = p \uhrc \bar{x}$.
Let $d = \dim_{\bar{y}}(p)$. 
Then there is a subsequence $y_{i_1}, \ldots, y_{i_d}$ (of distinct variables) such that if $j \in [m] \setminus \{i_1, \ldots, i_d\}$
then there is $i \in \{i_1, \ldots, i_d\}$ such that the formula $y_j = y_i$ belongs to $p$.
By reordering the sequence $\bar{y}$ we can assume that $i_1 = 1, \ldots, i_d = d$ (and hence, if $d < j \leq m$ then $y_j = y_i$ belongs
to $p$ for some $i \leq d$).
Let $p^*(\bar{x}, y_1, \ldots, y_d)$ be the set of all formulas $\varphi \in p(\bar{x}, \bar{y})$ such that 
all variables in $\varphi$ belong to $\rng(\bar{x}) \cup \{y_1, \ldots, y_d\}$.
It follows that $\dim_{(y_1, \ldots, y_d)}(p^*) = d$ and for every $\sigma$-structure $\mcA$ and every $\bar{a} \in A^{|\bar{x}|}$
then number of $\bar{b} \in A^{|\bar{y}|}$ such that $\mcA \models p(\bar{a}, \bar{b})$ equals the number of 
$\bar{b} \in A^d$ such that $\mcA \models p^*(\bar{a}, \bar{b})$.
Lemma~\ref{the number of realizations in A' for y-bar} implies that there are
$\gamma \in [0, 1]$ and $c > 0$ such that for all large enough $n$,
\begin{align*}
\mbbP^{\mcA'}\big(\{\mcA \in \mbW^{\mcA'} : &\text{ $\mcA$ is $(p^*, q, \gamma - 11^{d - 1}\cdot 5\varepsilon^*)$-saturated}\\
 &\text{and $(p^*, q, \gamma + 11^{d + 1}\cdot 5\varepsilon^*)$-unsaturated} \}\big) \ \geq \ 
1 - e^{-cn}.
\end{align*}
By the construction of $p^*$ the above lower bound on the probability also holds if $p^*$ is replaced by $p$.
\qed

\begin{defi}\label{definition of Y-n}{\rm
For every $n$, let $\mbY_n$ be the set of all $\mcA \in \mbW^{\mbY'_n}$ such that
whenever $p(\bar{x}, \bar{y})$ and $q(\bar{x})$ are complete atomic $\sigma$-types with
$|\bar{x}\bar{y}| \leq k$, $0 < \dim_{\bar{y}}(p) = d$, $q \subseteq p$ and
$\gamma$ is the number associated to $p$ and $q$ as in 
Corollary~\ref{all members of Y-n are sufficiently saturated}, then 
$\mcA$ is $(p, q, \gamma - 11^{|\bar{y}| - 1}\cdot 5\varepsilon^*)$-saturated and
$(p, q, \gamma + 11^{|\bar{y}| - 1}\cdot 5\varepsilon^*)$-unsaturated.
}\end{defi}

\begin{lem}\label{Y-n has large probability}
There is a constant $c > 0$ such that for all sufficiently large  $n$,
$\mbbP_n\big(\mbY_n\big) \geq \big(1 - e^{-c n}\big) \big(1 - \delta'(n)\big)$.
\end{lem}

\proof
There are, up to changing variables, only finitely many atomic $\sigma$-types $p(\bar{x})$ such that $|\bar{x}| \leq k$.
It follows from 
Corollary~\ref{all members of Y-n are sufficiently saturated}
that there is a constant $c > 0$ such that for all large enough $n$ and
all $\mcA' \in \mbY'_n$,
\[
\mbbP_n^{\mcA'}\big(\mbY_n \cap \mbW^{\mcA'}\big) \geq 1 - e^{-c n}.
\]
Since $\mbY_n \subseteq \mbW^{\mbY'_n}$ we have
$\mbbP_n(\mbY_n) = \mbbP_n\big(\mbY_n \ | \ \mbW^{\mbY'_n}\big) \mbbP_n\big(\mbW^{\mbY'_n}\big)$.
By Lemma~\ref{P and P' agree when constraints speak only about W'},
$\mbbP_n(\mbW^{\mbY'_n}) = \mbbP'_n(\mbY'_n)$ and
by Lemma~\ref{P(X given Y) equals P_Y(X cut Y)} we have
$\mbbP_n(\mbY_n \ | \ \mbW^{\mbY'_n}) = \mbbP^{\mbY'_n}(\mbY_n \cap \mbW^{\mbY'_n})$.
Hence
$\mbbP_n(\mbY_n) = \mbbP^{\mbY'_n}(\mbY_n \cap \mbW^{\mbY'_n}) \mbbP'_n\big(\mbY'_n\big)$.
Then, reasoning similarly as in the proof of Lemma~\ref{P(p) is a product} (using~(\ref{first basic equality about restricted distributions})),  
we get
\begin{align*}
&\mbbP^{\mbY'_n}\big(\mbY_n \cap \mbW^{\mbY'_n}\big) \ = \
\sum_{\mcA' \in \mbY'_n} \mbbP^{\mbY'_n}\big(\mbY_n \cap \mbW^{\mcA'}\big) \ = \
\sum_{\mcA' \in \mbY'_n}\sum_{\mcA \in \mbY_n \cap \mbW^{\mcA'}}\mbbP^{\mbY'_n}(\mcA) = \\
&\sum_{\mcA' \in \mbY'_n}\sum_{\mcA \in \mbY_n \cap \mbW^{\mcA'}}
\frac{\mbbP'_n(\mcA')}{\mbbP'_n(\mbY'_n)} \mbbP^{\mcA'}(\mcA) \ = \
\sum_{\mcA' \in \mbY'_n}\frac{\mbbP'_n(\mcA')}{\mbbP'_n(\mbY'_n)}\sum_{\mcA \in \mbY_n \cap \mbW^{\mcA'}}
\mbbP^{\mcA'}(\mcA) = \\
&\sum_{\mcA' \in \mbY'_n}\frac{\mbbP'_n(\mcA')}{\mbbP'_n(\mbY'_n)}\mbbP^{\mcA'}\big(\mbY_n \cap \mbW^{\mcA'}\big) 
\ \geq \ 
\sum_{\mcA' \in \mbY'_n}\frac{\mbbP'_n(\mcA')}{\mbbP'_n(\mbY'_n)}
\big(1 - e^{-c n}\big) = \big(1 - e^{-c n}\big).
\end{align*}
Using Part~(2) of Assumption~\ref{inductive assumptions} we get
\[
\mbbP_n(\mbY_n) \ = \ \mbbP^{\mbY'_n}\big(\mbY_n \cap \mbW^{\mbY'_n}\big) \mbbP'_n(\mbY'_n) \ \geq \
\big(1 - e^{-c n}\big) \big(1 - \delta'(n)\big). \qedhere
\]

\begin{prop}\label{finishing the induction step}
Let $\varepsilon = 11^{k-1} \cdot 5\varepsilon^*$.
Then there is a function $\delta : \mbbN \to \mbbR^{\geq 0}$ such that if 
$\sigma'$, $\varepsilon'$, $\delta'$, $\mbW'_n$, $\mbY'_n$ and  $\mbbP'_n$ are replaced by
$\sigma$, $\varepsilon$, $\delta$, $\mbW_n$, $\mbY_n$ and  $\mbbP_n$, respectively,
then parts~(1) -- (4) of Assumption~\ref{inductive assumptions} hold.
\end{prop}

\proof
From Lemma~\ref{Y-n has large probability} it follows that there is a function
$\delta : \mbbN \to \mbbR^{\geq 0}$ such that $\lim_{n\to\infty}\delta(n) = 0$ and 
$\mbbP_n(\mbY_n) \geq 1 - \delta(n)$ for all sufficiently large $n$.
Hence parts~(1) and~(2) of 
Assumption~\ref{inductive assumptions}
hold if $\mbbP'_n$, $\mbY'_n$ and $\delta'$ are replaced by $\mbbP_n$, $\mbY_n$ and $\delta$, respectively.

Let $\varepsilon = 11^{k - 1}\cdot 5\varepsilon^*$.
By Corollary~\ref{probability of p converges to P(p)},
Part~(3) of 
Assumption~\ref{inductive assumptions}
holds if $\sigma'$, $\mbbP'_n$, $\mbW'_n$ and $\varepsilon'$ are replaced by $\sigma$, $\mbbP_n$, $\mbW_n$ and $\varepsilon$, respectively.
By Corollary~\ref{all members of Y-n are sufficiently saturated}
and Definition~\ref{definition of Y-n},
Part~(4) of Assumption~\ref{inductive assumptions}
holds if $\sigma'$, $\mbY'_n$ and $\varepsilon'$ are replaced by $\sigma$, $\mbY_n$ and $\varepsilon$, respectively.
\qed

\begin{cor}\label{inductive assumption holds for all k and epsilon}
For all $k \in \mbbN^+$ and all $\varepsilon > 0$ there are $\delta : \mbbN^+ \to \mbbR^{\geq 0}$ and 
$\mbY_n \subseteq \mbW_n$, for $n \in \mbbN^+$, such that the following hold:
\begin{enumerate}
\item[(1)] $\lim_{n\to\infty} \delta(n) = 0$.

\item[(2)] $\mbbP_n(\mbY_n) \geq 1 - \delta(n)$ for all sufficiently large $n$.

\item[(3)] For every complete atomic $\sigma$-type $p(\bar{x})$ with $|\bar{x}| \leq k$
there is a number which we denote 
$\msfP(p(\bar{x}))$, or just $\msfP(p)$, such that for all sufficiently large $n$ and all $\bar{a} \in [n]^{|\bar{x}|}$ 
which realize the identity fragment of $p$,
\[
\big| \mbbP_n\big(\{\mcA \in \mbW_n : \mcA \models p(\bar{a})\}\big) \ -  \ \msfP(p(\bar{x})) \big| \ \leq \ \varepsilon.
\]

\item[(4)] For every complete atomic $\sigma$-type $p(\bar{x}, \bar{y})$ with $|\bar{x}\bar{y}| \leq k$ and
$0 < \dim_{\bar{y}}(p(\bar{x}, y)) = d$,
if $q(\bar{x}) = p \uhrc \bar{x}$, then there is $\alpha \in [0, 1]$ such that, for all sufficiently large $n$, every
$\mcA \in \mbY_n$ is $(p, q, \alpha - \varepsilon)$-saturated and $(p, q, \alpha + \varepsilon)$-unsaturated.
\end{enumerate}
\end{cor}

\proof
This follows from 
Proposition~\ref{finishing the induction step}
because $k \in \mbbN^+$ and $\varepsilon' > 0$ in the argument until
Proposition~\ref{finishing the induction step}
are arbitrary and (for any fixed $k$) the choice of $\varepsilon$ in 
Proposition~\ref{finishing the induction step} 
tends to zero as $\varepsilon'$ tends to zero.
\qed

\begin{cor}\label{concluding elimination of aggregation functions}
If all aggregation functions in $\varphi(\bar{x}) \in PLA^+(\sigma)$ are strongly admissible, 
then $\varphi(\bar{x})$ is asymptotically equivalent (with respect to $\mbbG$) to a basic probability formula.
\end{cor}

\proof
According to 
Corollary~\ref{inductive assumption holds for all k and epsilon},
for every $k \in \mbbN^+$ and every  $\varepsilon > 0$ there are $\delta : \mbbN^+ \to \mbbR^{\geq 0}$ and 
$\mbY_n \subseteq \mbW_n$, for $n \in \mbbN^+$, such that~(1), (2) and~(4) of that corollary hold.
This means that Assumption~\ref{assumption about saturation} holds.
Hence Proposition~\ref{saturation implies elimination of strongly admissible functions}
implies that if all aggregation functions in $\varphi(\bar{x}) \in PLA^+(\sigma)$ are strongly admissible, then
$\varphi(\bar{x})$ is asymptotically equivalent to a basic probability formula.
\qed
\\

\noindent
With corollaries~\ref{inductive assumption holds for all k and epsilon}
and~\ref{concluding elimination of aggregation functions}
we have completed the proof of
Theorem~\ref{elimination of strongly admissible aggregation functions}.
Informally speaking, the next corollary tells that for every $PLA^+(\sigma)$-network $\mbbG$ that uses only 
strongly admissible aggregation functions there is a $PLA^+(\sigma)$-network $\tilde{\mbbG}$ which uses no
aggregation functions at all and is such that every query defined by a $PLA^+(\sigma)$-formula
with only strongly admissible aggregation functions has the same asymtotic probability whether
we compute it with $\mbbG$ or with $\mbbG'$.

\begin{cor}[aggregation-free networks]\label{equivalence to an aggregation-free network} 
Let $\sigma$ be a finite relational signature and let
$\mbbG$ be a $PLA^+(\sigma)$-network
such that every probability formula of $\mbbG$ 
contains only strongly admissible aggregation functions.
Let $(\mbbP_n : n \in \mbbN^+)$ be the sequence of probability distributions induced by $\mbbG$.

By Part~(i) of 
Theorem~\ref{elimination of strongly admissible aggregation functions},
for every $R \in \sigma$, the probability formula $\theta_R$ which is associated to $R$ by $\mbbG$
is asymptotically equivalent to a basic probability formula $\chi_R$ with respect to $\mbbG$.
Let $\tilde{\mbbG}$ be the $PLA^+(\sigma)$-network with the same underlying directed acyclic graph as $\mbbG$ and 
where, for each $R \in \sigma$, $\chi_R$ is the probability formula associated to $R$ by $\tilde{\mbbG}$.
Let $(\tilde{\mbbP}_n : n \in \mbbN^+)$ be the sequence of probability distributions induced by $\tilde{\mbbG}$.
Then the following hold:
\begin{enumerate}
\item[(i)] For every atomic $\sigma$-type $p(\bar{x})$, every $m \in \mbbN^+$ and every $\bar{a} \in [m]^{|\bar{x}|}$,
\[
\lim_{n\to\infty} \tilde{\mbbP}_n(p(\bar{a})) = \lim_{n\to\infty} \mbbP_n(p(\bar{a})).
\]
The common limit depends only on the probability formulas of $\tilde{\mbbG}$ and the common directed acyclic graph
of $\mbbG$ and $\tilde{\mbbG}$.

\item[(ii)] Let $p(\bar{x}, \bar{y})$ and $q(\bar{x})$ be complete atomic $\sigma$-types such that $q \subseteq p$.
Then there is $\alpha \in [0, 1]$, depending only on $p$, $q$, the common directed acyclic graph of $\mbbG$ and $\tilde{\mbbG}$
and the probability formulas of $\tilde{\mbbG}$ such that
\begin{align*}
&\lim_{n\to\infty} \tilde{\mbbP}_n\big( \{ \mcA \in \mbW_n : \mcA \text{ is $(p, q, \alpha - \varepsilon)$-saturated 
and $(p, q, \alpha + \varepsilon)$-unsaturated} \}\big)   = \\
&\lim_{n\to\infty} \mbbP_n\big( \{ \mcA \in \mbW_n : \mcA \text{ is $(p, q, \alpha - \varepsilon)$-saturated 
and $(p, q, \alpha + \varepsilon)$-unsaturated} \}\big) = 1.
\end{align*}

\item[(iii)] Let $\varphi(\bar{x}) \in PLA^+(\sigma)$ be such that every aggregation function in $\varphi$ is strongly admissible.
Then there is a basic probability formula $\psi(\bar{x})$ such that $\varphi(\bar{x})$ and $\psi(\bar{x})$ are asymptotically equivalent
with respect to $\mbbG$ and with respect to $\tilde{\mbbG}$.
Without loss of generality we may assume that $\psi(\bar{x})$ has the form $\bigwedge_{i=1}^k (\psi_i(\bar{x}) \to c_i)$ 
where each $\psi_i(\bar{x})$ is the conjuction of all formulas in a complete atomic $\sigma$-type $p_i(\bar{x})$ and that all 
complete atomic $\sigma$-types in the variables $\bar{x}$ are enumerated without repetition by $p_1, \ldots, p_k$.
If $I \subseteq [0, 1]$ is an interval such that no $c_i$ is an endpoint of $I$, then for every $m \in \mbbN^+$
and every $\bar{a} \in [m]^{|\bar{x}|}$,
\[
\lim_{n\to\infty} \big| \tilde{\mbbP}_n\big(\{\mcA \in \mbW_n : \mcA(\psi(\bar{a})) \in I \}\big) - 
\mbbP_n\big(\{\mcA \in \mbW_n : \mcA(\varphi(\bar{a})) \in I \} \big) \big| \ = \ 0,
\]
where $\tilde{\mbbP}_n\big(\{\mcA \in \mbW_n : \mcA(\psi(\bar{a}))\big)$ can be computed by only
using $\tilde{\mbbG}$ and $\psi$, hence in constant time with respect to the domain size.
\end{enumerate}
\end{cor}

\noindent
{\bf Proof sketch.}
(i) It suffices to consider complete atomic $\sigma$-types. 
Let $p(\bar{x})$ be a complete atomic $\sigma$-type.
An analysis of the proof leading to 
Corollary~\ref{inductive assumption holds for all k and epsilon}
shows that 
the limit $\lim_{n\to\infty} \mbbP_n(p(\bar{a}))$ depends only on
the basic probability formulas
$\chi_R$ first mentioned in Part~(5) of Assumption~\ref{inductive assumptions};
the crucial step where $\chi_R$ is used is
Lemma~\ref{uniformity of probabilities under qf}.
If a probability formula $\theta_R$, as in Part~(5) of Assumption~\ref{inductive assumptions},
is a basic probability formula then we can simply let $\chi_R$ be the same formula as $\theta_R$ to make sure that
Part~(5) of Assumption~\ref{inductive assumptions} holds.
Let $\mbbG$ and $\tilde{\mbbG}$ be as assumed. 
By the construction of $\tilde{\mbbG}$ it follows from what has been said that both limits
$\lim_{n\to\infty} \mbbP_n(p(\bar{a}))$ and $\lim_{n\to\infty} \tilde{\mbbP}_n(p(\bar{a}))$
depend only on the probability formulas $\chi_R$ of $\tilde{\mbbG}$, so the limits are equal.

(ii) If we analyze the proofs above,
in particular the proof of Lemma~\ref{the number of realizations in A'},
we see that the ``saturation number'' $\alpha$ depends only on limits of the form
$\lim_{n\to\infty} \mbbP_n(p(\bar{a}))$ where $p(\bar{x})$ is an atomic $\sigma$-type.
By Part~(i) we get the same limit if $\mbbP_n$ is replaced by $\tilde{\mbbP}_n$ and therefore the conclusion follows.

(iii) Part~(ii) implies that the saturation conditions stated by 
Assumption~\ref{assumption about saturation}
are satisfied by both $(\mbbP_n : n \in \mbbN^+)$ and $(\tilde{\mbbP}_n : n \in \mbbN^+)$.
Therefore Proposition~\ref{saturation implies elimination of strongly admissible functions}
applies to both $(\mbbP_n : n \in \mbbN^+)$ and $(\tilde{\mbbP}_n : n \in \mbbN^+)$.
So if $\varphi(\bar{x}) \in PLA^+(\sigma)$ has only strongly admissible aggregation functions then there are
basic probability formulas $\psi(\bar{x})$ and $\tilde{\psi}(\bar{x})$ such that 
$\varphi(\bar{x}) \sim_{\mbbG} \psi(\bar{x})$ and $\varphi(\bar{x}) \sim_{\tilde{\mbbG}} \tilde{\psi}(\bar{x})$.
An analysis of the proof of Proposition~\ref{saturation implies elimination of strongly admissible functions} shows that
the constructions of $\psi$ and $\tilde{\psi}$ depend only on the ``saturation numbers'' $\alpha$ mentioned in Part~(ii) of this corollary.
Since these saturation numbers are the same (as stated in Part~(ii)) for $(\mbbP_n : n \in \mbbN^+)$ and $(\tilde{\mbbP}_n : n \in \mbbN^+)$
it follows that the constructed $\psi$ and $\tilde{\psi}$ are the same.
So $\varphi(\bar{x}) \sim_{\mbbG} \psi(\bar{x})$ and $\varphi(\bar{x}) \sim_{\tilde{\mbbG}} \psi(\bar{x})$.

Suppose, without loss of generality,  that $\psi(\bar{x})$ is $\bigwedge_{i=1}^k (\psi_i(\bar{x}) \to c_i)$
where each $\psi_i(\bar{x})$ is a conjunction of the formulas in a complete atomic $\sigma$-type $p_i(\bar{x})$
and that all complete atomic $\sigma$-types are enumerated without repetition by $p_1, \ldots, p_k$.
By reordering if necessary we may assume that $c_1, \ldots, c_l \in I$ and $c_i \notin I$ if $i > l$.
Then 
\[
\mbbP_n\big(\{\mcA \in \mbW_n : \mcA(\psi(\bar{a})) \in I \} \big) = \mbbP_n\bigg(\bigvee_{i=1}^l \psi_i(\bar{a}) \bigg)=
\sum_{i=1}^l \mbbP_n\big(\psi_i(\bar{a})\big)
\]
and by Part~(i) the same holds if `$\mbbP_n$' is replaced by `$\tilde{\mbbP}_n$'.
Since $\varphi(\bar{x}) \sim_{\mbbG} \psi(\bar{x})$ we get
\[
\lim_{n\to\infty} \big| \tilde{\mbbP}_n\big(\{\mcA \in \mbW_n : \mcA(\psi(\bar{a})) \in I \}\big) - 
\mbbP_n\big(\{\mcA \in \mbW_n : \mcA(\varphi(\bar{a})) \in I \} \big) \big| \ = \ 0. \qedhere
\]

\section{Almost sure elimination of safe $CPL$-formulas when the distribution is induced
by a $coPLA^+$-network}\label{elimination of safe CPL-formulas}

\noindent
In this section we prove a  quantifier elimination result,
Theorem~\ref{quantifier elimination of safe formulas},
for the class of ``safe'' \emph{CPL}-formulas with 
respect to a sequence of probability distributions $\mbbP$ induced by a $coPLA^+(\sigma)$-network $\mbbG$, that is a 
$PLA^+(\sigma)$-network that (in its probability formulas) uses only strongly admissible, or continuous, 
aggregation functions.
Unlike the lifted Bayesian networks considered in 
Section~\ref{Lifted Bayesian networks}, such $\mbbG$ can
induce $\mbbP$ for which with high likelihood some (or all) relations in $\sigma$ are sparse.
For example, as explained in Example~\ref{example of PLA-networks},
if $\sigma$ contains a binary relation $R$ then, for any fixed $\alpha \in (0, 1)$, a $coPLA^+(\sigma)$-network can express that the probability that $R(x, y)$ holds is $n^{-\alpha}$ (independently of other edges) 
where $n$ is the cardinality of the domain.
Therefore $coPLA^+(\sigma)$-networks can induce the
probability distributions studied by Shelah and Spencer in \cite{SS} in relation to first-order logic.

$CPL$ (see Definition~\ref{definition of CPL}) 
is a natural query language since it can express queries about relative frequencies.
For example it can express the condition that the relative frequency of $\varphi_1(\bar{x})$
among tuples $\bar{x}$ that satisfy $\varphi_2(\bar{x})$ is at least $r$, or alternatively, at least
as large as the relative frequency of $\psi_1(\bar{x})$ among tuples $\bar{x}$ that satisfy $\psi_2(\bar{x})$.

Theorem~\ref{quantifier elimination of safe formulas} below provides a quantifier elimination result for such queries. 
For reasons that become clear below, its scope is restricted to 
formulas $\varphi(\bar{x}) \in CPL(\sigma)$ which are ``safe'' with respect to $\mbbG$. 
We begin with a discussion which motivates the definition of ``safe'' $CPL$-formulas further down.
As usual let $\sigma$ be a finite relational signature and $\mbW_n$ the set of all $\sigma$-structures
with domain $[n]$.
Let $\mbbG$ be a $coPLA^+(\sigma)$-network and
let $(\mbbP_n : n \in \mbbN^+)$ be the sequence of probability distributions induced by $\mbbG$.
Let $p(\bar{x}, \bar{y})$ and $q(\bar{x})$ be complete atomic $\sigma$-types such that $q \subseteq p$.
By Corollary~\ref{inductive assumption holds for all k and epsilon}~(4), there is $\alpha \in [0, 1]$, 
depending only on $p$, $q$ and $\mbbG$, such that for every $\varepsilon > 0$,
\[
\lim_{n\to\infty} \mbbP_n\big( \{ \mcA \in \mbW_n : \mcA \text{ is $(p, q, \alpha - \varepsilon)$-saturated 
and $(p, q, \alpha + \varepsilon)$-unsaturated} \}\big) \  = \ 1.
\]
We can call  $\alpha$ the {\em scaled saturation number} of the pair $(p, q)$, where ``scaled'' refers to the fact that
the $\bar{y}$-dimension of $p(\bar{x}, \bar{y})$ is taken into account in the definition of (un)saturation.

We can extend the idea of scaled saturation numbers to pairs $(\varphi(\bar{x}, \bar{y}), q(\bar{x}))$ where
$\varphi(\bar{x}, \bar{y}) \in CPL(\sigma)$ is quantifier-free and $q(\bar{x})$ is a complete atomic $\sigma$-type
such that $\forall \bar{x}, \bar{y} \big(\varphi(\bar{x}, \bar{y}) \rightarrow q(\bar{x})\big)$ is true in every finite $\sigma$-structure.
Then there are complete atomic $\sigma$-types $p_i(\bar{x}, \bar{y})$, $i = 1, \ldots, s$, 
such that $q \subseteq p_i$ for all $i$ and 
$\varphi(\bar{x}, \bar{y})$ is equivalent to $\bigvee_{i=1}^s p_i(\bar{x}, \bar{y})$ (where $p_i$ is identified with the conjunction
of formulas in $p_i$).
Let $d = \max\{\dim_{\bar{y}}(p_i) : i = 1, \ldots, s\}$. 
Without loss of generality, suppose that $p_1, \ldots, p_t$, $t \leq s$, enumerates, without repetition, all $p_i$ such that $\dim_{\bar{y}}(p_i) = d$.
Let $\alpha_i$ be the scaled saturation number of $(p_i, q)$ for each $i$.
If $d < |\bar{y}|$ then let $\alpha = 0$. 
Otherwise (i.e. if $d = |\bar{y}|$) let $\alpha = \alpha_1 + \ldots + \alpha_t$.
Call $\alpha$ the {\em saturation number} of the pair $(\varphi, q)$.
It now follows that
if $\varepsilon > 0$ and 
\[
\mbX^\varepsilon_n = 
\big\{ \mcA \in \mbW_n : \text{ if $\bar{a} \in q(\mcA)$ then }  
(\alpha - \varepsilon)n^{|\bar{y}|} \leq  |\varphi(\bar{a}, \mcA)| \leq (\alpha + \varepsilon)n^{|\bar{y}|} \big\}
\]
then 
$\lim_{n\to\infty} \mbbP_n(\mbX^\varepsilon_n) = 1$.

However, if 
$\psi(\bar{x}, \bar{y}), \theta(\bar{x}, \bar{y}) \in CPL(\sigma)$
then we can, in general, {\em not} guarantee that there is $\alpha \in [0, 1]$ such that for all
$\bar{a} \in [n]^{|\bar{x}|}$ and $\varepsilon > 0$,
the following probability converges to 1 as $n\to \infty$:
\[
\mbbP_n\big( \big\{\mcA \in \mbW_n : \text{ if $\theta(\bar{a}, \mcA) \neq \es$, then }
\big| |\psi(\bar{a}, \mcA) \cap \theta(\bar{a}, \mcA)| \big/ |\theta(\bar{a}, \mcA)| - \alpha \big| < \varepsilon \big\} \big).
\]
Therefore we cannot in general be sure that $\lim_{n\to\infty}\mbbP_n(\varphi(\bar{a}))$ exists if $\varphi(\bar{x})$ has the form
\[
\Big( \| \psi_1(\bar{x}, \bar{y}) \ | \ \theta_1(\bar{x}, \bar{y}) \|_{\bar{y}}  \ \geq \ 
\| \psi_2(\bar{x}, \bar{y}) \ | \ \theta_2(\bar{x}, \bar{y}) \|_{\bar{y}} \ + \ r \Big).
\]
This motivates considering $CPL$-formulas in which the value of terms of the form\\
$\| \psi(\bar{x}, \bar{y}) \ | \ \theta(\bar{x}, \bar{y}) \|_{\bar{y}}$ 
almost surely converges. This is made precise by the next two definitions.

\begin{defi}\label{definition of saturation number of a probability term} {\rm
Let $\psi(\bar{x}, \bar{y}), \theta(\bar{x}, \bar{y}) \in CPL(\sigma)$.\\
(i) We say that $\theta(\bar{x}, \bar{y})$ is {\em atomically $\bar{x}$-complete} if there is a complete atomic
$\sigma$-type $q(\bar{x})$ such that $\forall \bar{x}, \bar{y} (\theta(\bar{x}, \bar{y}) \rightarrow q(\bar{x}))$ is true
in every finite $\sigma$-structure.\\
(ii) We call
$\| \psi(\bar{x}, \bar{y}) \ | \ \theta(\bar{x}, \bar{y}) \|_{\bar{y}}$ a {\em conditional probability term} of $CPL(\sigma)$.\\
(iii) Suppose that $\theta(\bar{x}, \bar{y})$ is atomically $\bar{x}$-complete and that, for some 
complete atomic $\sigma$-type $q(\bar{x})$,
$\forall \bar{x}, \bar{y} (\theta(\bar{x}, \bar{y}) \rightarrow q(\bar{x}))$ is true in every finite $\sigma$-structure.
Let $\alpha \in [0, 1]$.
The {\em saturation number of the conditional probability term 
$\| \psi(\bar{x}, \bar{y}) \ | \ \theta(\bar{x}, \bar{y}) \|_{\bar{y}}$
exists and is $\alpha$} if,
for every $\varepsilon > 0$,
\begin{align*}
\lim_{n\to\infty}
\mbbP_n\bigg( \bigg\{ \mcA \in \mbW_n : &\text{ there is $\bar{a} \in [n]^{|\bar{x}|}$ such that $\mcA \models q(\bar{a})$ and } \\
&\bigg| \alpha - \frac{|\psi(\bar{a}, \mcA) \cap \theta(\bar{a}, \mcA)|}{|\theta(\bar{a}, \mcA)|} \bigg| > \varepsilon \bigg\} \bigg) \ = \ 0.
\end{align*}
}\end{defi}

\begin{defi}[Safe formula]\label{definition of safe formula}  {\rm 
A formula $\varphi(\bar{x}) \in CPL(\sigma)$ is {\em safe with respect to $\mbbG$} if the following hold:
\begin{enumerate}
\item $\varphi(\bar{x})$ does not contain $\forall$ or $\exists$.

\item For every subformula of $\varphi(\bar{x})$ of the form
\[
\Big( \| \psi_1(\bar{y}, \bar{z}) \ | \ \theta_1(\bar{y}, \bar{z}) \|_{\bar{z}}  \ \geq \ 
\| \psi_2(\bar{y}, \bar{z}) \ | \ \theta_2(\bar{y}, \bar{z}) \|_{\bar{z}} \ + \ r \Big)
\]
or
\[
\Big( r \ + \ \| \psi_1(\bar{y}, \bar{z}) \ | \ \theta_1(\bar{y}, \bar{z}) \|_{\bar{z}}  \ \geq \ 
\| \psi_2(\bar{y}, \bar{z}) \ | \ \theta_2(\bar{y}, \bar{z}) \|_{\bar{z}} \Big)
\]
and every complete atomic $\sigma$-type $q(\bar{y})$ such that
$\exists \bar{y}, \bar{z}, \bar{z}' (\theta_1(\bar{y}, \bar{z}) \wedge \theta_2(\bar{y}, \bar{z}') \wedge q(\bar{y}))$ 
has a finite model,
the saturation number of 
$\| \psi_i(\bar{y}, \bar{z}) \ | \ \theta_i(\bar{y}, \bar{z}) \wedge q(\bar{y})) \|_{\bar{y}}$
exists for $i = 1, 2$ and if it is denoted by $\alpha_i$, 
then $r \neq |\alpha_1 - \alpha_2|$.
\end{enumerate}
}\end{defi}

\noindent
Observe that if a formula is safe then every subformula of it is also safe.

\begin{rem}[Why first-order quantifiers are omitted in safe formulas]\label{remark on why first-order quantifiers are omitted} 
{\rm
As mentioned in Example~\ref{more about admissible functions}
the aggregation function $length_\alpha$, where $\alpha \in  (0, 1)$, is strongly admissible, or in other words, continuous.
Let $\sigma = \{R\}$ where $R$ is binary and let $\mbW_n$ be the set of all $\sigma$-structures with domain $[n]$.
We define a $coPLA^+(\sigma)$-network $\mbbG$ by 
letting  the probability formula $\theta_R(x, y)$ of $R$ be $length_\alpha(\psi(x, y, z) : z)$ where 
$\psi(x, y, z)$ is the formula $z = z$.
Let $(\mbbP_n : n \in \mbbN^+)$ be the sequence of probability distributions induced by $\mbbG$.
Then for every $n$, every $\mcA \in \mbW_n$ and all $a, b \in [n]$, $\mbbP_n(R(a, b)) = \mcA(\theta_R(a, b)) = 1/n^\alpha$,
where $\mcA$ is the unique $\es$-structure with domain $[n]$.
By a seminal result of Shelah and Spencer \cite[Theorem~2]{SS}, if $\alpha$ is rational then there is a first-order sentence $\varphi \in FO(\sigma)$ such that $\lim_{n\to\infty} \mbbP_n(\varphi)$ does not exist
(and recall that $FO(\sigma) \subseteq CPL(\sigma)$).
So if we want that ``safeness'' implies convergence we must omit the first-order quantifiers from safe formulas.

Strictly speaking, the result of Shelah and Spencer referred to is about undirected graphs.
But each $\mcA \in \mbW_n$ gives rise to an undirected graph by, for different $a, b \in [n]$ considering $\{a, b\}$ as an undirected
edge of the undirected graph induced by $\mcA$ if $\mcA \models a \neq b \wedge R(a, b) \wedge R(b, a)$.
In the sentence $\varphi$ constructed by Shelah and Spencer one then changes every subformula like $R(x, y)$ to 
`$x \neq y \wedge R(x, y) \wedge R(y, x)$'. Then this modified sentence $\varphi'$ will have the same probability with respect to $\mbbP_n$ as
$\varphi$ has with respect to the probability distribution on undirected graphs with vertex set $[n]$ which gives every
edge probability $1/n^{2\alpha}$ independently of the existence of other edges (and $2\alpha$ is rational if $\alpha$ is).
}\end{rem}

\noindent
The following technical lemma will be used in the proof of Theorem~\ref{quantifier elimination of safe formulas}
in combination with Lemma~\ref{elimination of one simple conditional quantifier}
below.

\begin{lem}\label{reduction to conditional quantifiers of simpler form}
Let $\psi_i(\bar{x}, \bar{y}), \theta_i(\bar{x}, \bar{y})) \in CPL(\sigma)$ for $i = 1, 2$.
Let $q_1(\bar{x}), \ldots, q_m(\bar{x})$ enumerate all complete atomic $\sigma$-types $q(\bar{x})$ such that
$\exists \bar{x}, \bar{y}, \bar{z} \big( \theta_1(\bar{x}, \bar{y}) \wedge \theta_2(\bar{x}, \bar{z}) \wedge q(\bar{x}) \big)$
has a finite model.
For every finite $\sigma$-structure $\mcA$ and every $\bar{a} \in [n]^{|\bar{x}|}$,
\begin{align}\label{the conditional formula holds}
&\mcA \models \Big( \| \psi_1(\bar{a}, \bar{y}) \ | \ \theta_1(\bar{a}, \bar{y}) \|_{\bar{y}}  \ \geq \ 
\| \psi_2(\bar{a}, \bar{y}) \ | \ \theta_2(\bar{a}, \bar{y}) \|_{\bar{y}} \ + \ r \Big) \\
&\text{if and only if} \nonumber \\
&\mcA \models \bigvee_{i=1}^m 
\Big( \| \psi_1(\bar{a}, \bar{y}) \ | \ \theta_1(\bar{a}, \bar{y}) \wedge q_i(\bar{a}) \|_{\bar{y}}  \ \geq \ 
\| \psi_2(\bar{a}, \bar{y}) \ | \ \theta_2(\bar{a}, \bar{y}) \wedge q_i(\bar{a}) \|_{\bar{y}} \ + \ r \Big) 
\label{the disjunction of simpler conditional formulas holds}
\end{align}
and similarly if `$+ \ r$' is moved to the left hand side of `$\geq$'.
\end{lem}

\proof
Let $\psi_i(\bar{x}, \bar{y}), \theta_i(\bar{x}, \bar{y})) \in CPL(\sigma)$ for $i = 1, 2$ and
let $q_1(\bar{x}), \ldots, q_m(\bar{x})$ enumerate all complete atomic $\sigma$-types $q(\bar{x})$ such that
$\exists \bar{x}, \bar{y}, \bar{z} \big( \theta_1(\bar{x}, \bar{y}) \wedge \theta_2(\bar{x}, \bar{y}) \wedge q(\bar{x}) \big)$
has a finite model.
Let $\mcA$ be a finite $\sigma$-structure and let $\bar{a} \in A^{|\bar{x}|}$.

First suppose that~(\ref{the conditional formula holds}) holds.
By the semantics of $CPL$, $\theta_1(\bar{a}, \mcA) \neq \es$ and $\theta_2(\bar{a}, \mcA) \neq \es$,
so $\mcA \models \exists \bar{y}, \bar{z} (\theta_1(\bar{a}, \bar{y}) \wedge \theta_2(\bar{a}, \bar{z}))$
and therefore there is $k$ such that $\mcA \models q_k(\bar{a})$. 
By~(\ref{the conditional formula holds}) and the semantics of $CPL$ we also have
\[
\frac{|\psi_1(\bar{a}, \mcA) \cap \theta_1(\bar{a}, \mcA)|}{|\theta_1(\bar{a}, \mcA)|} \ \geq \ 
\frac{|\psi_2(\bar{a}, \mcA) \cap \theta_2(\bar{a}, \mcA)|}{|\theta_2(\bar{a}, \mcA)|} + r.
\]
As  $\mcA \models q_k(\bar{a})$ the above implies that
\[
\frac{|\psi_1(\bar{a}, \mcA) \cap (\theta_1 \wedge q_k)(\bar{a}, \mcA)|}{|(\theta_1 \wedge q_k)(\bar{a}, \mcA)|} \ \geq \ 
\frac{|\psi_2(\bar{a}, \mcA) \cap (\theta_2 \wedge q_k)(\bar{a}, \mcA)|}{|(\theta_2 \wedge q_k)(\bar{a}, \mcA)|} + r.
\]
and hence (by the semantics of $CPL$) we get
\begin{equation}\label{a disjunct of the simpler formula holds}
\mcA \ \models \ \Big( \| \psi_1(\bar{a}, \bar{y}) \ | \ \theta_1(\bar{a}, \bar{y}) \wedge q_k(\bar{a}) \|_{\bar{y}}  \ \geq \ 
\| \psi_2(\bar{a}, \bar{y}) \ | \ \theta_2(\bar{a}, \bar{y}) \wedge q_k(\bar{a}) \|_{\bar{y}} \ + \ r \Big) 
\end{equation}
which implies~(\ref{the disjunction of simpler conditional formulas holds}).

Now suppose that~(\ref{the disjunction of simpler conditional formulas holds}) holds.
Then there is $k$ such that~(\ref{a disjunct of the simpler formula holds}) holds.
Hence 
\[
\mcA \models \exists \bar{y}, \bar{z} ((\theta_1 \wedge q_k)(\bar{a}, \bar{y}) \wedge (\theta_2 \wedge q_k)(\bar{a}, \bar{z}))
\]
and
\[
\frac{|\psi_1(\bar{a}, \mcA) \cap (\theta_1 \wedge q_k)(\bar{a}, \mcA)|}{|(\theta_1 \wedge q_k)(\bar{a}, \mcA)|} \ \geq \ 
\frac{|\psi_2(\bar{a}, \mcA) \cap (\theta_2 \wedge q_k)(\bar{a}, \mcA)|}{|(\theta_2 \wedge q_k)(\bar{a}, \mcA)|} + r.
\]
Since $(\theta_i \wedge q_k)(\bar{a}, \mcA) = \theta_i(\bar{a}, \mcA)$, for $i = 1, 2$,
it follows that
\[
\frac{|\psi_1(\bar{a}, \mcA) \cap \theta_1(\bar{a}, \mcA)|}{|\theta_1(\bar{a}, \mcA)|} \ \geq \ 
\frac{|\psi_2(\bar{a}, \mcA) \cap \theta_2(\bar{a}, \mcA)|}{|\theta_2(\bar{a}, \mcA)|} + r.
\]
so~(\ref{the conditional formula holds}) holds.

The case when `$+ \ r$' is to the left of `$\geq$' is proved similarly.
\qed

\begin{lem}\label{elimination of one simple conditional quantifier}
Suppose that $\varphi(\bar{x})$ has the form 
\begin{equation*}\label{the CPL-formula to be translated}
\Big( \| \psi_1(\bar{x}, \bar{y}) \ | \ \theta_1(\bar{x}, \bar{y}) \|_{\bar{y}}  \ \geq \ 
\| \psi_2(\bar{x}, \bar{y}) \ | \ \theta_2(\bar{x}, \bar{y}) \|_{\bar{y}} \ + \ r \Big)
\end{equation*}
Also suppose that, for some complete atomic $\sigma$-type $q(\bar{x})$, the sentences
$\forall \bar{x}, \bar{y} (\theta_1(\bar{x}, \bar{y}) \to q(\bar{x}))$ and
$\forall \bar{x}, \bar{y} (\theta_2(\bar{x}, \bar{y}) \to q(\bar{x}))$ hold in all finite $\sigma$-structures.
Furthermore, suppose that for $i = 1, 2$ the saturation number of
$\| \psi_i(\bar{x}, \bar{y}) \ | \ \theta_i(\bar{x}, \bar{y}) \|_{\bar{y}}$ exists and is $\alpha_i$.
Then:
\begin{enumerate}
\item If $\alpha_1 < \alpha_2 + r$ then 
$\varphi(\bar{x})$ is almost surely equivalent to~$\bot$.

\item If $\alpha_1 > \alpha_2 + r$ then 
$\varphi(\bar{x})$ is almost surely equivalent to $q(\bar{x})$.
\end{enumerate}
If $\varphi(\bar{x})$ is as above except that `$+ r$' is moved to the left of `$\geq$', 
then $\varphi(\bar{x})$ is almost surely equivalent to $\bot$ if 
$r + \alpha_1 < \alpha_2$,
and $\varphi(\bar{x})$ is almost surely equivalent to $q(\bar{x})$ if 
$r + \alpha_1 > \alpha_2$.
\end{lem}

\proof
The numbers $\alpha_1$ and $\alpha_2$ depend only on
$\theta_1, \theta_2, \psi_1, \psi_2, q$ and $\mbbG$.
Thus we can make a case distinction.
First suppose that 
\[
\alpha_1 < \alpha_2 + r.
\]
Let $\delta > 0$ be such that if 
$\alpha'_1$ and $\alpha'_2$
are within distance $\delta$ from $\alpha_1$ and $\alpha_2$, respectively, then
$\alpha'_1 < \alpha'_2 + r$.
Let $\mbX^\delta_n$ be the set of all $\mcA \in \mbW_n$ such that for every $\bar{a} \in [n]^{|\bar{x}|}$,
if $\mcA \models q(\bar{a})$, then, for $i = 1, 2$,
\begin{equation}\label{alpha-1 < alpha-2 + r}
\bigg| \alpha_i - \frac{|\psi_i(\bar{a}, \mcA) \cap \theta_i(\bar{a}, \mcA)|}{|\theta_i(\bar{a}, \mcA)|} \bigg| < \delta.
\end{equation}
Then $\lim_{n\to\infty}\mbbP_n(\mbX^\delta_n) = 1$ .

It now suffices to show that for all $\mcA \in \mbX^\delta_n$ and all $\bar{a} \in [n]^{|\bar{x}|}$,
$\mcA \not\models \varphi(\bar{a})$, because it then follows that $\varphi(\bar{x})$ is almost surely equivalent to $\bot$.
So let $\mcA \in \mbX^\delta_n$ and $\bar{a} \in [n]^{|\bar{x}|}$.

If $\mcA \not\models q(\bar{a})$ then $\theta_1(\bar{a}, \mcA) = \es$, so by the semantics of CPL, 
$\mcA \not\models \varphi(\bar{a})$.
Now suppose that $\mcA \models q(\bar{a})$.
By the choice of $\delta$, the definition of $\mbX^\delta_n$ and~(\ref{alpha-1 < alpha-2 + r}) we get
\[
\frac{|\theta_1(\bar{a}, \mcA) \cap \psi_1(\bar{a}, \mcA)|}{|\theta_1(\bar{a}, \mcA)|} \ < \ 
\frac{|\theta_2(\bar{a}, \mcA) \cap \psi_2(\bar{a}, \mcA)|}{|\theta_2(\bar{a}, \mcA)|} \ + \ r.
\]
Hence $\mcA \not\models \varphi(\bar{a})$ according to the semantics of $CPL$.
Now suppose that 
\[
\alpha_1 > \alpha_2 + r.
\]
Let $\delta > 0$ be such that if 
$\alpha'_1$ and $\alpha'_2$
are within distance $\delta$ from $\alpha_1$ and $\alpha_2$, respectively, then
$\alpha'_1 > \alpha'_2 + r$.
Let $\mbX^\delta_n$ be defined in the same way as in the previous case, so in particular
$\lim_{n\to\infty}\mbbP_n(\mbX^\delta_n) = 1$.
We will show that $\varphi(\bar{x})$ is almost surely equivalent to $q(\bar{x})$.
It suffices that prove that for all $\mcA \in \mbX^\delta_n$ and $\bar{a} \in [n]^{|\bar{x}|}$, 
$\mcA \models \varphi(\bar{a})$ if and only if $\mcA \models q(\bar{x})$.
Let $\mcA \in \mbX^\delta_n$ and $\bar{a} \in [n]^{|\bar{x}|}$.

If $\mcA \models \varphi(\bar{a})$ then, by the semantics of $CPL$, $\theta_1(\bar{a}, \mcA) \neq \es$
which (by the assumptions of the lemma) implies that $\mcA \models q(\bar{a})$.
Now suppose that $\mcA \models q(\bar{a})$.
By the choice of $\delta$ and definition of $\mbX^\delta_n$ it follows that,
for $i = 1, 2$,
\[
\frac{|\theta_1(\bar{a}, \mcA) \cap \psi_1(\bar{a}, \mcA)|}{|\theta_1(\bar{a}, \mcA)|} \ > \ 
\frac{|\theta_2(\bar{a}, \mcA) \cap \psi_2(\bar{a}, \mcA)|}{|\theta_2(\bar{a}, \mcA)|} \ + \ r.
\]
Hence $\mcA \models \varphi(\bar{a})$ according to the semantics of $CPL$.
The last statement, concerning the variant of $\varphi(\bar{x})$ where `$+ r$' 
is to the left of `$\leq$', is proved by straightforward modifications of the given arguments.
\qed

\begin{thm}\label{quantifier elimination of safe formulas}
Let $\mbbG$ be a $coPLA^+(\sigma)$-network.
If $\varphi(\bar{x}) \in CPL(\sigma)$ is safe with respect to $\mbbG$, 
then $\varphi(\bar{x})$ is almost surely equivalent to a quantifier-free formula.
\end{thm}

\proof 
Let $\varphi(\bar{x}) \in CPL(\sigma)$ be safe.
We use induction on the quantifier rank of formulas.
If $\varphi$ is quantifier-free we are done.
So suppose that $\varphi(\bar{x})$ is not quantifier-free and that every safe formula of lower quantifier rank than $\varphi$
is almost surely equivalent to a quantifier-free formula.
If $\varphi(\bar{x})$ has any of the forms 
$\neg \psi(\bar{x})$, $\psi(\bar{x}) \wedge \theta(\bar{x})$, $\psi(\bar{x}) \vee \theta(\bar{x})$, 
$\psi(\bar{x}) \to \theta(\bar{x})$ or $\psi (\bar{x}) \leftrightarrow \theta(\bar{x})$
and both $\psi(\bar{x})$ and $\theta(\bar{x})$ are almost surely equivalent to quantifier-free formulas,
then it clearly follows that $\varphi(\bar{x})$ is almost surely equivalent to a quantifier-free formula.

Suppose that $\varphi(\bar{x})$ has the form
\begin{equation*}
\Big( \| \psi_1(\bar{x}, \bar{y}) \ | \ \theta_1(\bar{x}, \bar{y}) \|_{\bar{y}}  \ \geq \ 
\| \psi_2(\bar{x}, \bar{y}) \ | \ \theta_2(\bar{x}, \bar{y}) \|_{\bar{y}} \ + \ r \Big).
\end{equation*}
Let $q_1(\bar{x}), \ldots, q_m(\bar{x})$ enumerate all complete atomic $\sigma$-types $q(\bar{x})$ such that
\[
\exists \bar{x}, \bar{y}, \bar{z} \big( \theta_1(\bar{x}, \bar{y}) \wedge \theta_2(\bar{x}, \bar{z}) \wedge q(\bar{x}) \big)
\]
has a finite model.
By Lemma~\ref{reduction to conditional quantifiers of simpler form}, 
$\varphi(\bar{x})$ is equivalent, in all finite $\sigma$-structures, to
\[
\varphi'(\bar{x}) \ := \ \bigvee_{k=1}^m 
\Big( \| \psi_1(\bar{x}, \bar{y}) \ | \ \theta_1(\bar{x}, \bar{y}) \wedge q_k(\bar{x}) \|_{\bar{y}}  \ \geq \ 
\| \psi_2(\bar{x}, \bar{y}) \ | \ \theta_2(\bar{x}, \bar{y}) \wedge q_k(\bar{x}) \|_{\bar{y}} \ + \ r \Big).
\]
Hence it suffices to prove that, for every $k = 1, \ldots, m$, 
\[
\varphi_k(\bar{x}) \ := \ \Big( \| \psi_1(\bar{x}, \bar{y}) \ | \ \theta_1(\bar{x}, \bar{y}) \wedge q_k(\bar{x}) \|_{\bar{y}}  \ \geq \ 
\| \psi_2(\bar{x}, \bar{y}) \ | \ \theta_2(\bar{x}, \bar{y}) \wedge q_k(\bar{x}) \|_{\bar{y}} \ + \ r \Big).
\]
is almost surely equivalent to a quantifier-free formula.
Since $\varphi(\bar{x})$ is assumed to be safe it follows that, for $i = 1, 2$, the saturation number of 
$\| \psi_1(\bar{x}, \bar{y}) \ | \ \theta_1(\bar{x}, \bar{y}) \wedge q_k(\bar{x}) \|_{\bar{y}}$ exists and if it is denoted by
$\alpha_i$, then $\alpha_1 \neq \alpha_2 + r$.
Now it follows from 
Lemma~\ref{elimination of one simple conditional quantifier} 
that $\varphi_k(\bar{x})$ is almost surely equivalent to a quantifier-free formula.
The case when $\varphi(\bar{x})$ has the form
\begin{equation*}
\Big( r \ + \ \| \psi_1(\bar{x}, \bar{y}) \ | \ \theta_1(\bar{x}, \bar{y}) \|_{\bar{y}}  \ \geq \ 
\| \psi_2(\bar{x}, \bar{y}) \ | \ \theta_2(\bar{x}, \bar{y}) \|_{\bar{y}} \Big)
\end{equation*}
is treated similarly.
\qed

\begin{cor}\label{convergence for sCPL}
Let $\mbbG$ be a $coPLA^+(\sigma)$-network.
If $\varphi(\bar{x}) \in CPL(\sigma)$ is safe with respect to $\mbbG$,
then, for every $m \in \mbbN^+$ and $\bar{a} \in [m]^{|\bar{x}|}$, $\lim_{n\to\infty}\mbbP_n(\varphi(\bar{a}))$ exists.
\end{cor}

\proof
Let $\mbbG$ be a $coPLA^+(\sigma)$-network and suppose that
$\varphi(\bar{x}) \in CPL(\sigma)$ is safe with respect to $\mbbG$.
By Theorem~\ref{quantifier elimination of safe formulas}
there is a quantifier-free $\varphi'(\bar{x}) \in CPL(\sigma)$ which is almost surely equivalent to $\varphi(\bar{x})$
with respect to $\mbbG$.
But then $\varphi'(\bar{x})$ is also an aggregation-free $PLA^+(\sigma)$-formula which only takes the values 0 or 1.
Now Corollary~\ref{corollary to main results} says that if $\bar{a} \in [m]^{|\bar{x}|}$ then 
$\lim_{n\to\infty}\mbbP_n(\varphi'(\bar{a}))$ exists. 
As $\varphi$ and $\varphi'$ are almost surely equivalent also 
$\lim_{n\to\infty}\mbbP_n(\varphi(\bar{a}))$ exists. 
\qed

\begin{cor}\label{sCPL formulas in G}
Let $\sigma$ be a finite relational signature and let
$\mbbG$ be a $PLA^+(\sigma)$-network
such that every probability formula of $\mbbG$ 
contains only strongly admissible aggregation functions. 
Let $\tilde{\mbbG}$ be the aggregation-free network from Corollary \ref{equivalence to an aggregation-free network} and 
let $(\tilde{\mbbP}_n : n \in \mbbN^+)$ be the sequence of probability distributions induced by $\tilde{\mbbG}$.
Let $\varphi(\bar{x})$ be an arbitrary \emph{CPL}-formula.
Then the following holds: 
\begin{enumerate}
\item[(i)] $\varphi(\bar{x})$ is safe with respect to $\mbbG$ if and only if it is safe with respect to $\tilde{\mbbG}$.
\item[(ii)] If $\varphi(\bar{x})$ is safe with respect to $\mbbG$, 
it is almost surely equivalent to a quantifier-free $\varphi'(\vec{x})$ with respect to $\mbbG$ if and only if 
it is almost surely equivalent  to $\varphi'(\bar{x})$ with respect to $\tilde{\mbbG}$.
\item[(iii)] If $\varphi(\bar{x})$ is safe with respect to $\mbbG$, then for every $m \in \mbbN^+$ and $\bar{a} \in [m]^{|\bar{x}|}$, 
\[
\lim_{n\to\infty}\mbbP_n(\varphi(\bar{a})) =  \lim_{n\to\infty}\tilde{\mbbP}_n(\varphi(\bar{a})).
\]
\end{enumerate}
\end{cor}

\proof
By Corollary  \ref{equivalence to an aggregation-free network}(ii)  saturation numbers are the same whether they are calculated with respect to $\mbbG$ or $\tilde{\mbbG}$. Since safety of a \emph{CPL}-formula depends only on the saturation numbers,
this immediately implies Point (i). 
Similarly, the formulas $\varphi'$ and $\varphi_k$ and their quantifier-free equivalents in the proof of Theorem \ref{quantifier elimination of safe formulas} depend only on $\varphi$ and the saturation numbers, which coincide for $\mbbG$ and $\tilde{\mbbG}$.
Since the limit computed in Corollary \ref{convergence for sCPL} depends only on the quantifier-free formula, this in turn implies that the limits coincide, whether they are computed in $\mbbP_n$ or in $\tilde{\mbbP}_n$. 
\qed

\section{Relative asymptotic expressivity of inference frameworks}\label{inference frameworks}

\noindent
In this section we tie together 
results of this article and in \cite{Kop20, KW1, SS} about sequences of probability distributions
defined by different formalisms and queries defined by different logics.
We do this by introducing the notion of ``inference framework'' and 
by comparing inference frameworks with the notion of ``asymptotically at least as expressive''.
Informally, an inference framework is a
set of pairs $(\mbbP, L)$
where $\mbbP$ is a sequence of probability distributions and $L$ is a logic (which may depend on $\mbbP$).
In all concrete cases considered here, an inference framework $\mbF$ will be defined by a particular formalism 
(and usually with particular restrictions on it) for defining sequences of probability distributions and
a logic $\mcL$ such that for each pair $(\mbbP, L) \in \mbF$, $L$ is a sublogic of $\mcL$ determined by $\mbbP$
(although often $L = \mcL$).
Theorem~\ref{partial order of inference frameworks},
illustrated by Figure~\ref{map of inference frameworks},
describes the relationships, with respect to the notion ``asymptotically at least as expressive'', 
between a number of inference frameworks
which have implicitly been considered in this article and in \cite{Kop20, KW1, SS}.

As usual we let $\sigma$ be a finite relational signature and, for each $n \in \mbbN^+$, 
$\mbW_n$ denotes the set of all $\sigma$-structures with domain $[n]$.
However, for simplicity, in some notation (like $\mbW_n$) we do not explicitly show the dependence on $\sigma$.
{\em We also assume that $\sigma$ is nonempty}, which is of course the interesting case. 
(All previous results in this article hold also for empty $\sigma$.)

\begin{defi}\label{definition of inference framework}{\rm
An  {\em inference framework (for $\sigma$)} is a class $\mbF$ of pairs $(\mbbP, L)$ where 
$L$ is a logic (for $\sigma$) and
$\mbbP = (\mbbP_n : n \in \mbbN^+)$ where each $\mbbP_n$ is a probability distribution on $\mbW_n$.
}\end{defi}

\begin{defi}\label{definition of relative expressibility of inference frameworks} {\rm
Let $\mbF$ and $\mbF'$ be inference frameworks for $\sigma$. \\ 
(i) We write $\mbF \preccurlyeq \mbF'$ and say that $\mbF'$ is {\em asymptotically at least as expressive as}
$\mbF$ 
if for every $(\mbbP, L) \in \mbF$ there is $(\mbbP', L') \in \mbF'$ such that 
$\mbbP \sim_{tv} \mbbP'$ and for every $\varphi(\bar{x}) \in L$ there is $\varphi'(\bar{x}) \in L'$ such that 
$\varphi'(\bar{x})$ is asymptotically equivalent to $\varphi(\bar{x})$ with respect to $\mbbP$.\\
(ii) By $\mbF \simeq \mbF'$ we mean that 
$\mbF \preccurlyeq \mbF'$ and $\mbF' \preccurlyeq \mbF$
and if this is the case we may say that  $\mbF$ and $\mbF'$ are {\em asymptotically equally expressive}.\\
(iii) By $\mbF \prec \mbF'$ we mean that $\mbF \preccurlyeq \mbF'$
and  $\mbF' \not\simeq \mbF$
}\end{defi}

\noindent
As we discussed in Section~\ref{Introduction} one can ask whether our notion of ``asymptotically at least as expressive''
is the most appropriate one. Especially, one can question why we require that $\mbbP$ and $\mbbP'$ in the definition above
are total variation equivalent although the logics $L$ and $L'$ involved may not be able to ``define'' all subsets of $\mbW_n$.
An important reason for our choice is that we think that ``asymptotically at least as expressive'' should be a transitive notion
(as shown in Lemma~\ref{transitivity of at least as expressive} below) 
and all other candidates of such a notion that we have considered, 
for example by weakening the assumption about total variation equivalence, are not transitive.

\begin{rem}\label{remark on at least as expressive} {\rm
According to the definition of `$\preccurlyeq$', if
$\mbF \preccurlyeq \mbF'$, $(\mbbP, L) \in \mbF$,
 and $\varphi(\bar{x}) \in L$,
then there are  $(\mbbP', L') \in \mbF'$ and $\varphi'(\bar{x}) \in L'$ such that 
$\mbbP \sim_{tv} \mbbP'$ and 
$\varphi'(\bar{x})$ and $\varphi(\bar{x})$ are asymptotically equivalent {\em with respect to $\mbbP$}.
It follows straightforwardly from the definitions of `$\sim_{tv}$' and `asymptotic equivalence' that
$\varphi'(\bar{x})$ is also asymptotically 
equivalent to $\varphi(\bar{x})$ with respect to $\mbbP'$, thus establishing a ``symmetry'' between
$\mbbP$ and $\mbbP'$.
}\end{rem}

\begin{lem}[Transitivity of $\preccurlyeq$]\label{transitivity of at least as expressive} 
Suppose that $\mbF$, $\mbF'$ and $\mbF''$ are inference frameworks.
If $\mbF \preccurlyeq \mbF'$ and $\mbF' \preccurlyeq \mbF''$, then
$\mbF \preccurlyeq \mbF''$.
\end{lem}

\proof
Suppose that $\mbF \preccurlyeq \mbF'$ and $\mbF' \preccurlyeq \mbF''$.
Let $(\mbbP, L) \in \mbF$ and let $\varphi(\bar{x}) \in L$.
By assumption there are  $(\mbbP', L') \in \mbF'$ and $\varphi'(\bar{x}) \in L'$ such that
$\mbbP \sim_{tv} \mbbP'$ and $\varphi'(\bar{x}) \in L'$ is
asymptotically equivalent to $\varphi(\bar{x})$ with respect to $\mbbP$.
By Remark~\ref{remark on at least as expressive},
$\varphi(\bar{x})$ and $\varphi'(\bar{x})$ are asymptotically equivalent with respect to $\mbbP'$.
By assumption there are $(\mbbP'', L'') \in \mbF''$ and $\varphi'' \in L''$ such that
$\mbbP' \sim_{tv} \mbbP''$ and $\varphi'(\bar{x}) \in L'$ is
asymptotically equivalent to $\varphi'(\bar{x})$ with respect to $\mbbP'$.
From $\mbbP \sim_{tv} \mbbP' \sim_{tv} \mbbP''$ it straightforwardly follows (from the defintion of $\sim_{tv}$) that
$\mbbP \sim_{tv} \mbbP''$. 
Since $\varphi(\bar{x})$ and $\varphi'(\bar{x})$ are asymptotically equivalent with respect to $\mbbP'$
and $\varphi'(\bar{x})$ and $\varphi''(\bar{x})$ are asymptotically equivalent with respect to $\mbbP'$ it follows 
(straightforwardly from the definition of ``asymptotic equivalence'') 
that $\varphi(\bar{x})$ and $\varphi''(\bar{x})$ are asymptotically equivalent with respect to $\mbbP'$.
Since $\mbbP \sim_{tv} \mbbP'$ it follows that $\varphi(\bar{x})$ and $\varphi''(\bar{x})$ are asymptotically
equivalent with respect to $\mbbP$.
This proves that $\mbF \preccurlyeq \mbF''$.
\qed
\\

\noindent
The following lemma will be used in the proof of
Theorem~\ref{partial order of inference frameworks}:

\begin{lem}\label{coPLAN properly includes qfLBN}
(i) There is a sequence of probability distributions $\mbbP$
which is induced by a $coPLA$-network and such that for every sequence of probability distributions $\mbbP'$
that is induced by a noncritial lifted Bayesian network $\mbbP \not\sim_{tv} \mbbP'$.\\
(ii) There are a sentence
$\varphi \in FO$, a sentence $\psi \in aPLA$ and a sequence of probability distributions
$(\mbbP_n : n \in \mbbN^+)$ induced by a $coPLA$-network such that 
$\lim_{n\to\infty}\mbbP_n(\varphi) = 
\lim_{n\to\infty}\mbbP_n(\{\mcA \in \mbW_n : \mcA(\psi) = 1 \})$ exist and is neither 0 nor 1.
\end{lem}

\proof
Let $R \in \sigma$ have arity $k$.
Let $\mbbG$ be a $PLA(\sigma)$-network such that the underlying DAG has no edges at all
and if $Q \in \sigma$ and $Q \neq R$, then $\theta_Q$ (the probability formula associated to $Q$) is `$0$', so
the probability of $Q(\bar{b})$ (for $\bar{b}$ with length matching the arity of $Q$) is zero.
Also let $F : [0, 1]^{<\omega} \to [0,1]$ be defined by $F(\bar{r}) = 1/|\bar{r}|^k$ and let $\theta_R(\bar{x})$,
where $\bar{x} = (x_1, \dots, x_k)$ and $\bar{y} = (y_1, \ldots, y_k)$, be the formula 
\[
F(\bigwedge_{i=1}^k (x_i = x_i) \wedge \bigwedge_{i=1}^k (y_i = y_i) : \bar{y} : 
\bigwedge_{i \neq j} (x_i \neq x_j)
\bigwedge_{i=1}^k\bigwedge_{j=1}^k
(x_i \neq y_j) \wedge 
\bigwedge_{i \neq j} (y_i \neq y_j)).
\]
It is straightforward to see that $F$ is strongly admissible so $\mbbG$ is a $coPLA$-network.
Let $(\mbbP_n : n \in \mbbN^+)$ be the sequence of probability distributions induced by $\mbbG$.
By the definition of $\mbbP_n$ it follows that if $\bar{a} \in [n]^k$ is a sequence of 
different elements, then
\[
\mbbP_n(R(\bar{a})) = \frac{1}{(n-k)(n-k-1) \ldots (n-2k+1)} \sim \frac{1}{n^k}.
\]
Define the random variable $X_n : \mbW_n \to \mbbR$ by letting $X_n(\mcA)$ be the number of $\bar{a} \in [n]^k$
such that $\mcA \models R(\bar{a})$.
Then $X_n$ is the sum of $(n-k)(n-k-1) \ldots (n-2k+1)$ independent $\{0, 1\}$-valued random variables 
where each one has probability
\[
\frac{1}{(n-k)(n-k-1) \ldots (n-2k+1)}
\]
of being 1.
By a rough version of the Poisson approximation \cite[Equation~(1)]{Nov} we get
\[
\big| \mbbP_n(X_n \geq 1) - (1 - e^{-1}) \big| \ \leq \ C/n^k
\]
where $C > 0$ is a constant.
It follows that $\lim_{n\to\infty}\mbbP_n(X_n \geq 1) = 1 - e^{-1}$.
Let $\varphi$ denote the first-order sentence 
$\exists \bar{x} \big(\bigwedge_{i \neq j} (x_i \neq x_j) \wedge R(\bar{x})\big)$ 
and
let $\psi$ denote the sentence 
\[
\max\big(R(\bar{x}) : \bar{x} : \bigwedge_{i\neq j} (x_i \neq x_j) \big)
\]
which belongs to $aPLA$ (because max is admissible).
Observe that for every finite structure $\mcA$, $\mcA(\psi)$ is either 0 or 1, and $\mcA(\psi) = 1$ if and only if $\mcA \models \varphi$.
Now let
\[
\mbX_n = \{ \mcA \in \mbW_n : \mcA \models \varphi\}.
\]
Then $\lim_{n\to\infty}\mbbP_n(\mbX_n) = 1 - e^{-1}$ so Part~(ii) is proved.
Let $(\mbbP'_n : n \in \mbbN^+)$ be induced by a noncritical lifted Bayesian network.
Since $\varphi$ is noncritical (because every first-order formula is noncritical),
Theorem~\ref{main result on quantifier elimination} 
implies that
$\mbbP'_n(\mbX_n)$ converges to either 0 or 1.
As $\lim_{n\to\infty}\mbbP_n(\mbX_n) = 1 - e^{-1}$ we get $(\mbbP_n : n \in \mbbN^+) \not\sim_{tv} (\mbbP'_n : n \in \mbbN^+)$ so Part~(i) is proved.
\qed
\\

\noindent
In the next definition we use notation for logics that was introduced in Definitions~\ref{definition of first-order formulas}
and~\ref{restrictions of PLA+}.
Also recall the informal discussion about {\em (non)critical $CPL$-formulas} just before 
Theorem~\ref{main result on quantifier elimination} and
Definition~\ref{definition of safe formula} of safe $CPL$-formulas.
If $\mbbG$ is a lifted Bayesian network and every aggregation formula of $\mbbG$ is noncritical with respect to $\mbbG$
then we say that $\mbbG$ is a {\em noncritical lifted Bayesian network}.
If all aggregation formulas of $\mbbG$ are quantifier-free then we say that $\mbbG$ is a 
{\em quantifier-free lifted Bayesian network}.

\begin{defi}[Concrete inference frameworks]\label{concrete inference frameworks}  {\rm
We define notation for some concrete inference frameworks via the table below which should be understood as follows.
The first column gives the name of the inference framework and the second column describes the pairs
$(\mbbP, L)$ that belong to the inference framework named on the same row. 
For notational simplicity the (arbitrary nonempty finite relational) signature $\sigma$ is suppressed in the notation.

}\end{defi}

\begin{center}
\renewcommand{\arraystretch}{1.3} 
\begin{tabularx}{\textwidth}{|>{\centering\arraybackslash}p{4cm}|X|}
\hline 
name  & contains all $(\mbbP, L)$ such that $\mbbP$ is induced by \\
\hline
$(\mb{qfLBN}, \mb{qfFO})$ & a quantifier-free lifted Bayesian network and $L = qfFO$ \\ 
 \hline
$(\mb{ncLBN}, \mb{FO})$ & a noncritical lifted Bayesian network and $L = FO$ \\
 \hline
$(\mb{ncLBN}, \mb{CPL})$ & a noncritical lifted Bayesian network and $L = CPL$ \\
\hline
$(\mb{ncLBN}, \mb{ncCPL})$ & a noncritical lifted Bayesian network $\mbbG$ and $L$ is the set of all \newline
  noncritical $CPL$-formulas with respect to $\mbbG$ \\
 \hline
$(\mb{afPLAN}, \mb{qfFO})$ & a $\text{\em afPLA}$-network and $L = qfFO$ \\
\hline
$(\mb{ncLBN},  \mb{afPLA})$ & a noncritical lifted Bayesian network and $L = \text{\em afPLA}$ \\
\hline
$(\mb{ncLBN},  \mb{aPLA})$ & a noncritical lifted Bayesian network and $L = aPLA$ \\
 \hline
$(\mb{aPLAN},  \mb{aPLA})$ & an $aPLA$-network and $L = aPLA$ \\
 \hline
$(\mb{aPLAN^+},  \mb{aPLA^+})$ & an $aPLA^+$-network and $L = aPLA^+$ \\
\hline
$(\mb{coPLAN^+},  \mb{qfFO})$ & a $coPLA^+$-network and $L = qfFO$ \\
 \hline
$(\mb{coPLAN^+},  \mb{FO})$ & a $coPLA^+$-network and $L = FO$ \\
 \hline
$(\mb{coPLAN^+},  \mb{sCPL})$ & a $coPLA^+$-network $\mbbG$ and $L$ is the set of all safe $CPL$-formulas \newline
  with respect to $\mbbG$ \\
 \hline
 $(\mb{coPLAN^+},  \mb{afPLA})$ & a $coPLA^+$-network and $L = \text{\em afPLA}$ \\
\hline
$(\mb{coPLAN^+},  \mb{coPLA^+})$ & a $coPLA^+$-network and $L = coPLA^+$ \\
\hline
\end{tabularx}
\renewcommand{\arraystretch}{1} 
\end{center}

\bigskip

\begin{thm}[Relative asymptotic expressivity of inference frameworks]\label{partial order of inference frameworks} \phantom{x}
\begin{enumerate}
\item $(\mb{afPLAN}, \mb{qfFO}) \simeq (\mb{qfLBN}, \mb{qfFO}) \simeq (\mb{ncLBN}, \mb{FO})$ \\
$\simeq (\mb{ncLBN}, \mb{ncCPL}) \prec (\mb{ncLBN}, \mb{CPL})$.

\item $(\mb{ncLBN}, \mb{ncCPL}) \prec (\mb{ncLBN}, \mb{afPLA}) \simeq (\mb{ncLBN}, \mb{aPLA})$.

\item $(\mb{afPLAN}, \mb{afPLA}) \simeq (\mb{ncLBN}, \mb{afPLA}) \simeq (\mb{ncLBN}, \mb{aPLA})$ \\
$\prec (\mb{aPLAN}, \mb{aPLA}) \preccurlyeq (\mb{aPLAN^+}, \mb{aPLA^+})$.

\item $(\mb{coPLAN^+}, \mb{FO}) \prec (\mb{aPLAN^+}, \mb{aPLA^+})$.

\item $(\mb{ncLBN}, \mb{ncCPL}) \prec (\mb{coPLAN^+}, \mb{qfFO}) \simeq (\mb{coPLAN^+}, \mb{sCPL})$ \\
$\prec (\mb{coPLAN^+}, \mb{FO})$.

\item $(\mb{coPLAN^+}, \mb{sCPL}) \prec (\mb{coPLAN^+}, \mb{afPLA}) \simeq (\mb{coPLAN^+}, \mb{coPLA^+})$ \\
$\prec (\mb{aPLAN^+}, \mb{aPLA^+})$.

\item $(\mb{afPLAN}, \mb{afPLA}) \prec (\mb{coPLAN^+}, \mb{coPLA^+})$.

\item $(\mb{ncLBN}, \mb{aPLA})$ and $(\mb{coPLAN^+}, \mb{FO})$ are incomparable with respect to $\preccurlyeq$.\\
$(\mb{coPLAN^+}, \mb{coPLA^+})$ and $(\mb{coPLAN^+}, \mb{FO})$ are incomparable with respect to $\preccurlyeq$.\\
$(\mb{ncLBN}, \mb{aPLA})$ and $(\mb{coPLAN^+}, \mb{sCPL})$ are incomparable with respect to $\preccurlyeq$.

\end{enumerate}
The main contents of this theorem are illustrated by  Figure~\ref{map of inference frameworks} (in the introduction).
\end{thm}

\proof
(1) It follows easily from the definitions that for every {\em afPLA}-network there is a quantifier-free 
lifted Bayesian network which induces the same distributions, and vice versa.
Therefore
$(\mb{afPLAN}, \mb{qfFO}) \simeq (\mb{qfLBN}, \mb{qfFO})$.
Since every first-order formula is noncritical with respect to every $CPL$-network we have 
$(\mb{ncLBN}, \mb{FO}) \preccurlyeq (\mb{ncLBN}, \mb{ncCPL})$.
Hence, to show the two remaining statements about `$\simeq$' it suffices to show that
$(\mb{ncLBN}, \mb{ncCPL}) \preccurlyeq (\mb{qfLBN}, \mb{qfFO})$.
But this follows from
theorem~\ref{main result on quantifier elimination}.

Clearly,
$(\mb{ncLBN}, \mb{ncCPL}) \preccurlyeq (\mb{ncLBN}, \mb{CPL})$.
In \cite[Remark~3.18]{Kop20} (which generalizes and example of Keisler and Lotfallah \cite[Proposition~3.1]{KL})
it is shown that a lifted Bayesian network $\mbbG$ and $CPL$-sentence $\varphi$ (which is critical with respect to $\mbbG$)
exist such that if $(\mbbP_n : n \in \mbbN^+)$ is induced by $\mbbG$, then $\mbbP_n(\varphi)$ does not converge as $n\to\infty$.
It follows from 
Theorem~\ref{main result on quantifier elimination} that
$(\mb{ncLBN}, \mb{ncCPL}) \prec (\mb{ncLBN}, \mb{CPL})$.

(2) We clearly have $qfFO \leq \text{\em afPLA}$ so $(\mb{ncLBN}, \mb{qfFO}) \preccurlyeq (\mb{ncLBN}, \mb{afPLA})$.
Since for example `$1/2$' is an $afPLA$-formula such that $\mcA(1/2) = 1/2$ for every finite structure $\mcA$, it follows that
$(\mb{ncLBN}, \mb{qfFO}) \prec (\mb{ncLBN}, \mb{afPLA})$. 
By Part~(1) we now get 
$(\mb{ncLBN}, \mb{ncCPL}) \prec (\mb{ncLBN}, \mb{afPLA})$.
By 
Theorem~\ref{elimination of aggregation functions for a lifted BN}
we also get 
$(\mb{ncLBN}, \mb{afPLA}) \simeq (\mb{ncLBN}, \mb{aPLA})$.

(3) From Part~(2) we have
$(\mb{ncLBN}, \mb{afPLA}) \simeq (\mb{ncLBN}, \mb{aPLA})$.
Theorem~\ref{main result on quantifier elimination}~(ii)
states that if $\mbbG$ is a noncritical lifted Bayesian network, then there is 
a quantifier-free lifted Bayesian network $\mbbG'$ such that the sequences of distributions induced
by $\mbbG$ and $\mbbG'$ are asymptotically total variation equivalent.
Since for every quantifier-free lifted Bayesian network there is an $afPLA$-network which induces the same 
sequence of distributions we get
$(\mb{ncLBN}, \mb{afPLA}) \simeq (\mb{afPLAN}, \mb{afPLA})$ and thus
$(\mb{afPLAN}, \mb{afPLA}) \simeq (\mb{ncLBN}, \mb{aPLA})$.
It follows that in order to show that 
$(\mb{ncLBN}, \mb{aPLA}) \prec (\mb{aPLAN}, \allowbreak \mb{aPLA})$
it suffices to show that
$(\mb{afPLAN}, \mb{afPLA}) \prec (\mb{aPLAN}, \mb{aPLA})$.

Clearly $(\mb{afPLAN}, \mb{afPLA}) \preccurlyeq (\mb{aPLAN}, \mb{aPLA})$.
Suppose towards a contradiction that $(\mb{aPLAN}, \mb{aPLA}) \preccurlyeq (\mb{afPLAN}, \mb{afPLA})$.
By Lemma~\ref{coPLAN properly includes qfLBN},
there are a sentence $\psi \in \mb{aPLA}$ and $(\mbbP_n : n \in \mbbN^+)$ induced by
an $aPLA$-network
such that $\lim_{n\to\infty}\mbbP_n(\{\mcA \in \mbW_n : \mcA(\psi) = 1\})$ exists and is neither 0 nor 1.
According to the assumption there is a aggregation-free $PLA$-sentence $\psi'$ which is asymptotically equivalent to $\psi$
with respect to $(\mbbP_n : n \in \mbbN^+)$. As $\psi'$ is aggregation-free it follows that there is $c \in [0, 1]$ such that
$\mcA(\psi') = c$ for {\em every} finite structure $\mcA$. 
But this contradicts that $\psi$ and $\psi'$ are asymptotically equivalent.
Finally, we clearly have 
$(\mb{aPLAN}, \mb{aPLA}) \preccurlyeq (\mb{aPLAN^+}, \mb{aPLA^+})$.

(4) The first-order quantifiers $\exists$ and $\forall$ can be expressed in $aPLA^+$ by using
the aggregation functions max and min.
Therefore $FO \leq aPLA^+$.
We also have $coPLA^+ \leq aPLA^+$, so
$(\mb{coPLAN^+}, \mb{FO}) \preccurlyeq (\mb{aPLAN^+}, \mb{aPLA^+})$.
It remains to prove that 
\[
(\mb{aPLAN^+}, \mb{aPLA^+}) \not\preccurlyeq (\mb{coPLAN^+}, \mb{FO}).
\]
But this follows because, for example, `$1/2$' is a sentence of $aPLA$
which takes the value $1/2$ in every finite structure while first-order sentences can only take the values 0 and 1.

(5) We get $(\mb{coPLAN^+}, \mb{qfFO}) \simeq (\mb{coPLAN^+}, \mb{sCPL})$ 
from Theorem~\ref{quantifier elimination of safe formulas}.
From~(1) we get
$
(\mb{ncLBN}, \mb{ncCPL}) \simeq (\mb{qfLBN}, \mb{qfFO}).
$
Since for every quantifier-free lifted Bayesian network there is a $coPLA^+$-network which induces
the same sequence of distributions
we get $(\mb{qfLBN}, \mb{qfFO}) \preccurlyeq (\mb{coPLAN^+}, \mb{qfFO})$.
So to prove that $(\mb{ncLBN}, \mb{ncCPL}) \prec$ \\ 
$(\mb{coPLAN^+}, \mb{qfFO})$ it 
suffices to show that 
\[
(\mb{qfLBN}, \mb{qfFO}) \prec (\mb{coPLAN^+}, \mb{qfFO}).
\]
But since every quantifier-free lifted Bayesian network is a noncritical lifted Bayesian network
this follows from 
Lemma~\ref{coPLAN properly includes qfLBN}.
It remains to prove that
$(\mb{coPLAN^+}, \mb{sCPL}) \prec (\mb{coPLAN^+}, \mb{FO})$.
As $(\mb{coPLAN^+}, \mb{qfFO}) \simeq (\mb{coPLAN^+}, \mb{sCPL})$ 
it suffices to prove that
$(\mb{coPLAN^+}, \mb{qfFO}) \prec (\mb{coPLAN^+}, \mb{FO})$.\\
Clearly 
$(\mb{coPLAN^+}, \mb{qfFO}) \preccurlyeq (\mb{coPLAN^+}, \mb{FO})$ so it remains to show that
\[
(\mb{coPLAN^+}, \mb{FO}) \not\preccurlyeq (\mb{coPLAN^+}, \mb{qfFO}).
\]
By Lemma~\ref{coPLAN properly includes qfLBN} there is a first-order
sentence $\varphi$ and $(\mbbP_n : n \in \mbbN^+)$ induced by a $coPLA$-network such that 
$\lim_{n\to\infty}\mbbP_n(\varphi)$ exists and is neither 0 nor 1.
If $(\mb{coPLAN^+}, \mb{FO}) \preccurlyeq (\mb{coPLAN^+}, \mb{qfFO})$ then there is
 a quantifier-free first-order sentence $\varphi'$
which is asymptotically equivalent to $\varphi$ with respect to $(\mbbP_n : n \in \mbbN^+)$, 
which (as the sentences are $\{0, 1\}$-valued) means that they
are almost surely equivalent with respect to $(\mbbP_n : n \in \mbbN^+)$.
But $\varphi'$ can only be $\bot$ or $\top$ so it means that $\lim_{n\to\infty}\mbbP_n(\varphi)$ is either 0 or 1,
a contradiction.

(6) From Theorem~\ref{quantifier elimination of safe formulas} we get
$(\mb{coPLAN^+}, \mb{sCPL}) \simeq (\mb{coPLAN^+}, \mb{qfFO})$.
Since every quantifier-free first-order formula is equivalent to an 
aggregation-free $PLA$-formula we get $(\mb{coPLAN^+}, \mb{qfFO}) \preccurlyeq (\mb{coPLAN^+}, \mb{afPLA})$.
If $\varphi \in \text{\em afPLA}$ is the formula `$1/2$', then for every $(\mbbP_n : n \in \mbbN^+)$ induced by a $coPLA$-network
there is {\em no} quantifier-free first-order sentence $\varphi'$ which is asymptotically equivalent to $\varphi$, because
first-order formulas only take the values 0 or 1.
It follows that
$(\mb{coPLAN^+}, \mb{qfFO}) \prec (\mb{coPLAN^+}, \mb{afPLA})$.
This together with $(\mb{coPLAN^+}, \mb{sCPL}) \simeq (\mb{coPLAN^+}, \mb{qfFO})$ gives
$(\mb{coPLAN^+}, \mb{sCPL}) \prec (\mb{coPLAN^+}, \mb{afPLA})$.
From Theorem~\ref{elimination of strongly admissible aggregation functions}
we get 
\[
(\mb{coPLAN^+}, \mb{ afPLA}) \simeq (\mb{coPLAN^+}, \mb{coPLA^+}).
\]
It remains to prove that 
$(\mb{coPLAN^+}, \mb{coPLA^+}) \prec (\mb{aPLAN^+}, \mb{aPLA^+})$.
Since 
\[(\mb{coPLAN^+}, \mb{afPLA}) \simeq (\mb{coPLAN^+}, \mb{coPLA^+})
\] 
it suffices to prove that
$(\mb{coPLAN^+}, \mb{afPLA}) \prec (\mb{aPLAN^+}, \mb{aPLA^+})$.
But this follows from Part~(ii) of 
Lemma~\ref{coPLAN properly includes qfLBN}
because every aggregation-free $PLA$-sentence has the same value in every structure.

(7) We clearly have $(\mb{afPLAN}, \mb{afPLA}) \preccurlyeq (\mb{coPLAN^+}, \mb{coPLA^+})$
so we only show that $(\mb{coPLAN^+}, \mb{coPLA^+}) \not\preccurlyeq (\mb{afPLAN}, \mb{afPLA})$.\\
But since $(\mb{afPLAN}, \mb{afPLA}) \simeq (\mb{ncLBN}, \mb{afPLA})$ by Part~(2) this follows from
Lemma~\ref{coPLAN properly includes qfLBN}.

(8) The statements
\begin{align*}
&(\mb{ncLBN}, \mb{aPLA}) \not\preccurlyeq (\mb{coPLAN^+}, \mb{FO}), \\
&(\mb{coPLAN^+}, \mb{coPLA^+}) \not\preccurlyeq (\mb{coPLAN^+}, \mb{FO}), \text{ and}\\
&(\mb{ncLBN}, \mb{aPLA})  \not\preccurlyeq (\mb{coPLAN^+}, \mb{sCPL})
\end{align*}
follow since $aPLA$ and $coPLA^+$
contain for example the sentence `$1/2$', the value of which is always $1/2$,
while sentences in $FO$ and in $CPL$ can only take the values 0 and~1.
The statements
\begin{align*}
&(\mb{coPLAN^+}, \mb{FO}) \not\preccurlyeq (\mb{ncLBN}, \mb{aPLA}), \\
&(\mb{coPLAN^+}, \mb{FO}) \not\preccurlyeq (\mb{coPLAN^+}, \mb{coPLA^+}), \text{ and}\\
&(\mb{coPLAN^+}, \mb{sCPL}) \not\preccurlyeq (\mb{ncLBN}, \mb{aPLA})
\end{align*}
follow from
Lemma~\ref{coPLAN properly includes qfLBN}
and the above proved fact that 
$(\mb{coPLAN^+}, \mb{afPLA}) \simeq \\ (\mb{coPLAN^+}, \mb{coPLA^+})$.
\qed

\section{Conclusion}\label{Conclusion}

\noindent
We introduced the probability logic $coPLA^+$ which allows for probability formulas built using strongly admissible aggregation functions,
which satisfy stronger continuity requirements than the admissible aggregation functions used in the $aPLA$-formulas
studied by \cite{KW1}.
The stricter requirements reduce expressivity, ruling out the classical existential and universal quantifiers and their 
multivalued counterparts, maximum and minimum.
However, $coPLA^+$ covers for example the arithmetic and geometric mean as aggregation functions, which can model a dependence of one relation on the relative frequency of another,
and can also model (directed versions of) the sparse random graphs studied by Shelah and Spencer \cite{SS}.
We showed that queries expressible in $coPLA^+$ are asymptotically equivalent to aggregation-free queries with respect to a given $coPLA^+$-network.
An analogous quantifier elimination result with respect to $coPLA^+$-networks was shown to hold for safe formulas of conditional probability logic,
which may include conditional relative frequency quantifiers but not classical universal or existential quantification.
As a special case we obtained convergence results for expressive probability logics  even over such random graphs where first-order formulas can have divergent probabilities, such as those studied by Shelah and Spencer \cite{SS}.
Finally, we integrated the new results obtained here and previous results in 
\cite{Kop20, KW1, SS} by introducing the notion of an inference framework.
We classified several inference frameworks related to the present work and to \cite{Kop20, KW1, SS}
by means of their ``relative asymptotic expressivity'' which is defined using the
transitive notion of one inference framework being asymptotically at least as expressive as another.

\section*{Acknowledgement}
Parts of this work were accomplished during a research visit of the second author to the University of Uppsala, made possible by the support of LMUexcellent, funded by the Federal
Ministry of Education and Research (BMBF) and the Free State of Bavaria under the Excellence Strategy of the Federal Government and the Länder.

\bibliography{koponen}

\newcommand{\etalchar}[1]{$^{#1}$}
\begin{thebibliography}{DRKNP16}

\bibitem[AS00]{AlonSpencer}
N.~Alon and J.~H. Spencer.
\newblock {\em {T}he {P}robabilistic {M}ethod}.
\newblock John Wiley \& Sons, 2 edition, 2000.

\bibitem[Ber08]{Ber}
M.~Bergmann.
\newblock {\em {A}n {I}ntroduction to {M}any-{V}alued and {F}uzzy
  {L}ogic:{S}emantics, {A}lgebras, and {D}erivation {S}ystems}.
\newblock Cambridge University Press, 2008.

\bibitem[BP98]{BP}
S.~Brin and L.~Page.
\newblock The anatomy of a large-scale hypertextual web search engine.
\newblock {\em Computer Networks and ISDN Systems}, 30:107--117, 1998.

\bibitem[Che52]{Chernoff}
H.~Chernoff.
\newblock A measure of the asymptotic efficiency for tests of a hypothesis
  based on the sum of observations.
\newblock {\em Annals of Mathemathical Statistics}, 23:493--509, 1952.

\bibitem[CM19]{CM}
F.~G. Cozman and D.~D. Maua.
\newblock The finite model theory of bayesian network specifications:
  Descriptive complexity and zero/one laws.
\newblock {\em International Journal of Approximate Reasoning}, 110:107--126,
  2019.

\bibitem[DRKNP16]{DKNP}
L.~De~Raedt, K.~Kersting, S.~Natarajan, and D.~Poole.
\newblock {\em {S}tatistical {R}elational {A}rtificial {I}ntelligence: {L}ogic,
  {P}robability, and {C}omputation}, volume~32 of {\em Synthesis Lectures on
  Artificial Intelligence and Machine Learning}.
\newblock Morgan \& Claypool Publishers, 2016.

\bibitem[Ebb85]{Ebb85}
H.-D. Ebbinghaus.
\newblock Extended logics: The general framework.
\newblock In J.~Barwise and S.~Feferman, editors, {\em Model-Theoretic Logics},
  pages 25--76. Springer-Verlag, 1985.

\bibitem[EF99]{EF}
H-D. Ebbinghaus and J.~Flum.
\newblock {\em {F}inite {M}odel {T}heory}, volume~2.
\newblock Springer-Verlag, Berlin Heidelberg New York, 1999.
\newblock \href {https://doi.org/10.1007/3-540-28788-4}
  {\path{doi:10.1007/3-540-28788-4}}.

\bibitem[FWF23]{WRF23}
D.~Ravdin F.~Weitk\"amper and R.~Fabry.
\newblock Statistical relational learning with scaled weight parameters.
\newblock In F.~Lissi E.~Bellodi and R.~Zese, editors, {\em Inductive Logic
  Programming - 32nd International Conference 2023}, pages 139---153. Springer,
  2023.

\bibitem[GT07]{GT}
L.~Getoor and B.~Taskar.
\newblock {\em {I}ntroduction to {S}tatistical {R}elational {L}earning}.
\newblock The MIT Press, 2007.

\bibitem[GVdBP21]{VKNP}
Sriraam~Natarajan Guy Van~den Broeck, Kristian~Kersting and David Poole,
  editors.
\newblock {\em An Introduction to Lifted Probabilistic Inference}.
\newblock The MIT Press, 2021.

\bibitem[Jae97]{Jae97}
Manfred Jaeger.
\newblock Relational {B}ayesian networks.
\newblock In Dan Geiger and Prakash~P. Shenoy, editors, {\em {UAI} 97:
  Proceedings of the Thirteenth Conference on Uncertainty in Artificial
  Intelligence, Brown University, Providence, Rhode Island, USA, August 1-3,
  1997}, pages 266--273. Morgan Kaufmann, 1997.
\newblock URL:
  \url{https://dslpitt.org/uai/displayArticleDetails.jsp?mmnu=1\&smnu=2\&article\_id=320\&proceeding\_id=13}.

\bibitem[Jae98]{Jae98a}
Manfred Jaeger.
\newblock Convergence results for relational {B}ayesian networks.
\newblock In Dan Geiger and Prakash~P. Shenoy, editors, {\em Proceedings.
  Thirteenth Annual IEEE Symposium on Logic in Computer Science (LICS 98)},
  pages 44--55, 1998.
\newblock \href {https://doi.org/10.1109/LICS.1998.705642}
  {\path{doi:10.1109/LICS.1998.705642}}.

\bibitem[JW02]{JW}
G.~Jeh and J.~Widom.
\newblock Simrank: A measure of structural-context similarity.
\newblock In D.~Hand, D.~A. Keim, and R.~NG, editors, {\em KDD'02: Proceedings
  of the eighth ACM SIGKDD international conference on Knowledge discovery and
  data mining}, pages 538--543. ACM Press, 2002.

\bibitem[KBK{\etalchar{+}}14]{KBKNP}
Seyed~Mehran Kazemi, David Buchman, Kristian Kersting, Sriraam Natarajan, and
  David Poole.
\newblock Relational logistic regression.
\newblock In Chitta Baral, Giuseppe~De Giacomo, and Thomas Eiter, editors, {\em
  Principles of Knowledge Representation and Reasoning: Proceedings of the
  Fourteenth International Conference, {KR} 2014, Vienna, Austria, July 20-24,
  2014}. {AAAI} Press, 2014.
\newblock URL: \url{http://www.aaai.org/ocs/index.php/KR/KR14/paper/view/8013}.

\bibitem[KDR07]{KdR}
K.~Kersting and L.~De~Raedt.
\newblock Bayesian logic programming.
\newblock In L.~Getoor and B.~Taskar, editors, {\em Introduction to Statistical
  relational learning}, page 291. MIT Press, Cambridge MA, 2007.

\bibitem[KL09]{KL}
H.~J. Keisler and W.~B. Lotfallah.
\newblock Almost everywhere elimination of probability quantifiers.
\newblock {\em The Journal of Symbolic Logic}, 74:1121--1142, 2009.

\bibitem[KMG15]{KMG}
A.~Kimmig, L.~Mihalkova, and L.~Getoor.
\newblock Lifted graphical models: a survey.
\newblock {\em Machine Learning}, 99(1):1--45, Apr 2015.
\newblock \href {https://doi.org/10.1007/s10994-014-5443-2}
  {\path{doi:10.1007/s10994-014-5443-2}}.

\bibitem[Kop20]{Kop20}
Vera Koponen.
\newblock Conditional probability logic, lifted {B}ayesian networks, and almost
  sure quantifier elimination.
\newblock {\em Theoretical Computer Science}, 848:1--27, Dec 2020.
\newblock \href {https://doi.org/10.1016/j.tcs.2020.08.006}
  {\path{doi:10.1016/j.tcs.2020.08.006}}.

\bibitem[KW23]{KW1}
V.~Koponen and F.~Weitkämper.
\newblock Asymptotic elimination of partially continuous aggregation functions
  in directed graphical models.
\newblock {\em Information and Computation}, 293:105061, 2023.
\newblock \href {https://doi.org/10.1016/j.ic.2023.105061}
  {\path{doi:10.1016/j.ic.2023.105061}}.

\bibitem[MBGS19]{MBGS}
H.~Mittal, A.~Bhardwaj, V.~Gogate, and P.~Singla.
\newblock Domain-size aware {M}arkov logic networks.
\newblock In Kamalika Chaudhuri and Masashi Sugiyama, editors, {\em The 22nd
  International Conference on Artificial Intelligence and Statistics, {AISTATS}
  2019, 16-18 April 2019, Naha, Okinawa, Japan}, volume~89 of {\em Proceedings
  of Machine Learning Research}, pages 3216--3224. {PMLR}, 2019.
\newblock URL: \url{http://proceedings.mlr.press/v89/mittal19a.html}.

\bibitem[Nov19]{Nov}
S.~Y. Novak.
\newblock On the accuracy of poisson approximation.
\newblock {\em Extremes}, 22(4):729--748, Dec 2019.
\newblock \href {https://doi.org/10.1007/s10687-019-00350-6}
  {\path{doi:10.1007/s10687-019-00350-6}}.

\bibitem[PBK{\etalchar{+}}14]{PBKKN}
D.~Poole, D.~Buchman, S.~M. Kazemi, K.~Kersting, and S.~Natarajan.
\newblock Population size extrapolation in relational probabilistic modelling.
\newblock In Umberto Straccia and Andrea Cal{\`{\i}}, editors, {\em Scalable
  Uncertainty Management - 8th International Conference, {SUM} 2014, Oxford,
  UK, September 15-17, 2014. Proceedings}, volume 8720 of {\em Lecture Notes in
  Computer Science}, pages 292--305. Springer, 2014.
\newblock \href {https://doi.org/10.1007/978-3-319-11508-5\_25}
  {\path{doi:10.1007/978-3-319-11508-5\_25}}.

\bibitem[SS88]{SS}
S.~Shelah and J.~Spencer.
\newblock Zero-one laws for sparse random graphs.
\newblock {\em Journal of the American Mathematical Society}, 1(1):97--115,
  2022/05/07/ 1988.
\newblock Full publication date: Jan., 1988.
\newblock \href {https://doi.org/10.2307/1990968} {\path{doi:10.2307/1990968}}.

\bibitem[Wei21]{Wei21}
F.~Weitk\"amper.
\newblock An asymptotic analysis of probabilistic logic programming, with
  implications for expressing projective families of distributions.
\newblock {\em Theory and Practice of Logic Programming}, 21(6):802–817,
  2021.
\newblock \href {https://doi.org/10.1017/S1471068421000314}
  {\path{doi:10.1017/S1471068421000314}}.

\bibitem[Wei24]{Wei24}
F.~Weitk\"amper.
\newblock Functional lifted bayesian networks.
\newblock In S.~Muggleton and A.~Tamaddoni-Nezhad, editors, {\em Proceedings of
  the 31st International Conference of Inductive Logic Programming 2022}, pages
  142--156. Springer, 2024.

\end{thebibliography}
\bibliographystyle{alphaurl}

\vfill

\end{document}